\documentclass[universe,article,submit,moreauthors,pdflatex,10pt,a4paper]{mdpi} 
\graphicspath{{Definitions/}}
\usepackage{xcolor}
\usepackage{mathalfa}
\usepackage{hyperref}
\usepackage{amsmath,amssymb}
\usepackage{multirow}
\DeclareFontFamily{U}{rcjhbltx}{}
\DeclareFontShape{U}{rcjhbltx}{m}{n}{<->rcjhbltx}{}
\DeclareSymbolFont{hebrewletters}{U}{rcjhbltx}{m}{n}
\let\aleph\relax\let\beth\relax
\let\gimel\relax\let\daleth\relax
\DeclareMathSymbol{\aleph}{\mathord}{hebrewletters}{39}
\DeclareMathSymbol{\beth}{\mathord}{hebrewletters}{98}
\DeclareMathSymbol{\gimel}{\mathord}{hebrewletters}{103}
\DeclareMathSymbol{\daleth}{\mathord}{hebrewletters}{100}
\DeclareMathSymbol{\lamed}{\mathord}{hebrewletters}{108}
\DeclareMathSymbol{\mem}{\mathord}{hebrewletters}{109}
\DeclareMathSymbol{\ayin}{\mathord}{hebrewletters}{96}
\DeclareMathSymbol{\tsadi}{\mathord}{hebrewletters}{118}
\DeclareMathSymbol{\qof}{\mathord}{hebrewletters}{113}
\DeclareMathSymbol{\shin}{\mathord}{hebrewletters}{152}
\preto{\abstractkeywords}{\nolinenumbers}
\def\aap{A\&A\/}
\def\apjs{ApJS}
\def\apj{ApJ}
\def\aj{AJ}
\def\nat{Nat}
\def\apjl{ApJL}
\def\apss{ApSS}
\def\mnras{MNRAS}

\def\l{$\lambda$}

\def\mbh{$M_{\rm BH}$\/}
\def\nh{$n_{\mathrm{H}}$\/}

\def\rfe{$R_{\rm FeII}$}

\def\feiiq{\rm Fe{\sc ii}$\lambda$4570\/}

\def\ltsima{$\; \buildrel < \over \sim \;$}
\def\ltsim{\lower.5ex\hbox{\ltsima}}  
\def\gtsima{$\; \buildrel > \over \sim \;$}

\def\gtsim{\lower.5ex\hbox{\gtsima}}

\def\civ{{\sc{Civ}}\/}

\def\cm3{cm$^{-3}$\/}
\def\hb{{\sc{H}}$\beta$\/}

\def\hbbc{{\sc{H}}$\beta_{\rm BC}$\/}

\def\mgii{{Mg\sc{ii}}$\lambda$2800\/}

\def\oiiiopt{{\sc{[Oiii]}}$\lambda\lambda$4959,5007\/}

\def\hei{{\sc{Hei}}$\lambda$5016\/}

\def\feiiopt{{Fe \sc{ii}}$_{\rm opt}$\/}
\def\feii{{Fe\sc{ii}}\/}

\def\fe{{\sc{Fe}}\/}

\def\fe76087{{\sc [Fe vii]}$\lambda$6087\/}
\def\oiii{{\sc [Oiii]}$\lambda$5007}

\def\kms{km~s$^{-1}$}

\def\ergss{erg\, s$^{-1}$\/}

\def\heiiopt{He{\sc ii}$\lambda$4686\/}
\def\heii{{\sc H}e{\sc ii}\/}
\def\hei{{\sc H}e{\sc i}\/}
\def\rb{$r_{\rm BLR}$\/}

\usepackage{enumerate}
\definecolor{darkorange}{rgb}{1,0.612,0}
\definecolor{aquamarine}{rgb}{0.498,1,0.8314}

\Title{Optical singly-ionized iron emission in radio-quiet and relativistically jetted active galactic nuclei}

\Author{{Paola Marziani}$^{1}$\orcidA{}, Marco Berton$^{2,3,4}$\orcidB{}\footnote{ESO fellow}, Swayamtrupta Panda$^{5,6,7}$\orcidC{}\footnote{CNPq fellow}, and Edi Bon$^{8}$\orcidD{}}  
\AuthorNames{Paola Marziani}
\address{$^{1}$ \quad National Institute for Astrophysics (INAF), Astronomical Observatory of Padova, IT-35122 Padova, Italy\\
$^{2}$ \quad European Southern Observatory (ESO), Alonso de C\'ordova 3107, Casilla 19, Santiago 19001, Chile\\
$^{3}$ \quad  Finnish Centre for Astronomy with ESO (FINCA), University of Turku, Vesilinnantie 5, FI-20014 University of Turku, Finland\\  
$^{4}$ \quad Aalto University Mets\"ahovi Radio Observatory, Mets\"ahovintie 114, FI-02540 Kylm\"al\"a, Finland\\
$^{5}$ \quad Center for Theoretical Physics, Polish Academy of Sciences, Al. Lotnik{\'o}w 32/46, 02-668 Warsaw, Poland\\
$^{6}$ \quad Nicolaus Copernicus Astronomical Center, Polish Academy of Sciences, ul. Bartycka 18, 00-716 Warsaw, Poland\\
$^{7}$ \quad Laborat\'orio Nacional de Astrof\'isica, R. dos Estados Unidos, 154 - Na\c{c}\~oes, Itajub\'a - MG, 37504-364, Brazil\\
$^{8}$ \quad Belgrade Astronomical Observatory{, Volgina 7, 11060} Belgrade, Serbia}
\corres{Correspondence: paola.marziani@inaf.it; Tel.: +39-0498293415}
\abstract{The issue of the difference between optical and UV properties of radio-quiet and radio-loud (relativistically "jetted") active galactic nuclei (AGN) is a long standing one,  related to the fundamental question of why a minority of powerful AGN possess strong radio emission due to relativistic ejections. This paper examines  a particular aspect: the singly-ionized iron emission in the spectral range 4400 — 5600 \AA, where the prominent H{\sc I} H$\beta$\ and \oiiiopt\  lines are  also observed. { We present a detailed  comparison of the relative intensity of \feii\ multiplets in the spectral types of the quasar main sequence where most jetted sources are found, and afterwards discuss radio-loud narrow-line Seyfert 1 (NLSy1) nuclei with $\gamma$-ray detection and with prominent \feii\ emission.} { An \feii\ template based on I Zw 1 provides an accurate representation of the optical \feii\ emission for RQ and, with some caveats, also for RL sources. }
{CLOUDY photoionization simulations indicate that the observed spectral energy distribution can account for the modest \feii\ emission observed in composite radio-loud spectra. However, spectral energy differences alone cannot account for the stronger \feii\ emission observed  in radio-quiet sources, for similar physical parameters. As for RL NLSy1s, they do not seem to behave like other RL sources, likely because of their different physical properties, that could be ultimately associated with a higher Eddington ratio. }}
\begin{document}

\section{Introduction}

The wide majority of active galactic nuclei (AGN) are characterised by the presence of broad and narrow optical and UV lines emitted by ionic species over a wide range of ionization potential $\chi$.\footnote{For an introduction to AGN  spectra and the  interpretation of their spectra in  physical terms, see e.g., \cite{netzer90,peterson97,osterbrockferland06}, and references therein.} Restricting the attention to broad lines, type-1 AGN\footnote{{ Type-2 AGN do not show broad permitted lines in natural light; they are believed to be mostly obscured type-1 and will not be further considered here because they lack the diagnostics offered by the broad lines measurements.}} spectra invariably show the same low ionization lines ($\chi \lesssim$ 20 eV) that include  HI Balmer lines (\hb, H$\alpha$), \mgii, the CaII {IR} Triplet, and \feii\ features, in addition to higher ionization lines (with CIV$\lambda$1549 as prototypical representative in the rest-frame UV domain). 

\begin{figure}[ht!]
\centering
\vspace{3cm}
\includegraphics[width=6.5cm]{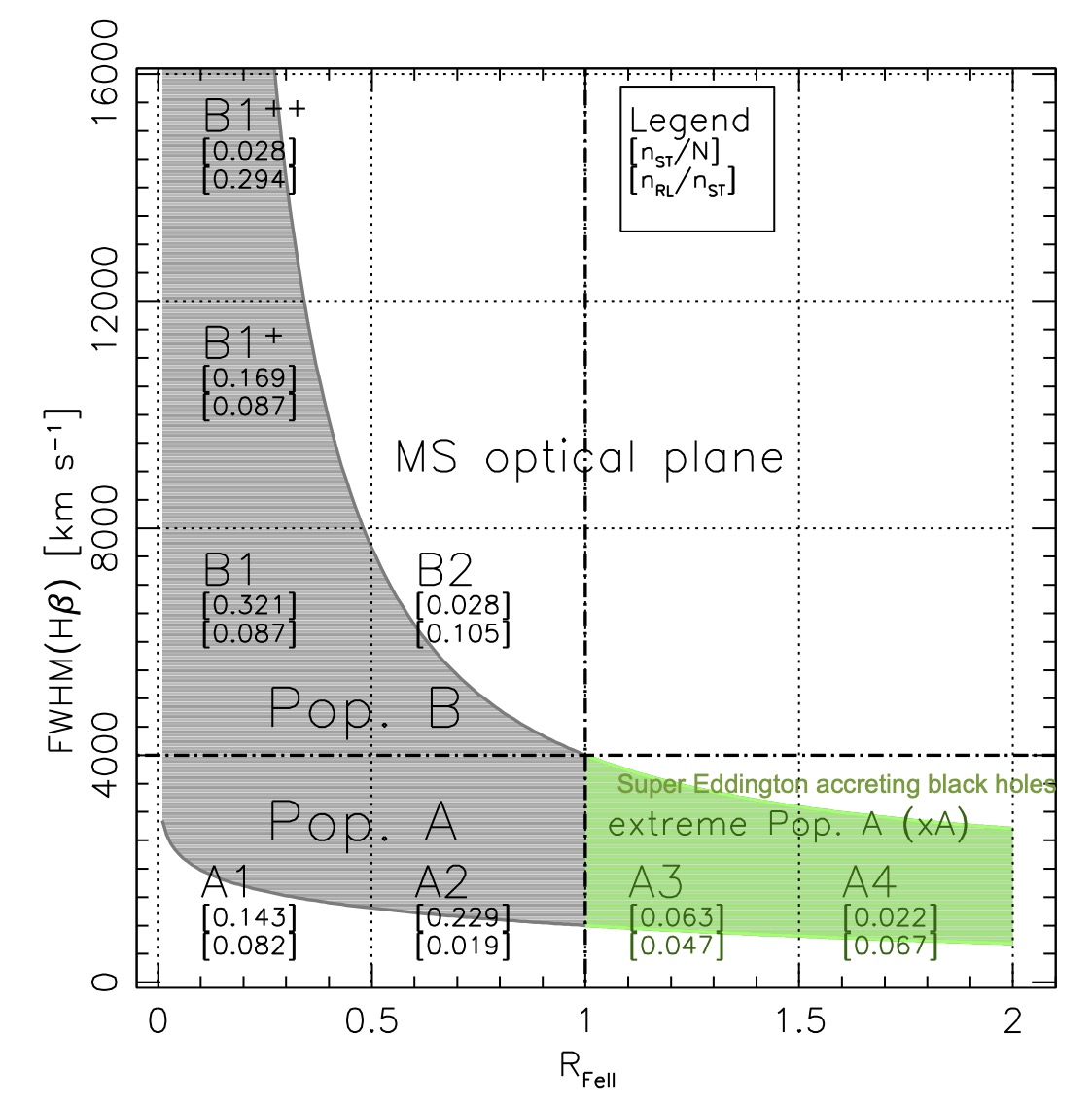}
\includegraphics[width=6.55cm]{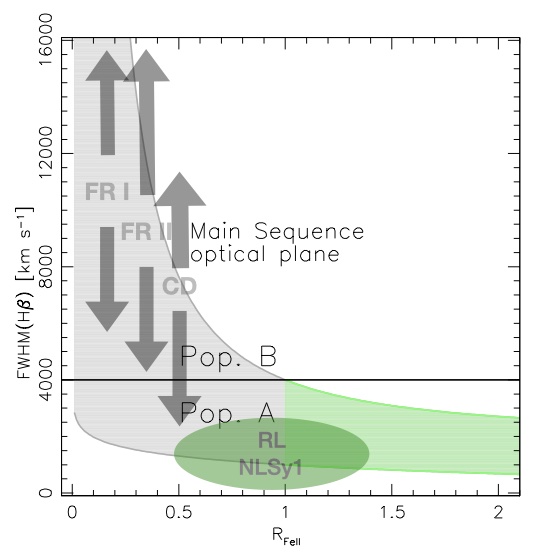}
\caption{Left: Sketch of the optical plane of the MS, FWHM \hb\ vs. \rfe. Numbers in square brackets provide, from top to bottom, the full sample fractional occupation, the fraction of RL (CD + FR-II) sources in each spectral bin. {Right: Same section of the optical plane,  but with a rough indication of the   FWHM domain spanned by FR-I, FR-II, and CD along with the location of the RL NLSy1 galaxies considered in this study (dark green area).}\label{fig:ms}}
\end{figure}

The relative intensities { of the emission lines} and their profiles do not scatter around an average, but instead change in a systematic way along the so-called quasar main sequence (MS, \cite{borosongreen92,sulenticetal00a,marzianietal01,shenho14}). The MS can be represented in a plane where the {full-width at half maximum (FWHM) of} H$\beta$\ is diagrammed against the parameter \rfe, defined as the intensity ratio between \feiiq\ and the broad component of \hb\ i.e., \rfe = F(\feiiq)/F(\hbbc).  Sketches or scatter plots  of the occupation of the quasar MS in this parameter plane have been often shown in the literature \citep[e.g.,][]{gaskell85,borosongreen92,shenho14,sulenticmarziani15}. Here we show a simple sketch of the quasar occupation in the { optical} parameter plane. Fig. \ref{fig:ms} identifies several spectral bins in the plane \citep{sulenticetal02}, and the domain of two populations: Population A, with FWHM \hb $\lesssim 4000$ \kms, and a Population B of broader sources. The rationale for the subdivision in two Populations has been provided in several papers \citep{sulenticetal00a,sulenticetal08,fraix-burnetetal17}. Population A objects are predominantly {sources with} high Eddington ratio, with evidence of powerful winds originating from the accretion disk \citep{leighlymoore04,richardsetal11,coatmanetal16,marzianietal16a,sulenticetal17,vietrietal17}. Conversely, Population B sources have lower Eddington ratio, and their line asymmetries are predominantly toward the red  \citep{marzianietal03b,marzianietal09}. The origin of the asymmetry is a topic of current debate:  partial obscuration and infall \citep{wangetal17}, and gravitational redshift \citep{corbin95,popovicetal95,zhengsulentic90,bonetal15,punslyetal20} are two processes that have been proposed. 

The importance of the \rfe\ parameter stems from the fact that the \feii\ emission extends from UV to the IR and can dominate the thermal balance of the low-ionization broad-line region (BLR) \citep{marinelloetal16}. \feii\ emission is self-similar in type-1 AGN but the relative intensity to \hb\ (parameterized by \rfe) can vary from undetectable to \rfe\ $\gtrsim 2$, with values larger than $\approx$ 2 being rare (less than 1\%\ in optically selected samples, \citep{marzianietal13a}).  The \rfe\ trend  of the MS in turn provides systematic constraints on the physical conditions of the line emitting gas \citep{marzianietal10,pandaetal18,pandaetal19}, and is related to several multi-frequency properties as well. For instance, \feii\ emission is also correlated with narrow-line properties {such as w}eak \feii\ implies strong \oiiiopt. \citet[][]{borosongreen92} formulated {this and the FWHM \hb -- \rfe} correlation in the context of the {Eigenvector 1} in a principal component analysis, the main forerunner of the quasar MS, although this result was known since much earlier time (see for instance Ref. \citep{steiner81}). {Furthermore,} strong \feii\ is associated with a high value of the soft-X ray photon index \citep{bolleretal96,wangetal96,sulenticetal00b,grupeetal01}, a tenet of the 4D Eigenvector 1 introduced by \citet{sulenticetal00b}. Nonetheless, a full understanding of the \feii\ emission remains a daunting task \citep{pandaetal18,pandaetal20,pandaetal20a,pandaetal21}. Even the main ionization mechanism is not fully clear. Early suggestions of a contribution of collisional ionization \citep[e.g.,][and references therein]{jolyetal08} have fallen into disfavor following the extensive monitoring of several objects \citep[][]{barthetal13,duetal18}: the \feiiopt\ emission responds to observed continuum variations with a delay larger than the one of \hb, but still fairly well defined. 

The MS has been instrumental to the establishment of a second major result, in addition to the realisation of the importance of \feii\ emission. 
RL sources are not uniformly distributed along the main sequence \citep{sulenticetal03,zamfiretal08}. In low-$z$ samples, they occupy the region of low \rfe\ and predominantly broad or very broad \hb. {The numbers in square brackets in {Fig.~\ref{fig:ms}} report the fraction of sources along the sequence, and the fraction of core-dominated (CD) and Fanaroff-Riley II (FR-II) sources in each spectral bin\footnote{{We consider here as radio loud (RL) only powerful jetted radio sources with radio-to-optical specific flux ratio $R \gtrsim 80$.}}. This excludes a population of radio-detected sources {that} are  likely associated with star formation, and are confined at the high-\rfe\ end of the main sequence \citep{bonzinietal15,caccianigaetal15,gancietal19}. CD and FR-II sources show systematic differences, in the sense that CD sources show stronger \feii\ emission and narrower Balmer line \citep{osterbrock77,mileymiller79,zamfiretal08}.}

In this small contribution we focus the attention on the \feii\ optical spectrum of type 1 AGN, and  we first provide a summary description of the optical \feii\ emission in the spectral region of \hb\ sample (Sect. \ref{feiilines}). The aim is not to solve the still-open problem of the \feii\ emission in type-1 AGN \citep{pandaetal18,pandaetal19}, but rather to establish whether the optical \feii\ emission might be different in {radio-quiet (RQ)} and RL AGN.  We then present an overview of the sample used for the analysis of the differences between RQ and RL sources, keeping the distinction between CD and FR-II among jetted sources. The comparison is carried out within the MS context, namely comparing sources that are RQ and RL belonging to the same spectral type (\S \ref{sample}), considering a “solid”  \feii\ template (i.e, a template with fixed relative line intensities), and a template in which the relative multiplet strengths are free to vary (\S \ref{fits}). The results (\S \ref{results}) confirm  a substantial equality for the \feii\ emission, within the non-trivial constraints imposed by {signal-to-noise ratio (S/N)} and resolution. Possible implications for the line emitting region structure in RQ and RL are briefly discussed, also with the help of photoionization computations (\S \ref{disc}).  

\section{The \feii\ emission lines}
\label{feiilines}
{The main \feii\ }optical lines in the spectral region between 4400 \AA\ and 5600 \AA\ are associated with 5 main multiplets. It is expedient to distinguish 3 main features: 
\begin{itemize}
    \item the blend on the blue side of \hb, made up of lines from the m37 and m38, and usually referred to as \feiiq, whose intensity is the \feii\ measure that enters in the definition of \rfe\ and applied in most papers following \citet{borosongreen92}. 
    \item m42, with two lines  appearing as satellite lines of \oiiiopt. The line at $\lambda$5018 might be affected by contamination from He {\sc i}$\lambda 5016$, and is strongly affected also by the red wing of \oiii. Given these difficulties the m42 lines were included in the fits but no result concerning this multiplet is considered;
    \item the lines of multiplets m48 and m49 that provide the bulk of the emission of the \feii\ blend on the red side of \hb\ (referred to as Fe{\sc ii}$\lambda$5270\ or as the \feii\ red blend). 
\end{itemize}
 
\begin{figure}[htp!]
\centering
\includegraphics[width=14cm]{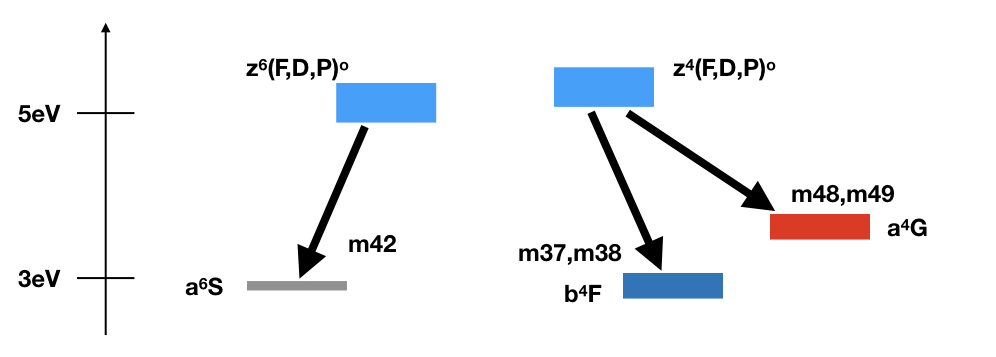}
\caption{Highly simplified Grotrian diagram of the energy levels leading to the strongest \feii\ transitions in the spectral range 4400 \AA\ — 5600 \AA. \label{fig:grotrian}}
\end{figure}

The transitions giving rise to these lines are schematically shown in the highly-simplified Grotrian diagram of Fig. \ref{fig:grotrian} \citep{phillips78a,osterbrockferland06,pradhannahar15}. 
 
The observed \feiiopt{}\  emission is produced by a combination of collisional and resonance-fluorescence processes \citep{verneretal99,verneretal04,sigutpradhan98,sigutpradhan03,pradhannahar15}. Collisional excitation, resonance fluorescence, in which a line or continuum photon is absorbed in a line transition, both play an important role.  Ly$\alpha$ fluorescence can populate the upper levels of Fig. \ref{fig:grotrian} via population of much higher energy level at $\approx 10$ eV, producing a cascade with two branches leading to the $z$\  $^6D$,$^6F$,$^6P$,$^4D$, $^4F$, and $^4P$\ terms between 4.8 and 5.6 eV responsible for the optical emission between 4500 and 5600 \AA. Collisional excitation, albeit significant \citep{phillips78b}, alone cannot explain the emission of the stronger \feii\  emitters.

\section{Sample}
\label{sample}

This paper considers two samples { of broad-line, type-1 sources}: 

\begin{itemize}
\item  a sample of bright quasars {from the Sloan Digital Sky Survey (SDSS) database} in the redshift range $0.4 \lesssim z \lesssim 0.7$\  \citep{marzianietal13,marzianietal13a}. Composite median spectra were extracted for the RQ, CD and FR-II radio classes.  A significant {number} of RL sources ({ out of the 680 quasars of the original sample}) is found only in the spectral bins A1, B1, B1$^+$: 8, 16, 9 CDs and 10, 23, 11 FR-IIs in each bin, respectively. All sources (save a borderline one) in the sample that are RL with extended emission satisfy the power criterion at  1.5 GHz $\log P \nu \gtrsim 31.7  $\ erg s$^{-1}$ Hz$^{-1}$ that separate FR-II sources from the lower luminosity FR-I \citep{deyoung02,kembhavinarlikar99}.\footnote{Both FR-I and FR-II sources are characterized by extended radio emission. FR-Is however show  lower radio surface brightness toward the outer extremities of the lobes, at variance with FR-II sources that are often described as ``edge brightened.'' According to the unification schemes \citep{urrypadovani95}, both classes are observed at relatively high inclination. Their optical spectra  can be of both type 1 and type 2 AGN.} In the last bin, however, the \feii\ intensity is consistent with 0, so that only sources in bin A1 and B1 are considered. In addition, in A1 FR-II the \feiiopt\ emission is too weak for a meaningful analysis.In conclusion, our bright quasar sample belongs in the following composite spectra: A1 RQ, A1 CD, B1 RQ, B1 CD, B1 FR-II; 
\item three individual relativistically jetted NLSy1s with $\gamma$-ray detections, that show more prominent \feii\ emission than the composites. The $\gamma$-ray detection supports the presence of a relativistic jet as the origin of the radio power for these sources \citep{komossaetal06}. They were selected because high-quality optical spectra were available for the three of them.
\end{itemize}

The three $\gamma$-ray sources are 1H 0323+342 ($\equiv$ B2 0321+33 $\equiv$ 1H 0323+342  $\equiv$ J03246+341), 3C 286 ($\equiv$J13311+305), and PKS 2004-447 ($\equiv$J20079--445). J03246 is the closest $\gamma$-ray NLSy1 ($z = 0.063$), hosted either by a spiral or an interacting late-type galaxy \citep{zhouetal06,antonetal08}, and it is the only known $\gamma$-NLSy1 showing a Fe K$\alpha$ emission line in X-rays \citep{kynochetal18}. This object was detected by the \textit{Fermi} Gamma-ray Space Telescope soon after its launch \citep{abdoetal09}. J13311 is a rather high redshift ($z \approx 0.850$) object, identified as a $\gamma$-ray source in all \textit{Fermi} catalogs (e.g., \citep{ackermannetal15}). Originally it was classified as a compact steep-spectrum (CSS) source based on its radio morphology \citep{peacockwall82}, and later as a NLSy1 from optical spectroscopy \citep{bertonetal17,yaokomossa21}. Also J20079 ($z \approx 0.240$) is classified as both a CSS and a NLSy1 \citep{oshlacketal01,galloetal06,bertonetal21} and, as J03246, it belongs to the first batch of NLSy1s detected by \textit{Fermi} \citep{abdoetal09}. The spectra of these three sources were obtained using the Asiago 1.22m Telescope for J03246 (full description of the data reduction in \citet{foschinietal15}), the SDSS archive for J13311, and the FORS2 instrument on the Very Large Telescope for J20079--445 \citep{bertonetal21}.

\section{Analysis}
\label{fits}

The analysis is mainly empirical  and, as mentioned,  there is no pretension to solve the problem of the optical \feii\ emission in AGN. Two approaches are followed in the measurement of the relative intensity of the multiplets: 
\begin{itemize}
\item the {modeling} of the spectrum using a "solid" \feii\ template, scaled and broadened to minimize the $\chi^2$\ in a multi-component fit. It is basically the one of \citet{borosongreen92} actualized with a higher resolution spectrum and a model of the \feii\ emission underlying \hb\ \citep{marzianietal09}. In this case, a multi-component fit was carried out including all known emission components, as detailed in several recent works \citep{sulenticetal15,negreteetal18,marzianietal19}. Specifically, the redward asymmetric \hb\ has been modeled by the use of three components {- a narrow, a broad and a very broad component} \citep[][and references therein]{wolfetal20}. After verification that the host galaxy spectral emission is not contributing significantly, the local continuum was fit with a power law;
\item the use of a "liquid" template that permits to change the relative intensity of the multiplets in the optical spectral range \citep{kovacevicetal10,kovacevicdojcinovicpopopvic15} \footnote{http://servo.aob.rs/FeII\_AGN/}. In this case, the continuum subtracted spectrum was fit to a set of variable multiplets and to \hb\ and \oiii\ profiles that were approximated as Gaussians. 
\end{itemize}

\section{Results}
\label{results}

\subsection{\feii\ emission comparison between RQ and RL}

\begin{figure}[t!]
\centering
\includegraphics[width=9.5 cm]{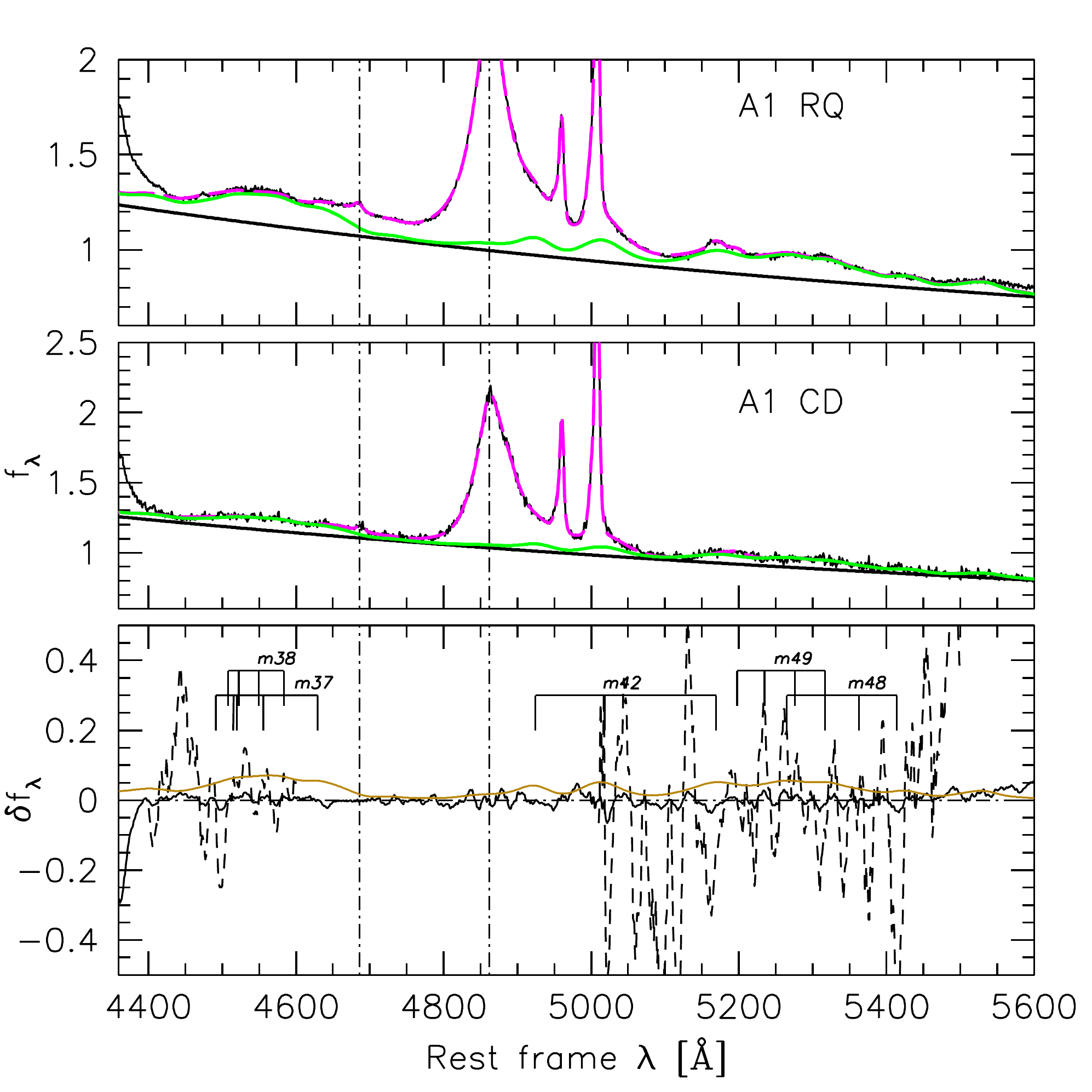}
\caption{From top to bottom: Non-linear $\chi^2$\ multi-component analysis with a solid \feii\ template of RQ, CD, and of the residuals in \feii\ emission for composites of spectral type A1. In the bottom panel the brown line traces the difference in the template (positive, since \feii\ is stronger in RQ than in CD). The black filled lines traces the difference between the observed \feii\ emission after rescaling of the CD \feii\ to obtain a minimum average $\sim 0$. The dashed line traces the relative difference, i.e., (\feii\ RQ - \feii\ CD)/\feii\ RQ.  Wavelengths of multiplet lines are from \citet{moore45}.  \label{fig:a1}}
\end{figure}   
 
\begin{figure}[t!]
\centering
\includegraphics[width=6.5 cm]{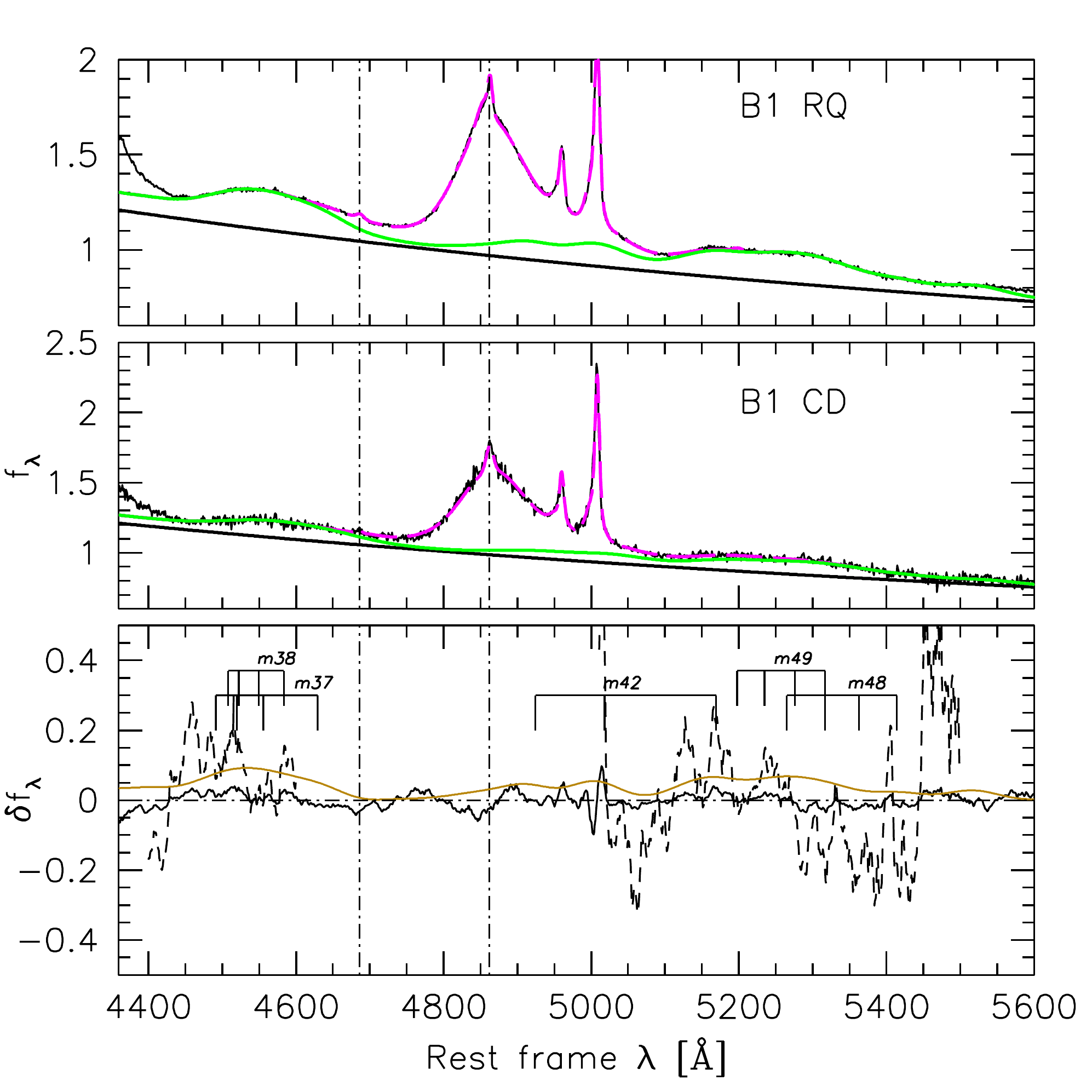}
\includegraphics[width=6.5 cm]{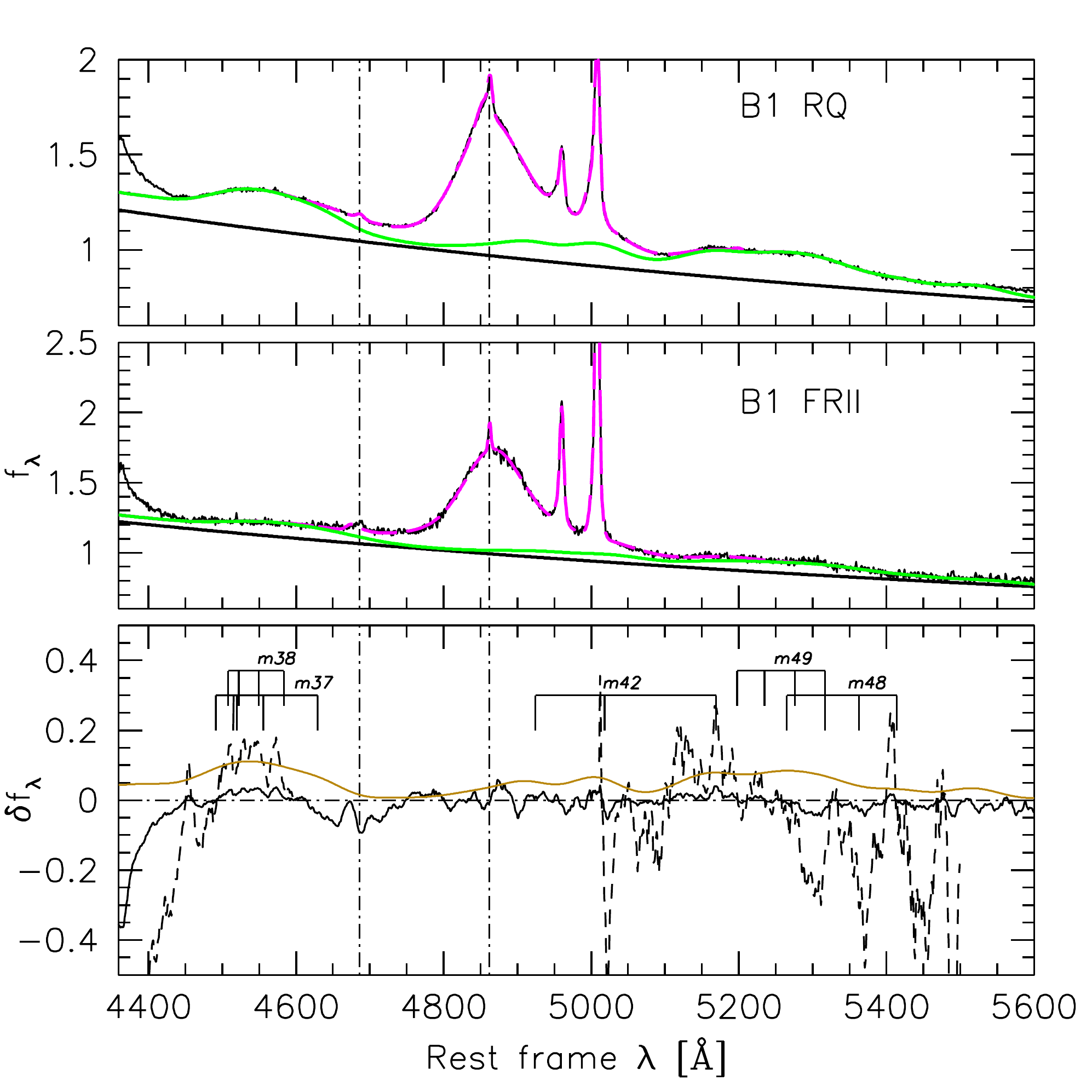} 
\caption{{In the left panel, comparison between RQ and CD composites of spectral type B1. In the right panel, comparison between B1 RQ, the same as in the left panel, and B1 FR-II}. {Bottom panels} as in Fig. \ref{fig:a1}.  }
\label{fig:b1} 
\end{figure}

The results of the application of the multi-component fit with the solid \feii\ template are shown in Figs. \ref{fig:a1} and \ref{fig:b1} for the spectral types A1 and B1, respectively. {In Fig.~\ref{fig:a1}, the only comparison we show is between A1 RQ and A1 CD sources, as there is no A1 FR-II composite.} {In Fig.~\ref{fig:b1}, the left panel shows the comparison between B1 RQ and B1 CD composites, while on the right the B1 RQ composites are compared to B1 FR-II}. The immediate result is that RL {(CD and FR-II)} composites show weaker \feii\ emission \citep{zamfiretal08}. The second immediate result \citep[already reported in][]{marzianietal03b} is the similar appearance of the \hb\ profiles. The \hb\ profiles of Figs. \ref{fig:a1} and \ref{fig:b1} confirm that the prominent redward asymmetry is not a prerogative of RL sources, but is a common feature in {\em both} RQ and RL sources, provided that they belong to Population B \citep{marzianietal96,sulenticetal08,punsly10}. 

The bottom panels {of Figs.~\ref{fig:a1} and \ref{fig:b1}} provide constraints on the nature of the difference between \feii\ emissions in the RQ and RL composite spectra. The brown line traces the difference in the templates, {which is positive since \feii\ is stronger in RQ than in CD and FR-II as well}. The black filled lines traces the difference between the observed \feii\ emission after rescaling of the RL \feii\ emission by a factor $k$\ to obtain an average of 0. The difference is computed between the observed \feii\ spectra, as follows:

\begin{equation}
\delta \mathrm{Fe}\textsc{ii}_{\lambda,\mathrm{i}} = f_{\lambda,\mathrm{rq}} - (\mathcal{M}_{\lambda,\mathrm{rq}} - \mathcal{F}e_{\lambda,\mathrm{rq}}) -   k_\mathrm{i} \cdot  [ f_{\lambda,\mathrm{i}} - (\mathcal{M}_{\lambda,\mathrm{i}} - \mathcal{F}e_{\lambda,\mathrm{i}})]\label{eq:dfe}
\end{equation}

\begin{equation}
\frac{\delta \mathrm{Fe}\textsc{ii}_{\lambda,\mathrm{i}}}{\mathcal{F}e_{\lambda,\mathrm{rq}}} = \frac{f_{\lambda,\mathrm{rq}} - (\mathcal{M}_{\lambda,\mathrm{rq}} - \mathcal{F}e_{\lambda,\mathrm{rq}}) -   k \cdot  [ f_{\lambda,\mathrm{i}} - (\mathcal{M}_{\lambda,\mathrm{i}} - \mathcal{F}e_{\lambda,\mathrm{i}})]}{\mathcal{F}e_{\lambda,\mathrm{rq}}} \label{eq:dfer}
\end{equation}

where the index $i$ refers to CD or FR-II, the $f_\lambda$\ to the observed flux as a function of wavelength, and the calligraphic symbols refer to model spectrum ($\mathcal M_\lambda$) and model \feii\ ($\mathcal{F}e)$ via the scaled and broadened template. Model parameters were obtained with a multi-component, non-linear minimum $\chi^2$\ technique, as implemented in IRAF \citep{kriss94}.

 {The  basic results of the spectral fits  are reported  in Table \ref{tab:basic}.} The Table columns list, in the  following order, the spectrum identification, the rest-frame equivalent width in \AA\ for \feiiq, \rfe, FWHM \hb, the spectral type defined on the basis of the reported \rfe\ and FWHM \hb, and the rescaling factor $k$.

Table \ref{tab:feii} lists the average of the absolute difference between {the observed iron} blends and the solid \feii\ template. The averages are computed over the full ranges 4434 \AA\ -- 4684 \AA\ (hereafter indicated as B {blend} or \feiiq), and 5100 \AA\ -- 5600 \AA\ ({hereafter R blend}) also used for the {\tt CLOUDY} photoionization simulations (Sect. \ref{cloudy}). The next columns yield the average and relative absolute differences  between RL and RQ after rescaling, for the restricted spectral region $\tilde{B}$\ between 4500 and 4590 \AA\  (roughly corresponding to multiplets 38 and 39) and $\tilde{R}$\ (between 5200 and 5330 \AA, with the strongest  features of  m48 and m49). The reason of this restriction is to avoid the contamination by \heii\ on the blue side of \hb. The \heiiopt\ line can have a significant effect and its estimation is based only from fit. 
 

\begin{table}
\begin{center}
\caption{Basic  results \label{tab:basic}}\scriptsize\tabcolsep=2pt
  \begin{tabular}{lccccc} \hline
Spectrum & W(\feiiq) & \rfe\  & FWHM \hb & ST &   $k$    \\
\hline
\multicolumn{6}{c}{Composite spectra}\\
\hline 
A1 RQ  & 31.6 &  0.291 &  3230 & A1 & \ldots   
 \\
A1 CD           & 16.2     &  0.397      &  3290  &  A1  & 1.815    \\
B1 RQ           &  44.1    &   0.414     &  6940  &  B1  & \ldots    \\
B1 CD           & 26.0     &   0.281     &  7060  &  B1  & 1.657      \\
B1 FR-II        & 21.5     &  0.184      &  6790  &  B1  & 1.900     \\
\feii\ template &   \ldots & \ldots      & 990    &   A  & \ldots    \\
\hline
\multicolumn{6}{c}{$\gamma$\ RL NLSy1s}\\
\hline 
1H 0323+342    &  50.1      & 1.03    & 1280       & A3 &  \ldots    \\
3C 286         & 33.5       & 0.792   & 3170       & A2 & \ldots     \\
PKS 2004-447   & 36.7       & 0.5455  & 1470       & A2 & \ldots     \\
\hline
 \end{tabular}
 \end{center}
\end{table} 

\begin{table}
\begin{center}
\caption{\feii\ results \label{tab:feii}}\scriptsize\tabcolsep=2pt
  \begin{tabular}{lccccccccccc} \hline
Spectrum   &   $\overline{f_\lambda-{\mathcal M_\lambda}(B)}^a$  & $\overline{f_\lambda-{\mathcal M_\lambda}(R)}^b$ & $\overline{\delta f_\lambda(\tilde{B})}^c$ & $\overline{\delta f_\lambda/f_\lambda(\tilde{B})}^c$ & $\overline{\delta f_\lambda(\tilde{R})}^c$ &  $\overline{\delta f_\lambda/f_\lambda(\tilde{R})}^c$ 
\\
\hline
\multicolumn{7}{c}{Composite spectra}\\
\hline 
A1 RQ     &   0.00146 $\pm$ 0.00942  & 0.00257 $\pm$ 0.01333             &\ldots &\ldots &\ldots &\ldots             \\
A1 CD     &   0.00014 $\pm$ 0.01265 & -0.00117 $\pm$ 0.01671    &   0.00244 &  -9.986E-05 &  0.0146 &  -0.00354   \\
B1 RQ     &   0.00149 $\pm$ 0.00665 & 0.00277 $\pm$ 0.00982  & \ldots  &  \ldots  &\ldots     & \ldots            \\
B1 CD &   -0.00188 $\pm$ 0.01698  &  9.44E-05$\pm$ 0.01745   &  0.01420 & 0.07467   &   -0.00403 &  -0.02626      \\
B1 FR-II  &   0.00297 $\pm$ 0.01554  &  0.0047$\pm$ 0.01574  & 0.03138    &   0.15740  & -0.00294  &  -0.01834    \\
\hline
\multicolumn{7}{c}{$\gamma$\ RL NLSy1s}\\
\hline 
1H 0323+342   &     0.00509 $\pm$ 0.05126 & 0.00949 $\pm$ 0.04280	  & -0.00598 & -0.0388  & 0.02072  & 0.1297       \\
3C 286        &   -0.00262 $\pm$ 0.03654  &  0.005087 $\pm$ 0.05126    &  -0.00304  & -0.0254  & 0.00516  & 0.0475     \\
PKS 2004-447  &   -0.00145 $\pm$ 0.02275   &  0.005514 $\pm$ 0.02115 & -0.0008   &  -0.0149 & 0.00866  & 0.1117 &       \\
\hline
  \end{tabular}
  \end{center}
\footnotesize{$^a$: Averages of the normalized flux difference between the continuum subtracted spectrum and the solid template in the range 4434--4684 \AA;  $^b$: same, in the range 5100--5600 \AA; $^c$: absolute and relative average difference following Eq. \ref{eq:dfe} and Eq. \ref{eq:dfer}, for the restricted B  and R ranges, respectively. }  
\end{table}

In the case of spectral type A1, the comparison between RQ and CD relative intensities of the multiplets is shown in the   panels of Fig. \ref{fig:a1}. {On the blue side,} the $\delta \mathrm{Fe}\textsc{ii}_{\lambda,\mathrm{i}}$\ is consistent with 0 over the range where most of the emission due to multiplets 37 and 38 is expected. Similar considerations apply to the red side emission; the main fluctuations according to Eq. \ref{eq:dfer} occur when \feii\ emission is weak. 

 
The comparison for the composite spectra of bin B1 is shown in Fig. \ref{fig:b1}. On the blue side of \hb\ in correspondence of the m37 and m38 emission there could be a slight excess for the CD composite, accompanied by a deficit on the red side of \hb\ where emission is associated with multiplets m48 and m49. The average relative difference is a few positive and negative percents  for the B and R regions, respectively (Table \ref{tab:feii}). 
Accepted as real the absolute difference is constrained within $\lesssim 3$ \%, and the relative difference is contained within $\pm$ 20 \%. The $\approx 3$\%\ average excess translates into an error in terms of $\delta$\rfe $\approx 0.03$, and in $W$ \feiiq\ the difference is also $\approx$ 3 \AA\ implying that there is no effect on the placement in the MS optical plane. To estimate the measurement errors on \rfe\ for the same S/N, equivalent width and FWHM of the lines, we used the quality parameter defined in an unrelated investigation, and derived from Monte Carlo simulations estimates of the errors for \hb\ and \rfe. The errors on \feii\ are the dominant one, and are $\delta$\rfe $\lesssim$ 0.05 for the RQ composites and  $\delta$\rfe $\approx$ 0.10 for the radio-loud ones.  
 

\begin{figure}[t!]
\centering
\includegraphics[width=4.2 cm]{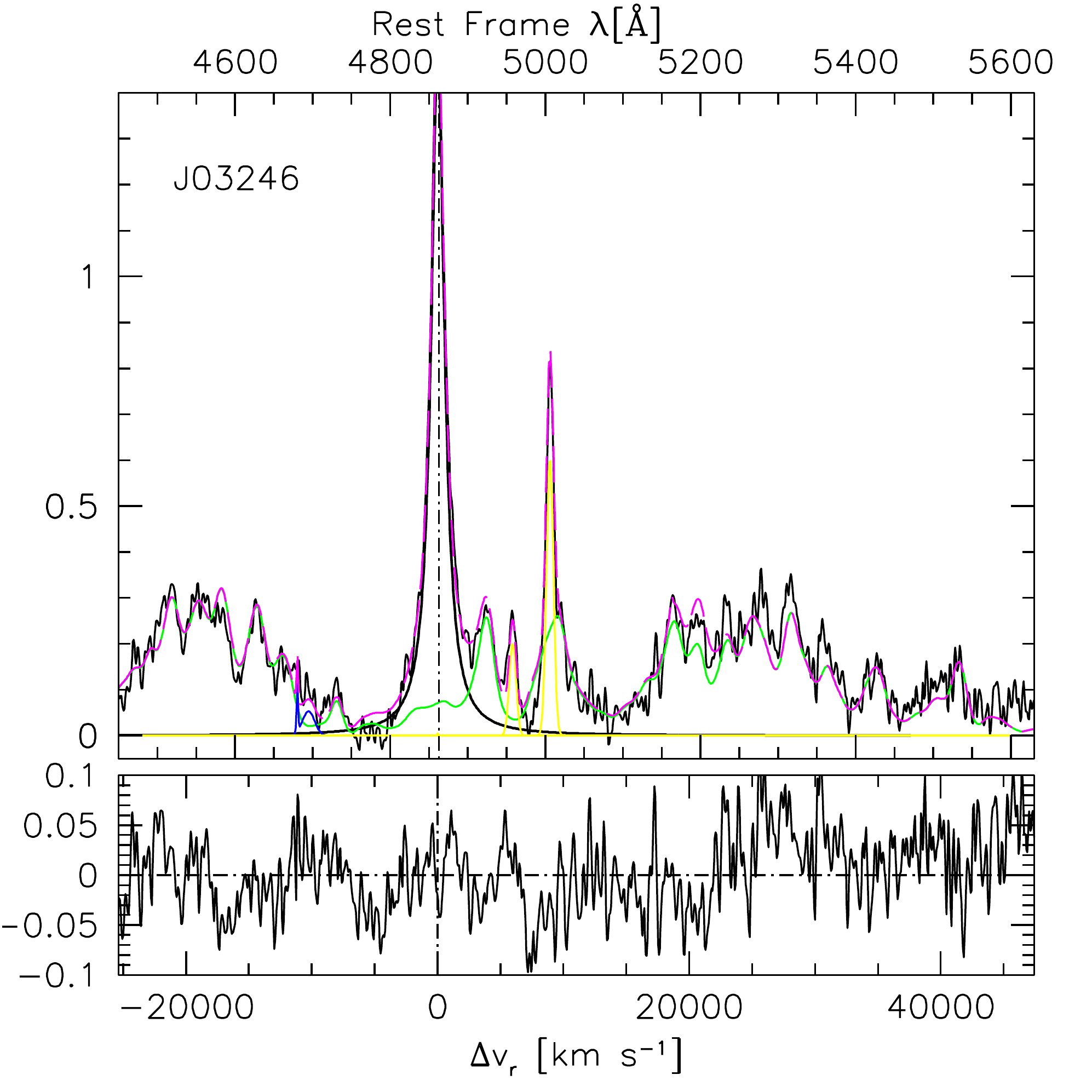}
\includegraphics[width=4.2 cm]{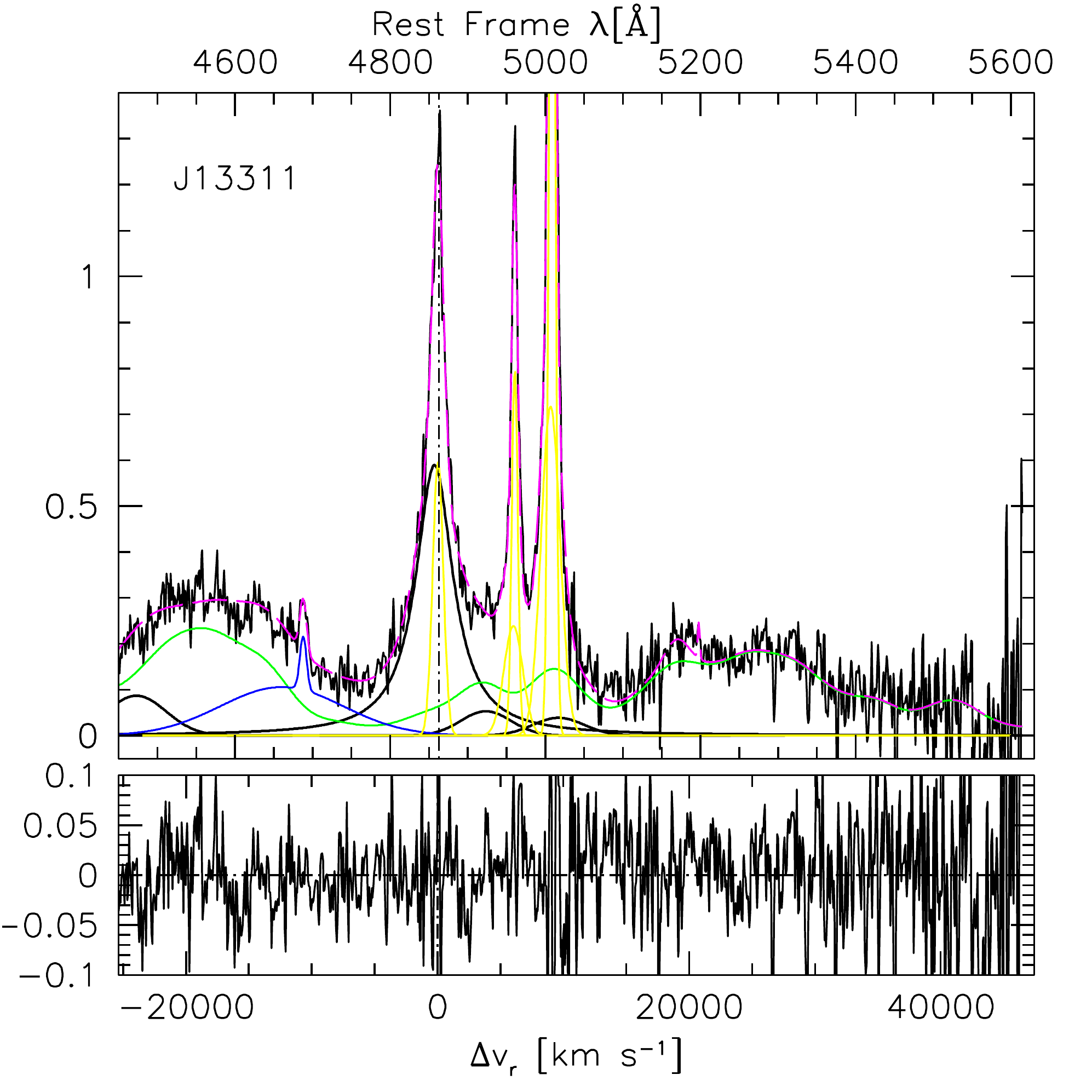} 
\includegraphics[width=4.2 cm]{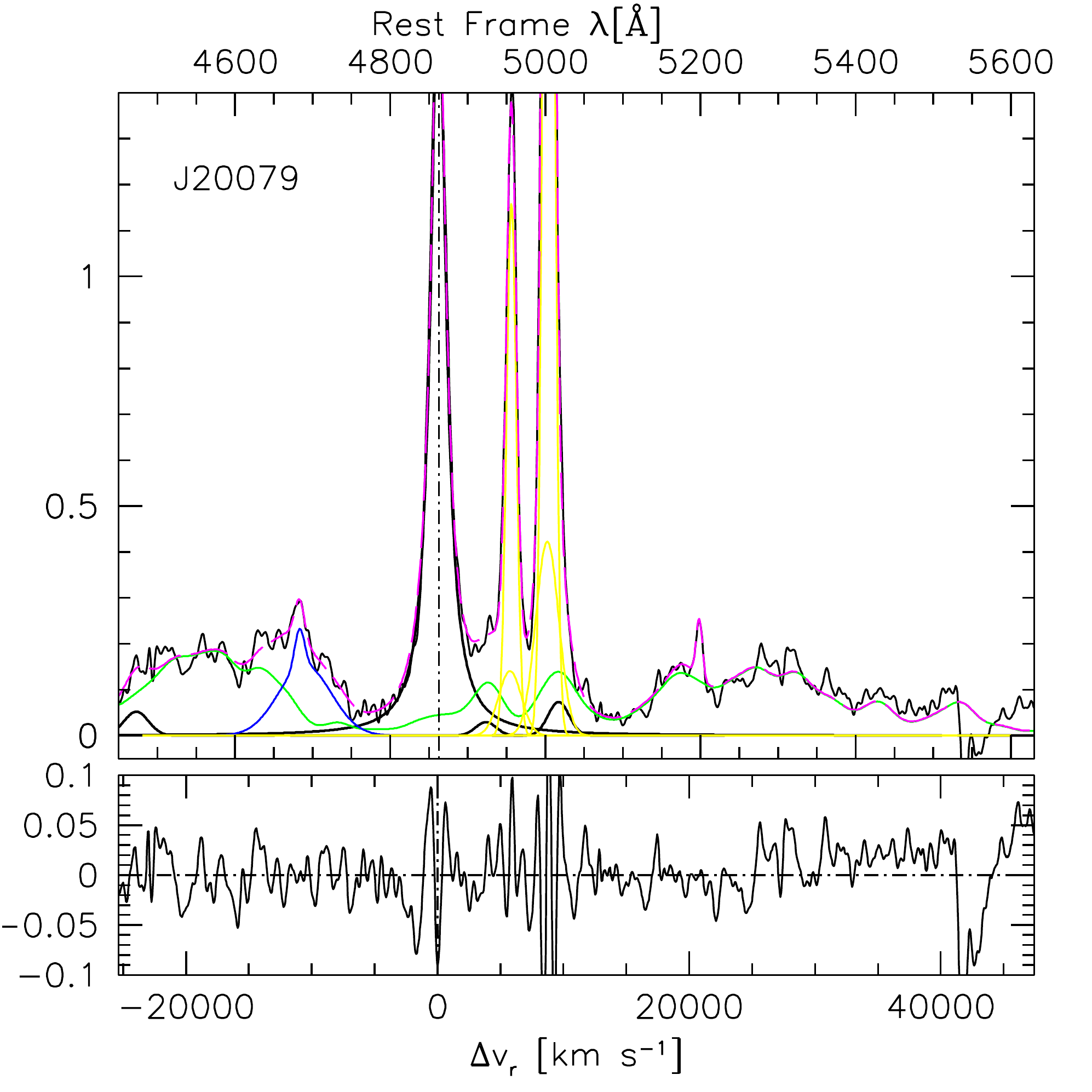}
\caption{Non-linear $\chi^2$\ multi-component analysis with a solid \feii\ template (green) of three jetted NLSy1s with $\gamma$-ray detections.
{ The spectra are shown after continuum subtraction. The thin filled line traces the observed spectrum. The dashed magenta line represents the model, and the thick black line the \hbbc. The \oiiiopt\ is shown in yellow. The lower panel show the residuals of the subtraction of the model from the observed spectrum. From left to right panel, J03246+341, J13311+305, and J20079-445.}\label{fig:gamma} } 
\end{figure}


\subsection{$\gamma$-ray detected RL NLSy1s}

We apply the same fitting techniques to three relativistically jetted NLSy1s with $\gamma$-ray detection. Fig. \ref{fig:gamma} shows that the solid template used for the fitting of the composites is providing a good agreement, in a case with higher values of \rfe\ and much narrower lines that is better posed to appreciate possible differences.  The fits are successful, and the residuals consistent with 0, with average systematic differences are $\lesssim$ 1 \%, and the rms scatter is $\lesssim 5$\%\ of the normalized continuum value at 5100 \AA. 

\begin{table}
\begin{center}
\caption{``Liquid'' \feii\  results \label{tab:liquid}}\scriptsize\tabcolsep=2pt
  \begin{tabular}{lcccccccccc} \hline
Spectrum    &  B/R{$^a$} &$\tilde{B}/\tilde{R}^b$& $\frac{m37+m38z}{m48+m49}^c$&  $\tilde{B}_\mathrm{k}/\tilde{R}_\mathrm{k}^d$  &    $\overline{\tilde{B} - \tilde{B}_\mathrm{k}}^e$ & $\overline{\tilde{R} - \tilde{R}_\mathrm{k}}^e$\\
\hline
\multicolumn{7}{c}{Composite spectra}\\
    \hline 
A1 RQ     &  0.993 & 0.994   & 0.82 & 0.974 & 0.00447 & -0.0182 &   \\            
A1 CD     &  1.141 & 1.045   &  0.72  & 0.942 & 0.0117 & 0.0075    \\
B1 RQ     &  0.884 &  0.891  &  0.97 & 0.899 & 0.00934 &  0.00419 \\
B1 CD     &  0.666 &  0.698  &  0.79  & 0.699 & -0.0375 & -0.0198  \\
B1 FR-II  &  0.685  &  0.726 & 0.65  & 0.890 & -0.051 & -0.0169       \\
\feii\ template     & 0.792  & 0.979 &   0.65 &  0.967  &  0.00264 & 0.00813    \\
\hline
\multicolumn{7}{c}{$\gamma$\ RL NLSy1s}\\
    \hline 
1H 0323+342   &  0.743 & 0.780     & 0.59 & 0.765  &    0.0048 & -0.00020       \\
3C 286        &  0.821   &  0.905 & 0.78 & 0.867   & 0.00591 & -0.00218       \\
PKS 2004-447  &    0.980 & 0.882 & 0.62 & 0.882 & 0.00093 & -0.0000466   \\
\hline
 \end{tabular}
 \end{center}
\footnotesize{$^a$: B/R ratio measured on the observed spectrum; $^b$: same as in the previous column, but $\tilde{B}/\tilde{R}$\ in the restricted range; $^c$: ratio between the  sum of {m}37+m38 and m48+m49 intensity obtained with the fits following \citet{kovacevicetal10}. Note that the ratio includes only the sums of these four multiplets, while   both the standard and the restricted range include additional \feii\ emission as well as a possible {residual} contribution of \hei\ and \heii\ lines. $^d$: ratio between the \feii\ emission of the B and R restricted range computed with the simulations of \citet{kovacevicetal10}.  $^e$: absolute  average difference  for the restricted B  and R ranges, respectively, using the {\em total } \feii\ emission following \citet{kovacevicetal10}.
}  
\end{table}

\subsection{Models with relative intensity of the multiplets free to vary} 
\label{free}
 
\begin{figure}[t!]
\centering
\includegraphics[width=8.5cm]{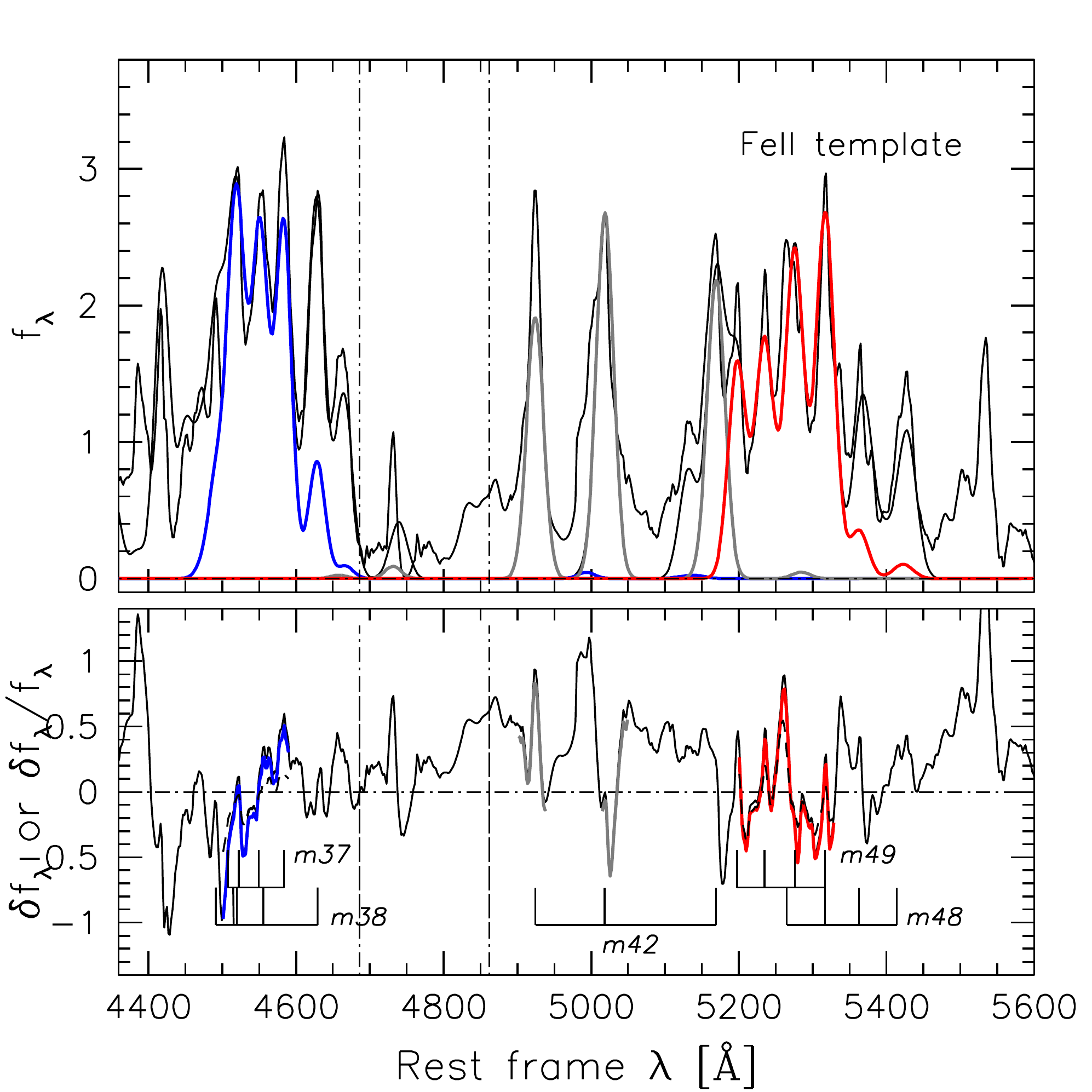}
\caption{Non-linear $\chi^2$\ multi-component analysis of the adopted solid \feii\ template with relative multiplet intensity free to vary. { The blue solid line represents the B blends, while the red solid line indicates the R blends.} }
\label{fig:templafe}
\end{figure}

We considered first the measurements based on the continuum-subtracted spectra in the full and restricted ranges (second and third columns of Table \ref{tab:liquid}), and we  utilized the web-based fitting tool\footnote{\href{http://servo.aob.rs/FeII_AGN}{http://servo.aob.rs/FeII$\_$AGN/uploads.php}} for implementing the scheme of   \citet{kovacevicetal10} for the optical \feii\ emission. The measurement of the ratio between the sum of the m37+m38, and  m48+m49 blends is reported in the fourth column of Table \ref{tab:liquid}. The fifth column reports the ratio $\tilde{B}_\mathrm{k}/\tilde{R}_\mathrm{k}$\ measured on the full spectral model of \citet{kovacevicetal10}. The last two columns report the average differences between the observed and the \citet{kovacevicetal10} modelling for the restricted ranges $ \tilde{B}_\mathrm{k}$\ and $\tilde{R}_\mathrm{k}$, respectively, and are meant to provide an estimate of how well the model \feii\   reproduces the actual observations. The agreement is very good with all cases having the average difference less than 0.05, and most less than 0.01. The same procedure was also applied to the solid \feii\ template used in the previous analysis (Fig. \ref{fig:templafe}). The values reported in Table \ref{tab:liquid} explain why the solid \feii\ template provides satisfactory results in all cases, and especially in the RQ cases. 

\begin{figure}[t!]
\centering
\includegraphics[width=4.4 cm]{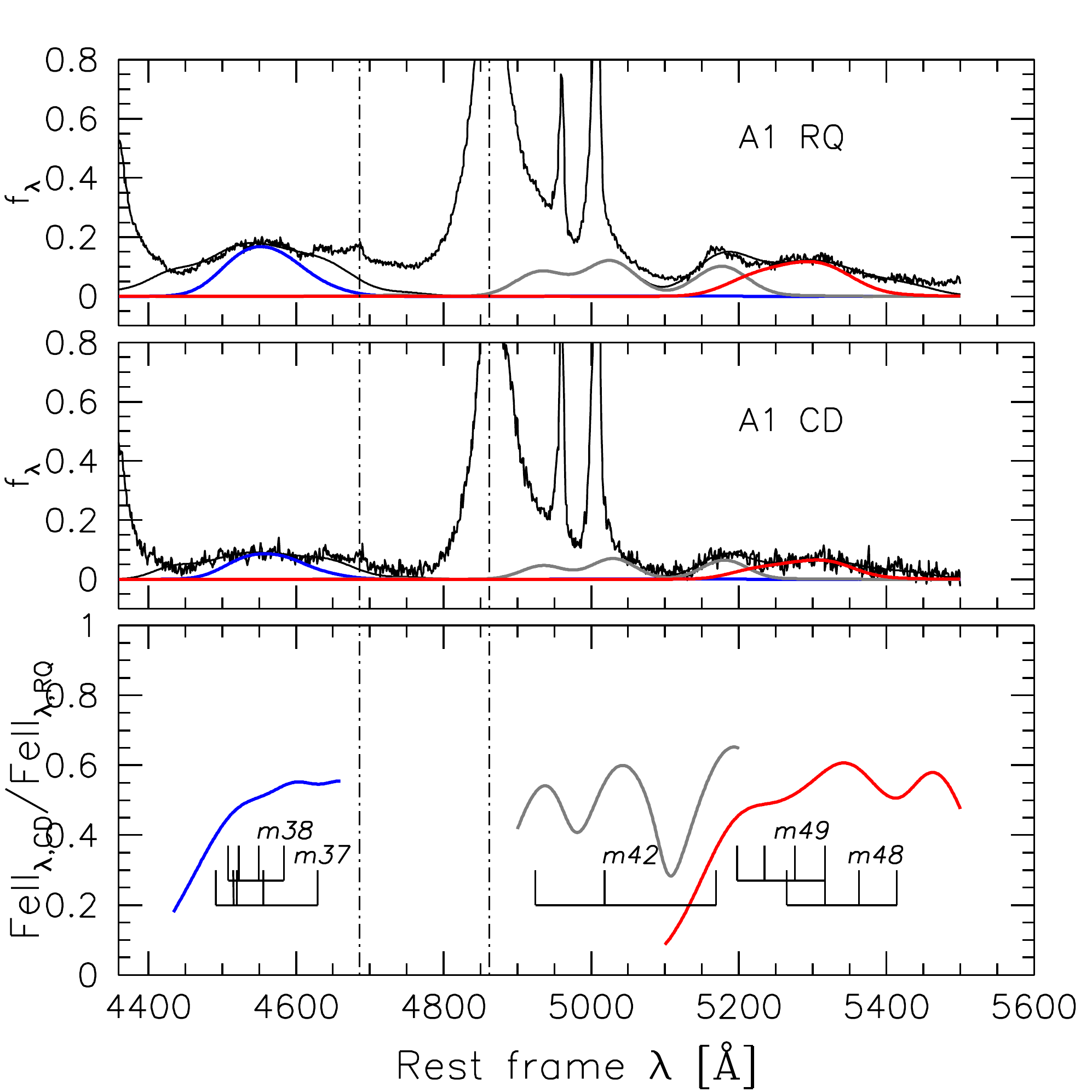}
\includegraphics[width=4.4 cm]{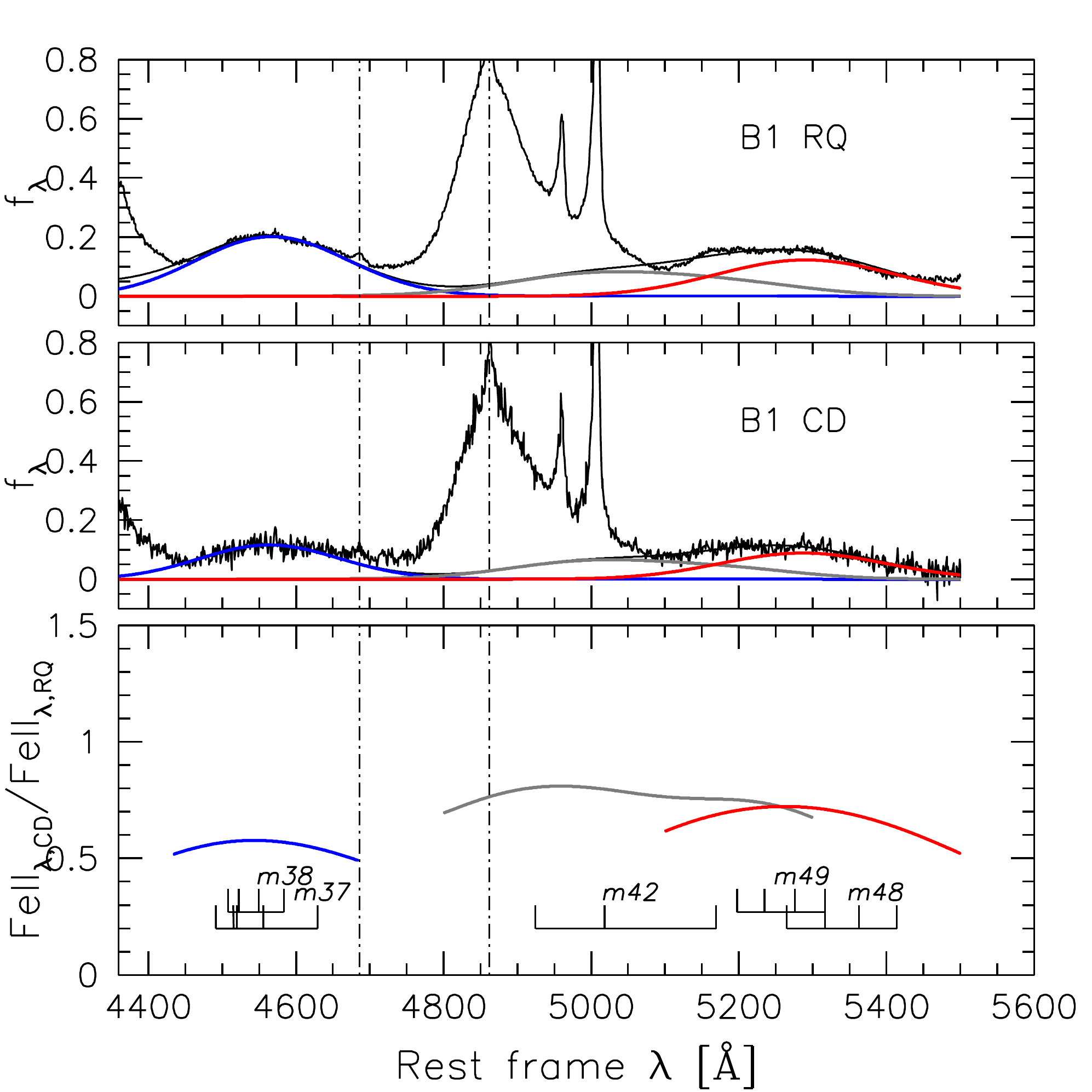}
\includegraphics[width=4.4 cm]{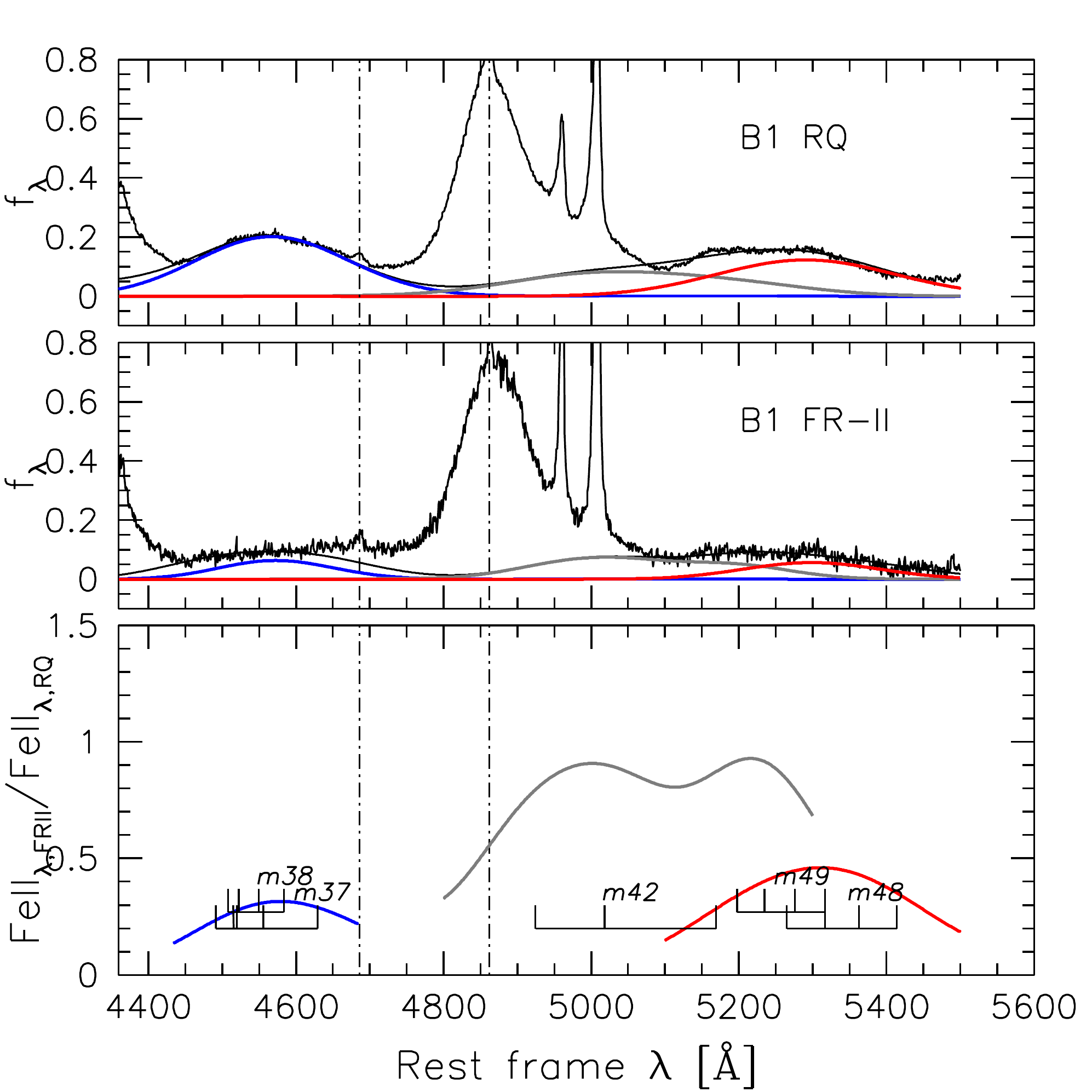}
\caption{Non-linear $\chi^2$\ multi-component analysis with a \feii\ template with relative multiplet intensity free to vary. {Colors as in Fig.~\ref{fig:templafe}}.
\label{fig:compfe}}
\end{figure}

We stress three main results: 

\begin{itemize}
    \item There is a significant difference (by a factor $\approx$2) between RQ and RL sources in the same spectral types (meaning similar mass, Eddington ratio $\lesssim$ 0.1, and luminosity). 
    \item The B/R ratio for the RQ A1 and B1 spectral types is $\approx 0.9 - 1.1$ (Fig. \ref{fig:compfe} and Table \ref{tab:liquid}). 
    \item The various measurements of the blue and {red} blend ratios suggest a somewhat higher values for the A1 and B1 RQ than for the RL sources ($\approx 0.9 -  1\ vs\ 0.7$). The effect is not strong: yet, it is apparent especially in Fig. \ref{fig:compfe}.   
\end{itemize}


It is advisable to consider that there could be effects not related to the {estimation of the} \feii\ intensity. {The main reason could be }a difference in the intensity of \heiiopt, even if \heiiopt\ is included in the fit. Other possibilities involve internal extinction {\citep{gaskell2017,baskin_laor2018}} affecting the \feii\ emitting region, limb-darkening \citep{marzianietal01}, incorrect continuum placement, contamination by lines other than \heii,  and blueshifted \hb\ emission. To overcome at least the problem of the \heiiopt\ uncertain contribution, and of the contribution of the \hb\ wings,  we defined a restricted range for B and R, $\tilde{B}$\ and $\tilde{R}$. The effect is however not significantly reduced (Table \ref{tab:feii}).



\begin{figure}[t!]
\centering
\includegraphics[width=4.4 cm]{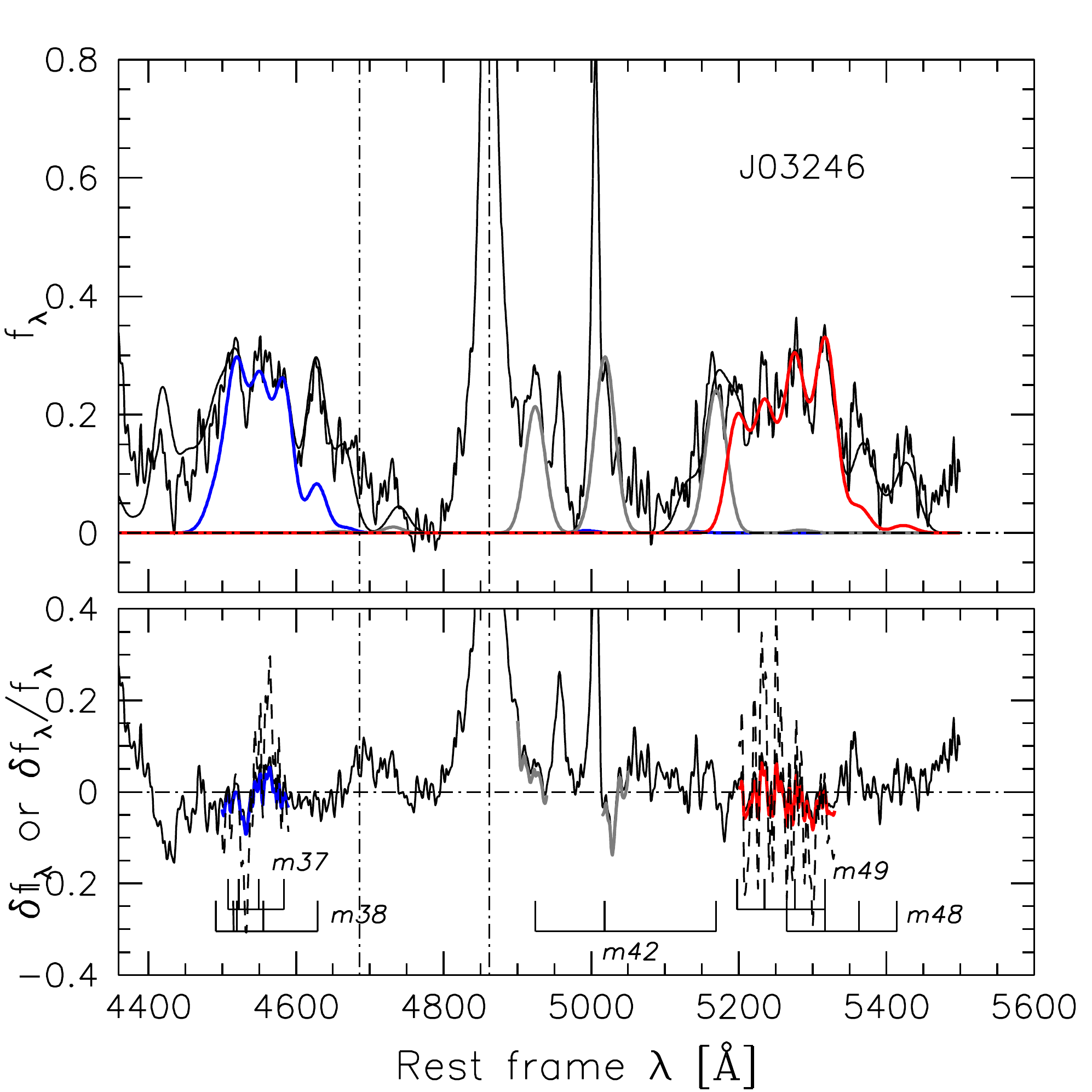}
\includegraphics[width=4.4 cm]{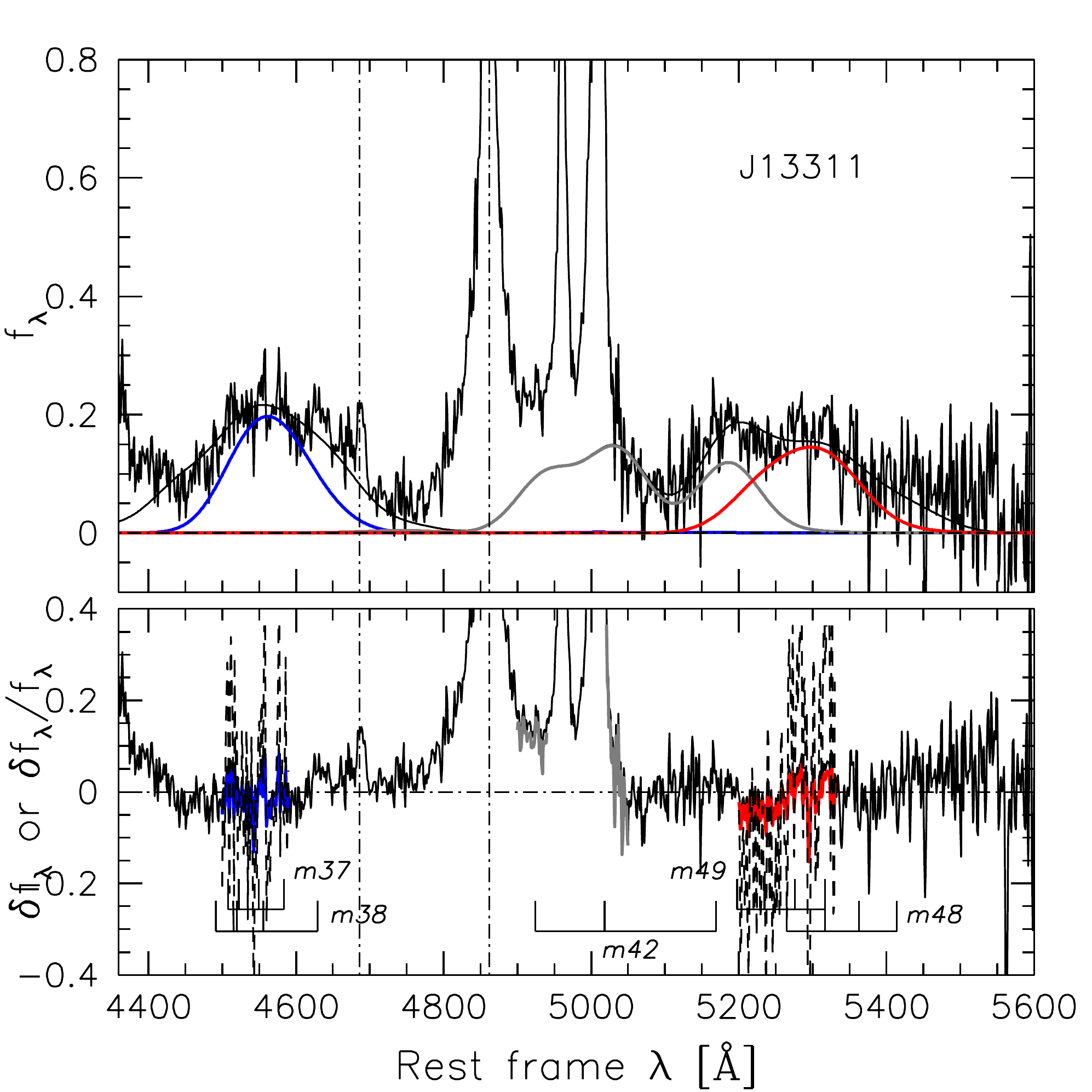}
\includegraphics[width=4.4 cm]{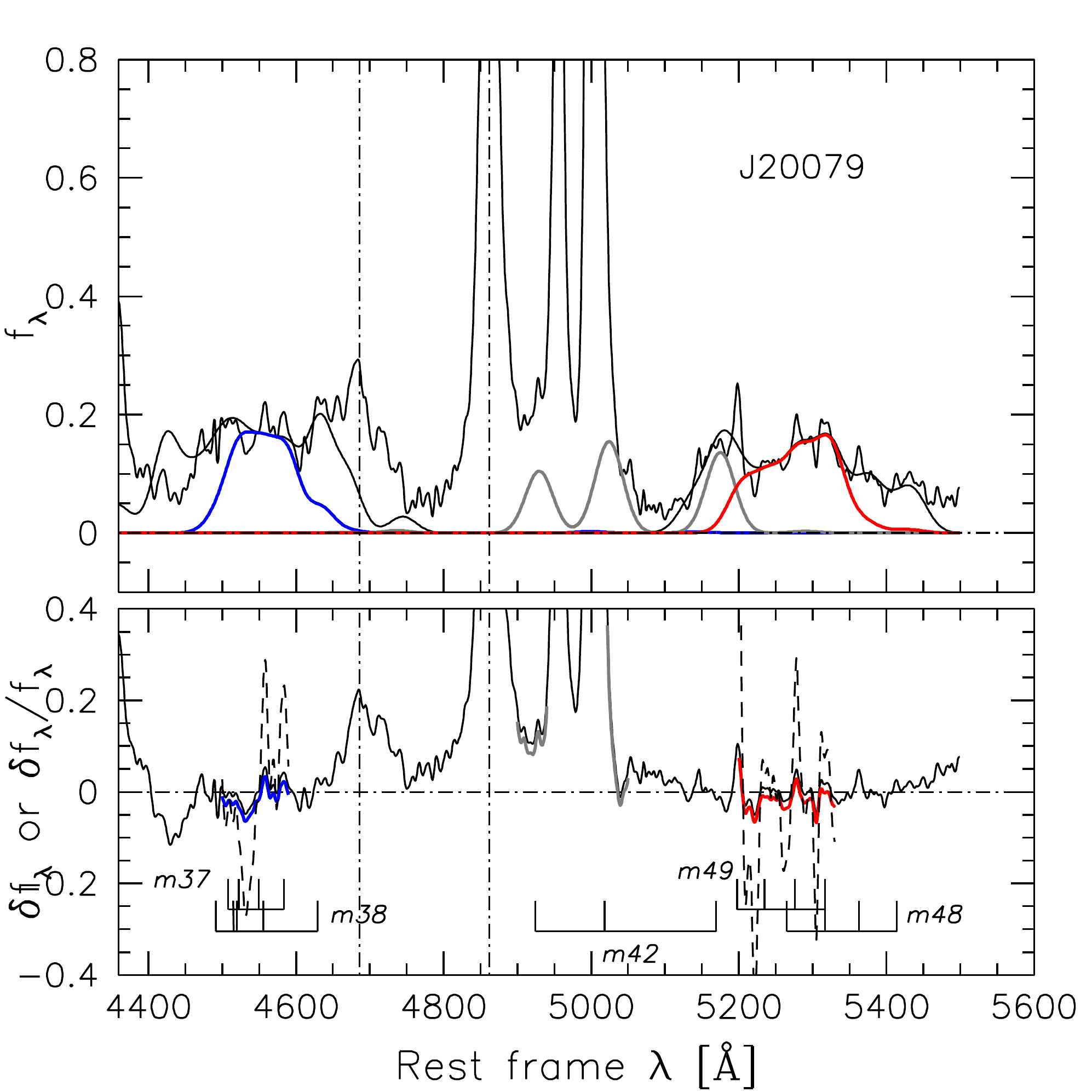}
\caption{Non-linear $\chi^2$\ multi-component analysis with a   \feii\ template  with relative multiplet intensity free to vary, from left to right for the A1 RQ vs. A1 CD, B1 RQ vs. B1 CD, B1 RQ vs. B1 FR-II  composites. The bottom panels show the ratios between the B, m42, and R for the three cases. {Colors as in Fig.~\ref{fig:templafe}.} \label{fig:gammafe}}
\end{figure}

 
 If we apply the same technique to the $\gamma-$detected sources, the intensity ratio between B and R is closer to the value of the template (Fig. \ref{fig:gammafe} and Table \ref{tab:liquid}) but not for 1H 0323+342.  The values reported in Table \ref{tab:liquid} confirm that the template is providing a good approximation for the RQ sources in bin A1 and B1. This is likely true { also} for spectral bins with higher \rfe\ (A2 and A3), as  the template is based on the spectrum of I Zw 1, a NLSy1  accreting at a high rate and especially emitting strong \feii, with \rfe\ slightly larger than 1, { and so of spectral type A3}\footnote{I Zw 1 is however not a very extreme object: the most extreme accretors show spectra similar to the one of PHL 1092 \citep{marinelloetal20a}.}. At the other end of the MS, \feii\ emission is faint and systematic differences might be not appreciable. So the issue of a possible, significant disagreement concerns only  $\approx 3$\%\ of all type 1 AGN. 
 

In summary, the use of  a solid \feii\ template appears fully justified in this work. There could be a genuine effect associated with radio-loudness, in the sense of a less prominent blue blend with respect to the red one. 

\section{Discussion}
\label{disc}


The composites   that we have considered in the present work all show modest \feii\ emission and satisfy the condition that \rfe $\lesssim 0.5$. In principle, on the context of photoionization, they do not pose a serious challenge to the conventional view of the of the {BLR} as a system of emitting clouds characterized by typical densities $n_\mathrm{H} \sim 10^{9.5}$\ cm$^{-3}$, and column density $N_\mathrm{c} = 10^{23}$\  [cm$^{-2}$] \citep{davidsonnetzer79}. Indeed,  physical parameters similar to these $n_\mathrm{H} \sim 10^{10 - 10.5}$\ cm$^{-3}$, along with  metallicity solar or slightly sub-solar, and a flattish spectral energy distribution {(SED)} \citep{laoretal97b,koristaetal97}, can account for the emission in the \rfe\ range of spectral bins A1 and B1, at BLR distances from the continuum sources consistent with {their} luminosity \citep{pandaetal19}. 

Thanks especially to the results of reverberation mapping and \cite{marinelloetal16}, the emitting region radius  was found to be a factor of 10 smaller than previously thought \cite{petersonetal04,peterson17}. The conventional view outlined in the previous paragraph may be only partially valid in Population A sources, as it is unable to account for the strong \feii\ emission in quasars \cite{collinsouffrindumont89,collinjoly00}. Different physical conditions are required for sources with stronger \feii\ emission: higher densities are needed to maintain the ionization parameter within reasonable limits. In spectral type A3, $n_\mathrm{H} \sim 10^{12}$\ cm$^{-3}$\ (strong \feii\ emission appears possible only if $n_\mathrm{H} \gtrsim 10^{10.5}$\ cm$^{-3}$, \citep{matsuokaetal08,martinez-aldamaetal15,pandaetal18}). High column density and super-solar metallicity provide a further boost to the \feii\ intensity and to the strength of the metal lines in the UV \citep{pandaetal18,sniegowskaetal21}.

\subsection{Photoionization computations}
\label{cloudy}
\begin{figure}
    \centering
    \includegraphics[width=0.75\hsize]{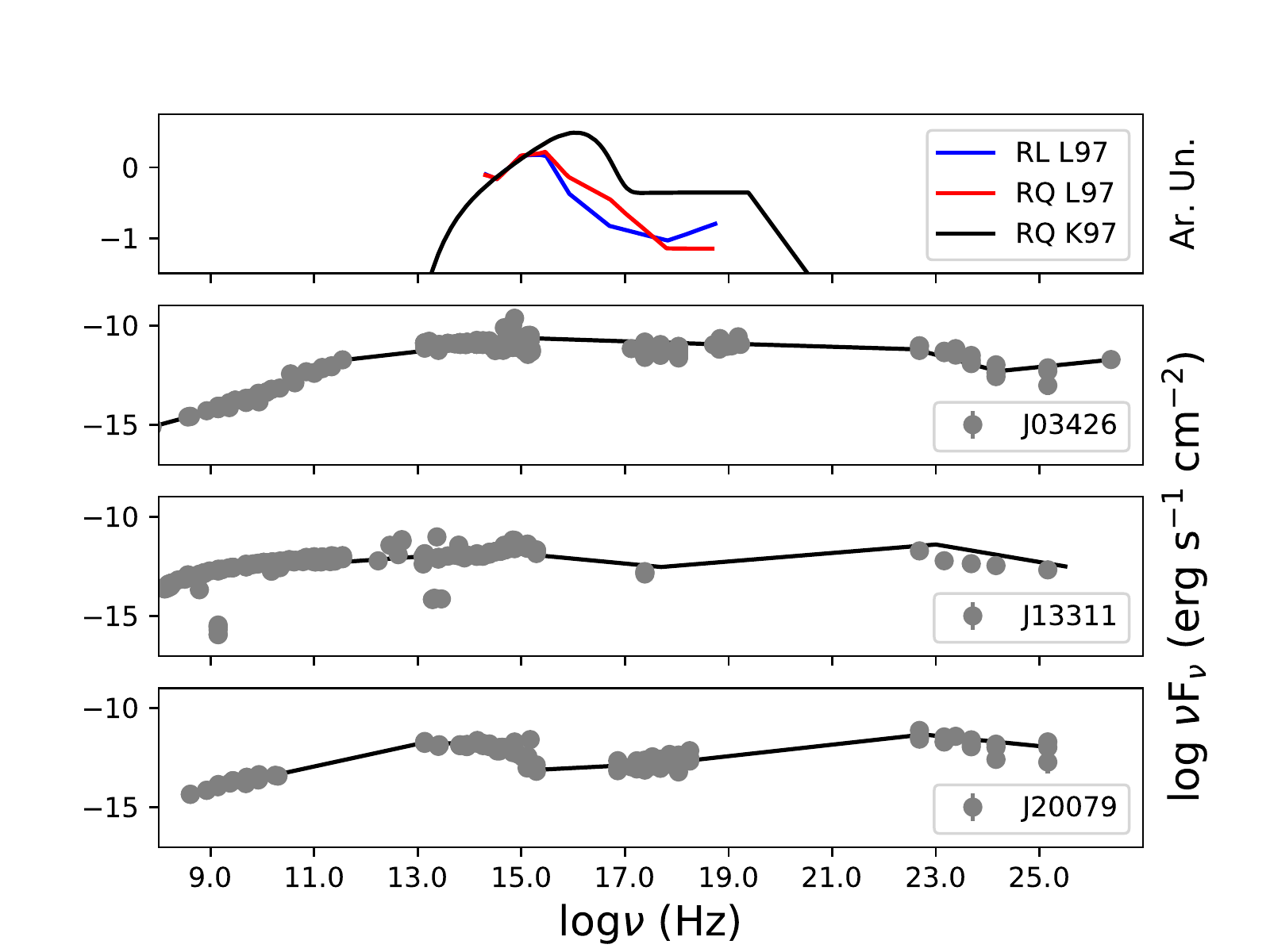}
    \caption{The SEDs used for photoionization computations. From top to bottom: 1) SED of RL and RQ sources from {\citet{laoretal97b}}, represented by the blue and red solid line, respectively{, and from \citet{koristaetal97} for RQ shown with black solid line. The y-axis for the upper panel is shown with arbitrary units (Ar. Un.) for the specific flux}; 2) SED of J03426. The observed data points are shown in grey, while the SED used in the models is represented by the black solid line, 3) as in (2) for J13311; 4) as in (2) for J20079. }
    \label{fig:sed}
\end{figure}

To ascertain the role of the SED in the strength of \feii\ we performed exploratory calculations of the RQ and RL differences, as well as for the $\gamma$-detected NLSy1s.  As a first attempt, {we assumed a} column density $N_\mathrm{c} = 10^{23}$\  cm$^{-2}$, solar metallicity, and 0 micro-turbulence. {We focused} on the exploration of the parameter ranges of {the density} $n_\mathrm{H}$ ($\sim 10^{9 - 11}$\ cm$^{-3}$), and {of} BLR radius {within} the range expected from the \citet{bentzetal13} scaling law, {corrected} for high-accretion rates (\citep{martinez-aldamaetal20,duwang19}; see also \citep{donofrioetal21}). We utilized {the} SEDs from \citet{laoretal97b} and \citet{koristaetal97}, which are most appropriate for bins A1 and B1, for RQ and RL separately (top panel of Fig. \ref{fig:sed}). The two RQ SEDs most likely bracket the distribution of SEDs in the two bins, as also assumed in previous work \citep{pandaetal19}.  For the composites, we consider a fixed luminosity $\lambda L_\lambda (5100$ \AA) $\approx 10^{45}$ \ergss, consistent with the values derived for the original sample \citep{marzianietal13a}. {For RL NLSy1s, we built a specific SED for each one of them by using} the multi-frequency data of the SED builder available at the Space Science Data Center of the Italian Space Agency\footnote{\href{https://tools.ssdc.asi.it/SED/}{https://tools.ssdc.asi.it/SED/}}. The {results} are shown in {the second, third, and fourth} panel of Fig. \ref{fig:sed}. 

\subsection{Composites RL and RQ}

\begin{figure}[t!]
\centering
\includegraphics[width=4.25  cm]{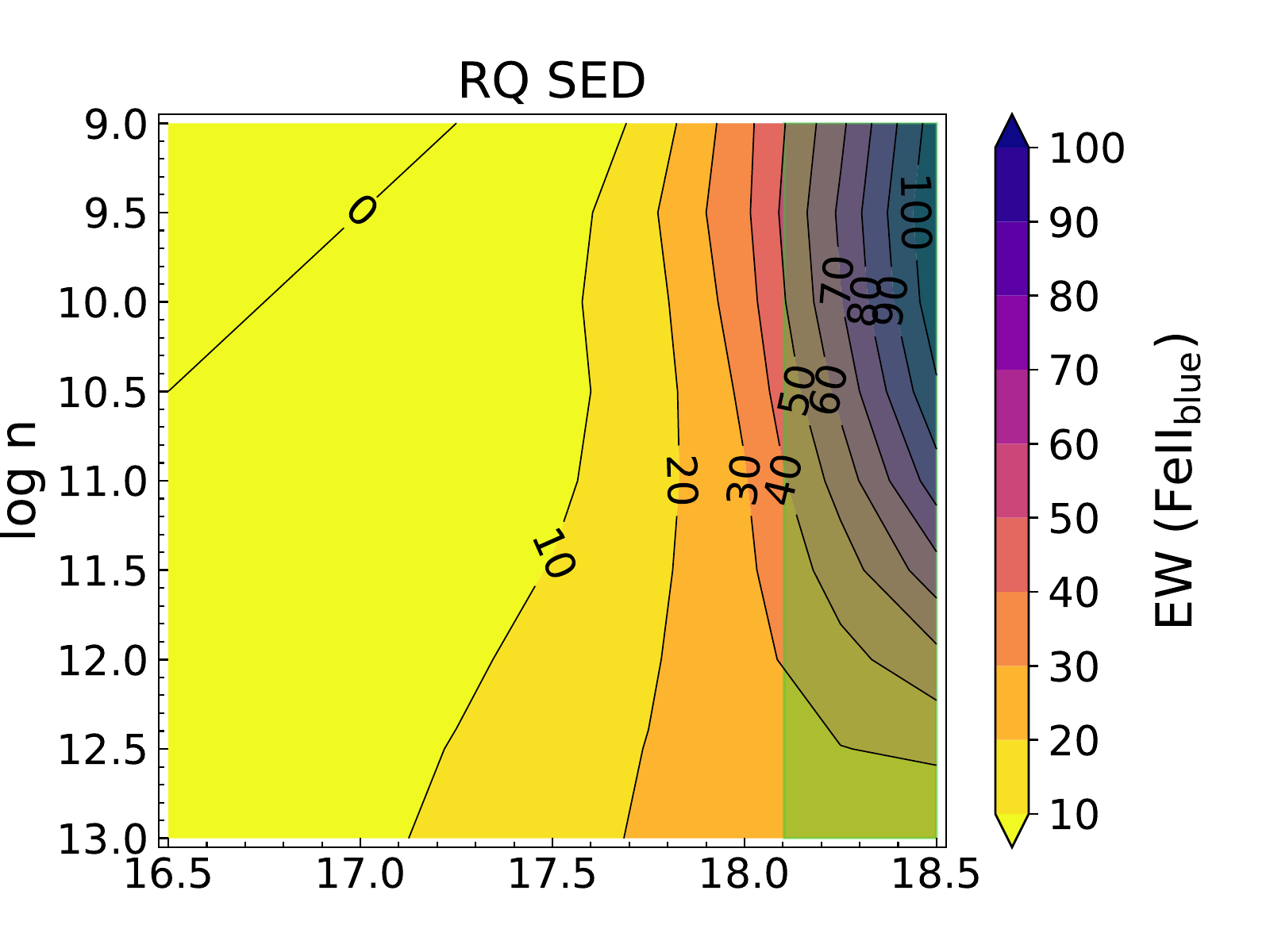}
\includegraphics[width=4.25  cm]{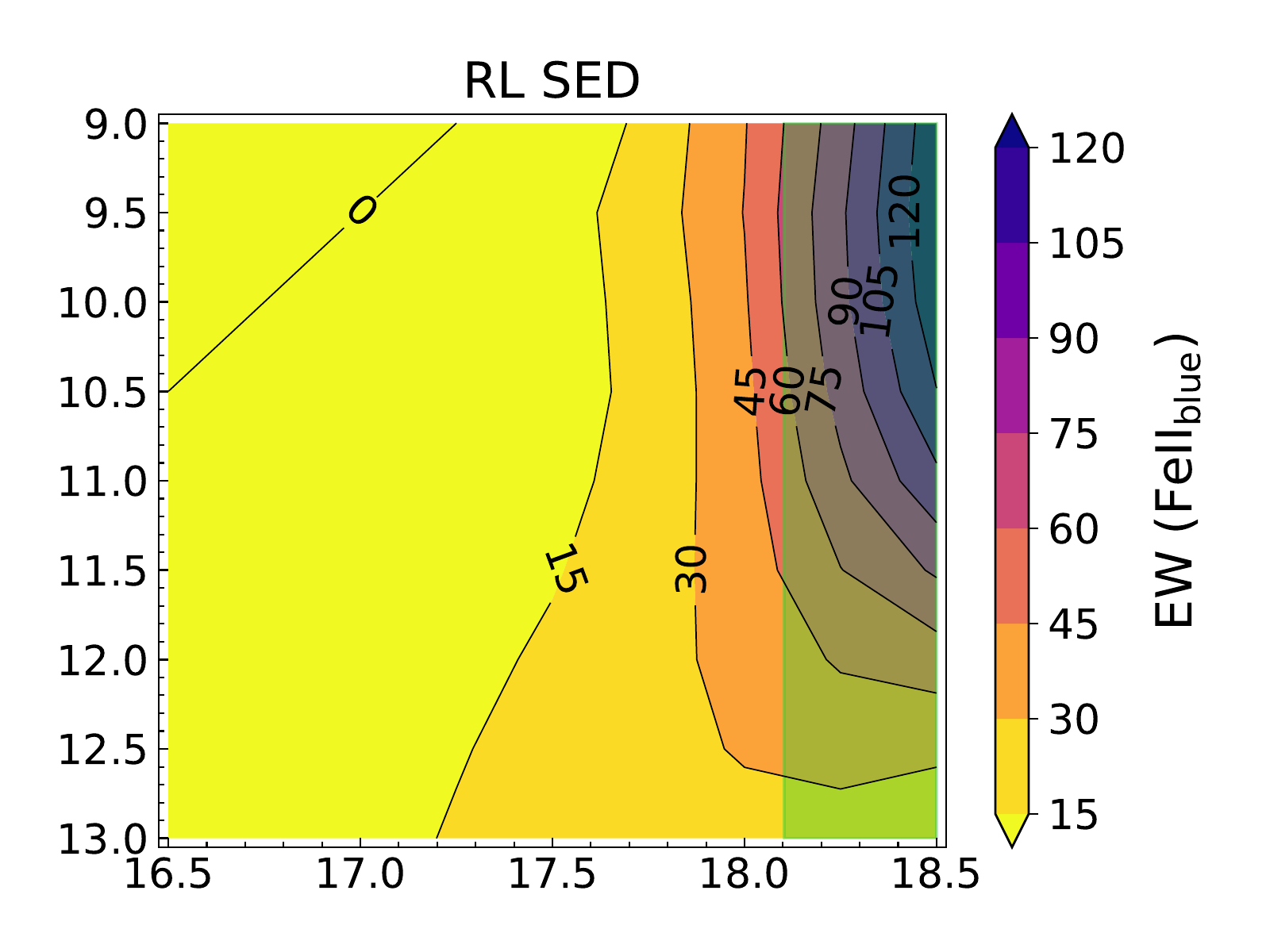}
\includegraphics[width=4.25  cm]{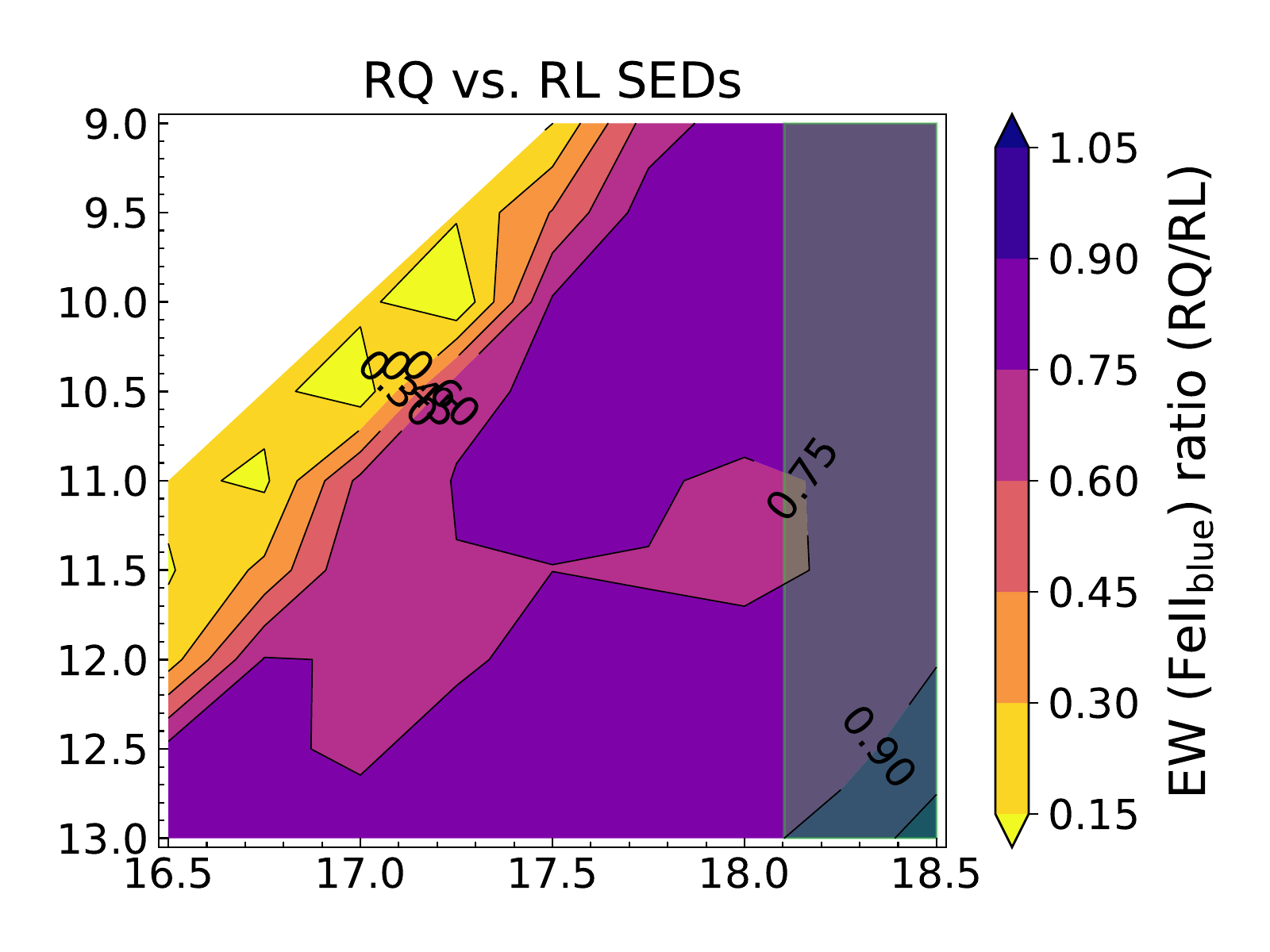}\\
\vspace{-0.25cm}
\includegraphics[width=4.25  cm]{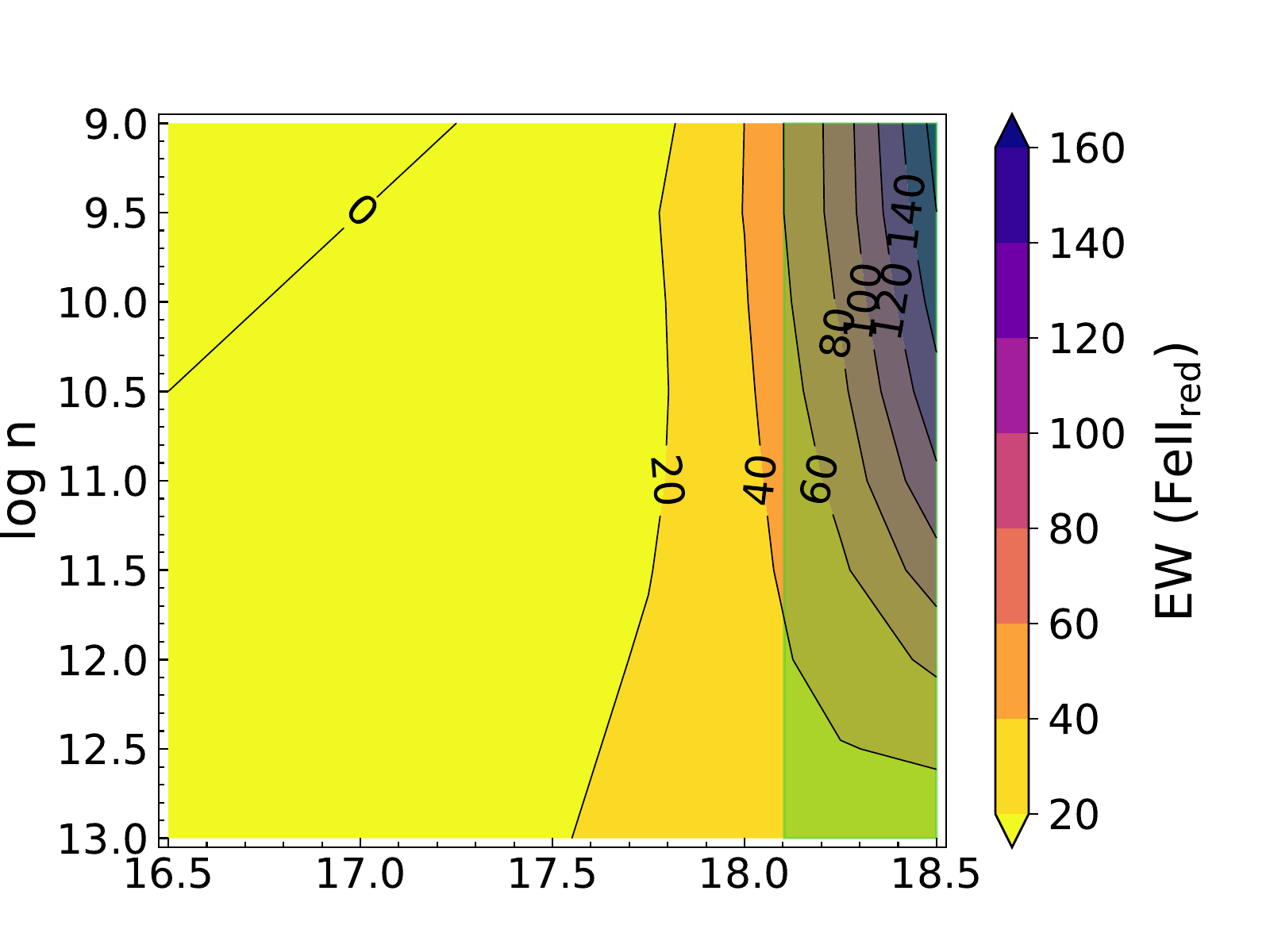}
\includegraphics[width=4.25  cm]{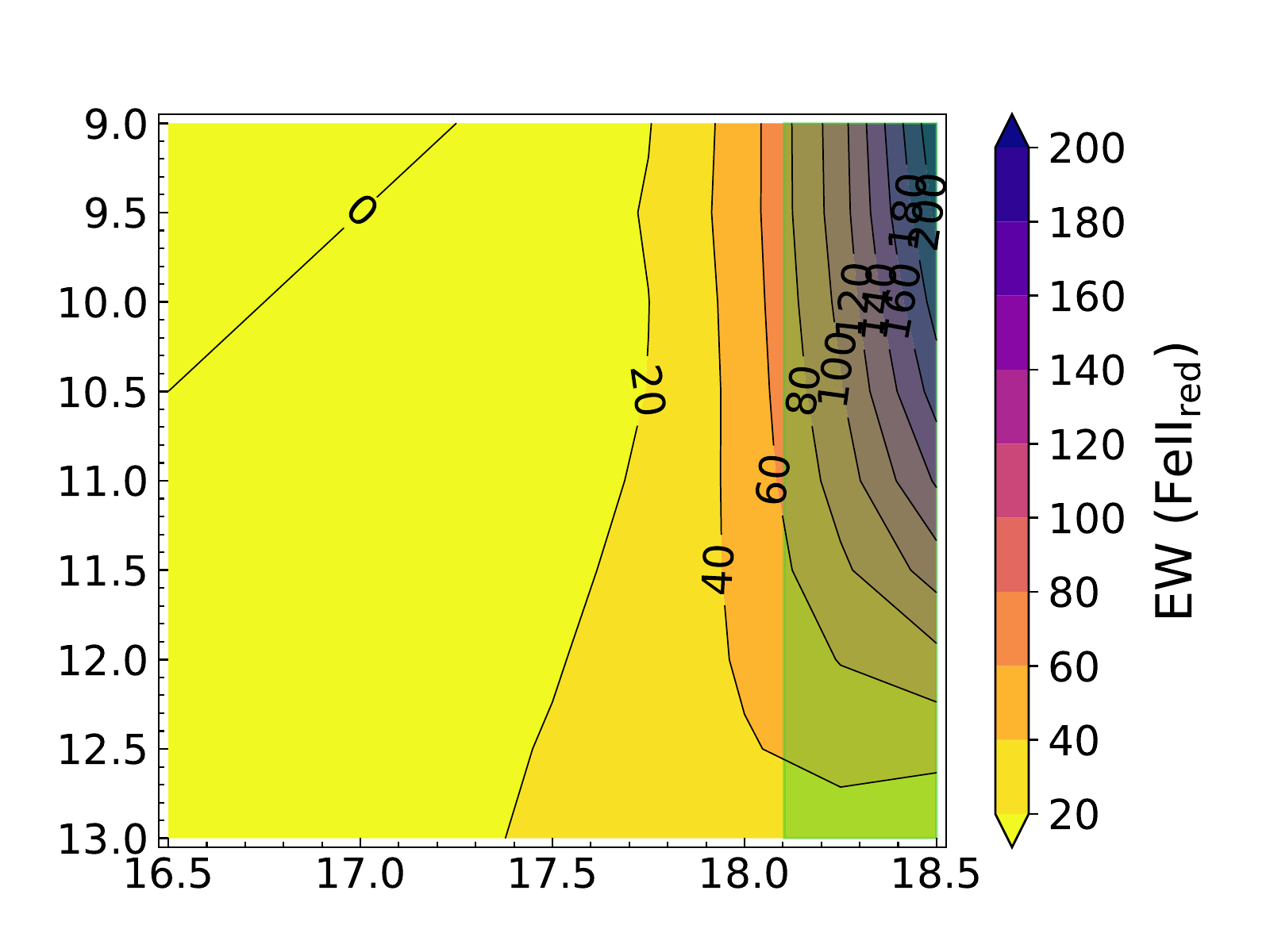}
\includegraphics[width=4.25  cm]{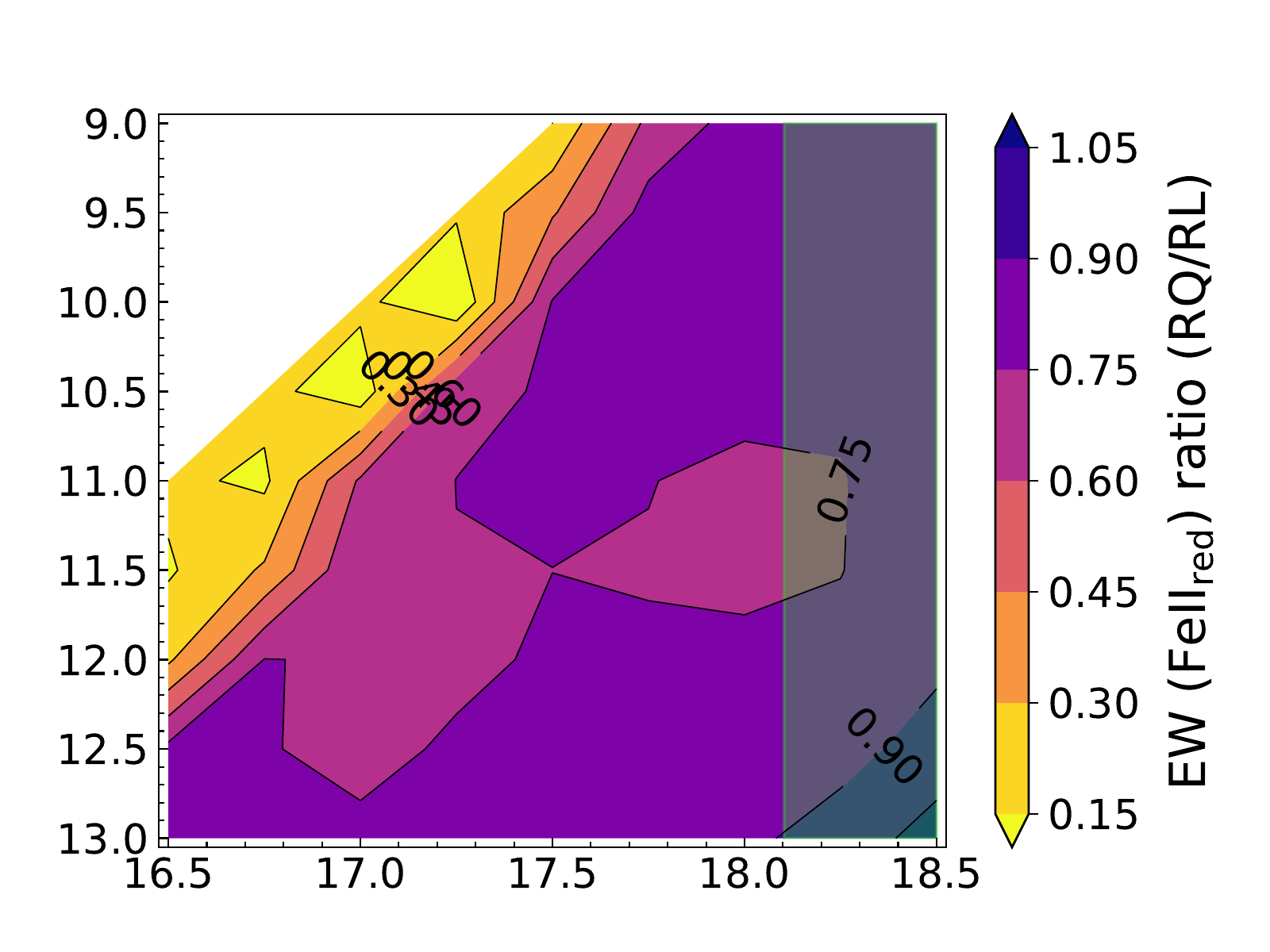}\\
\vspace{-0.25cm}
\includegraphics[width=4.25 cm]{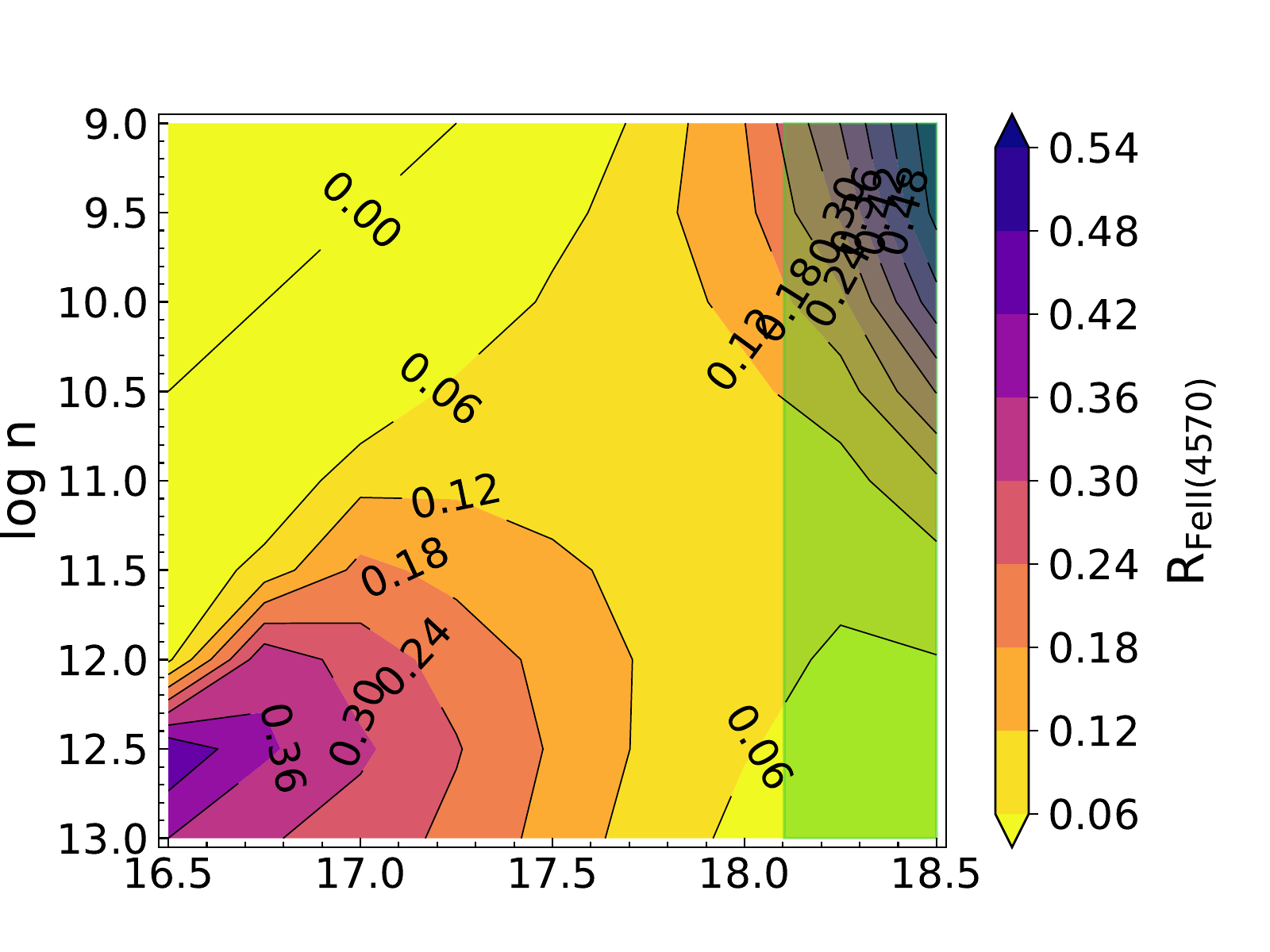}
\includegraphics[width=4.25  cm]{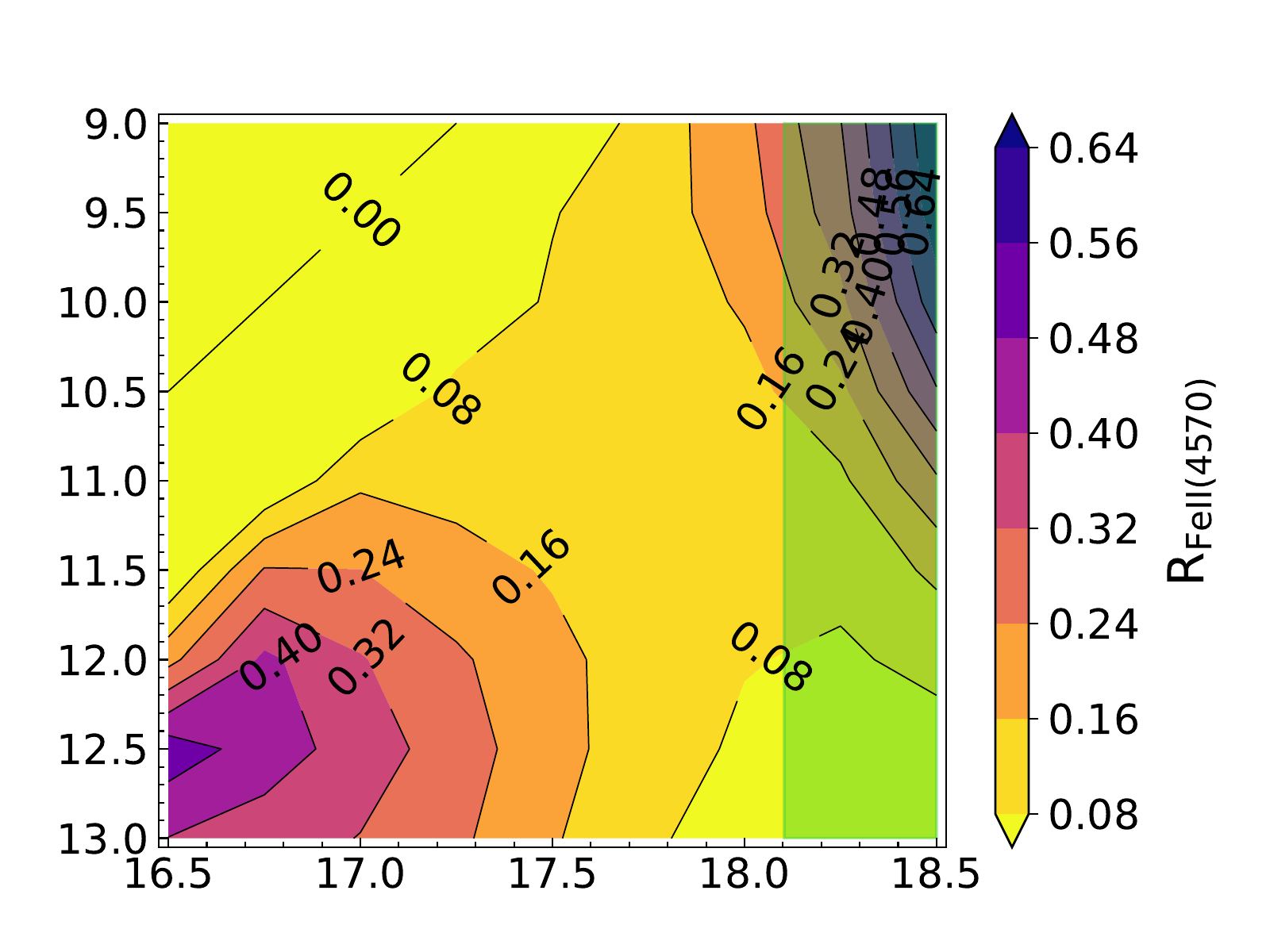}
\includegraphics[width=4.25  cm]{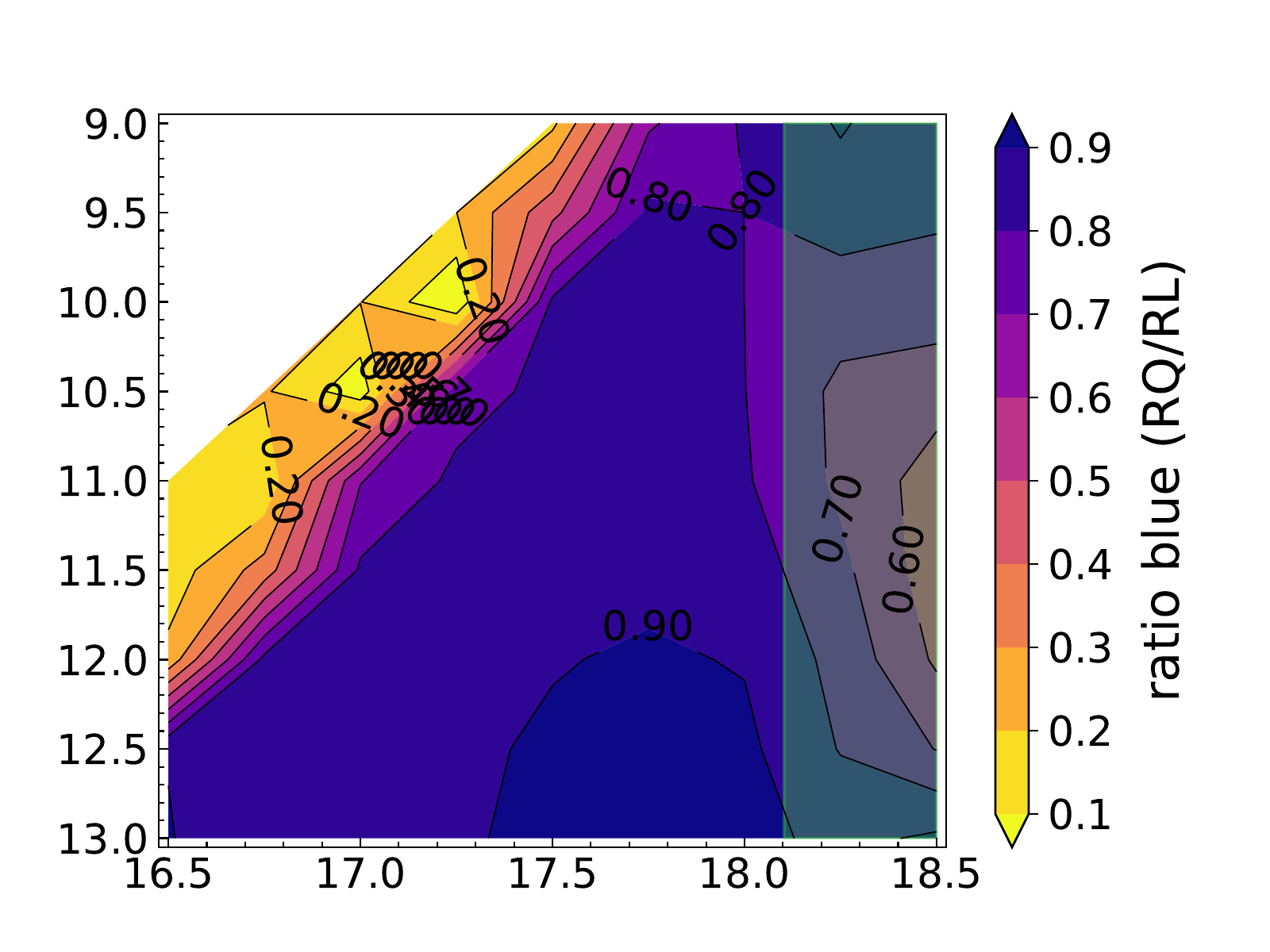}\\
\vspace{-0.25cm}
\includegraphics[width=4.25 cm]{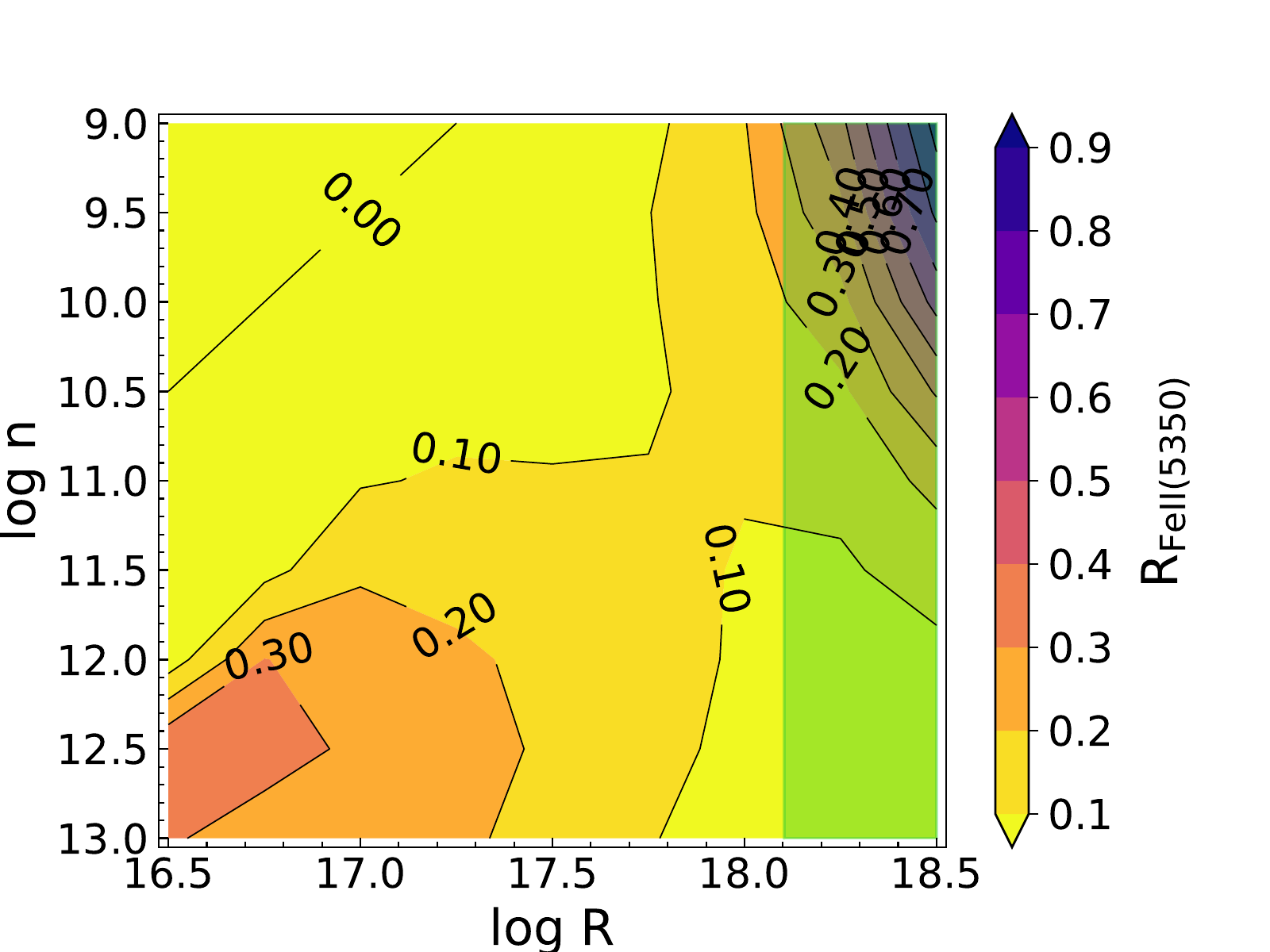}
\includegraphics[width=4.25  cm]{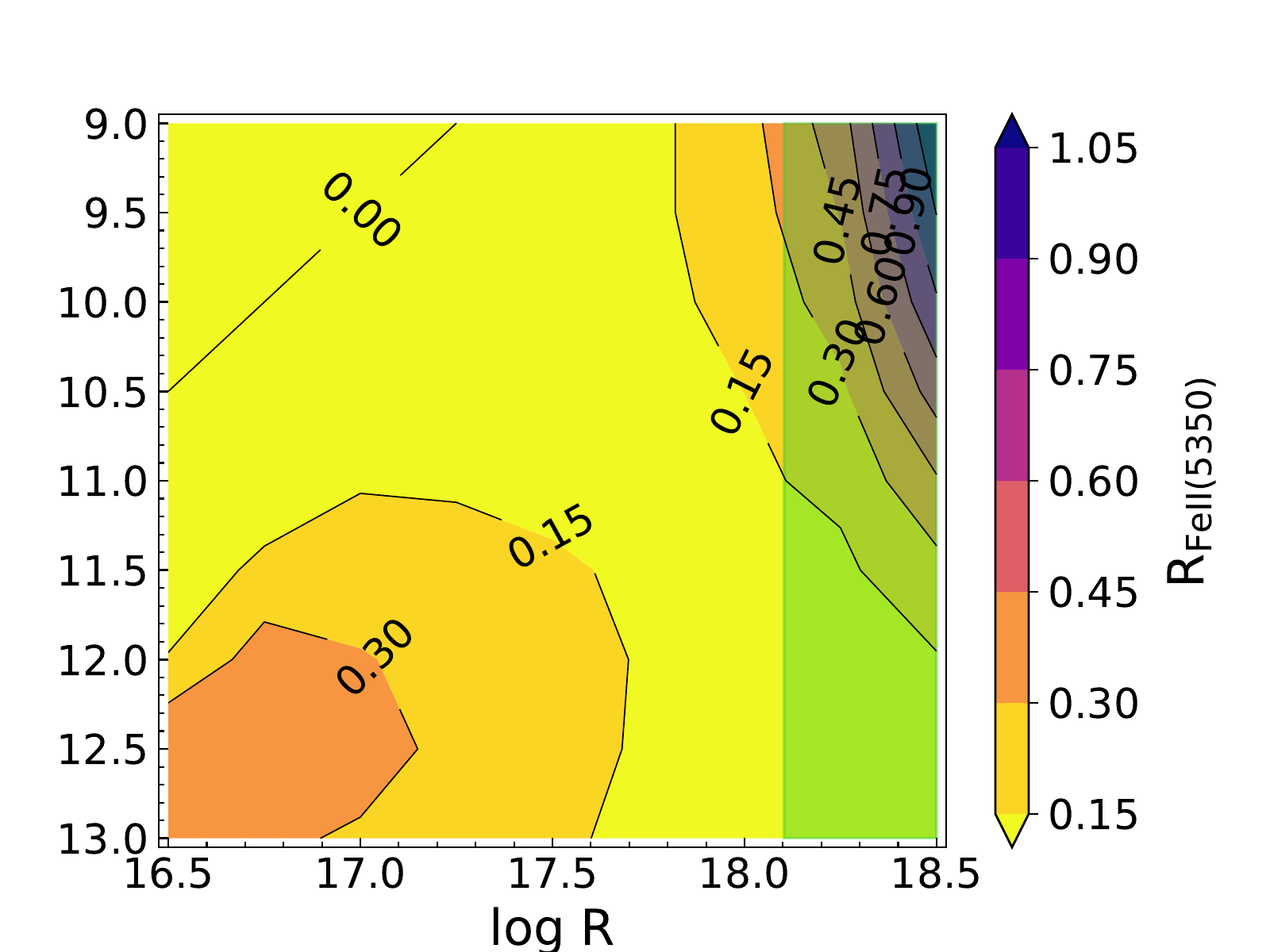}
\includegraphics[width=4.25  cm]{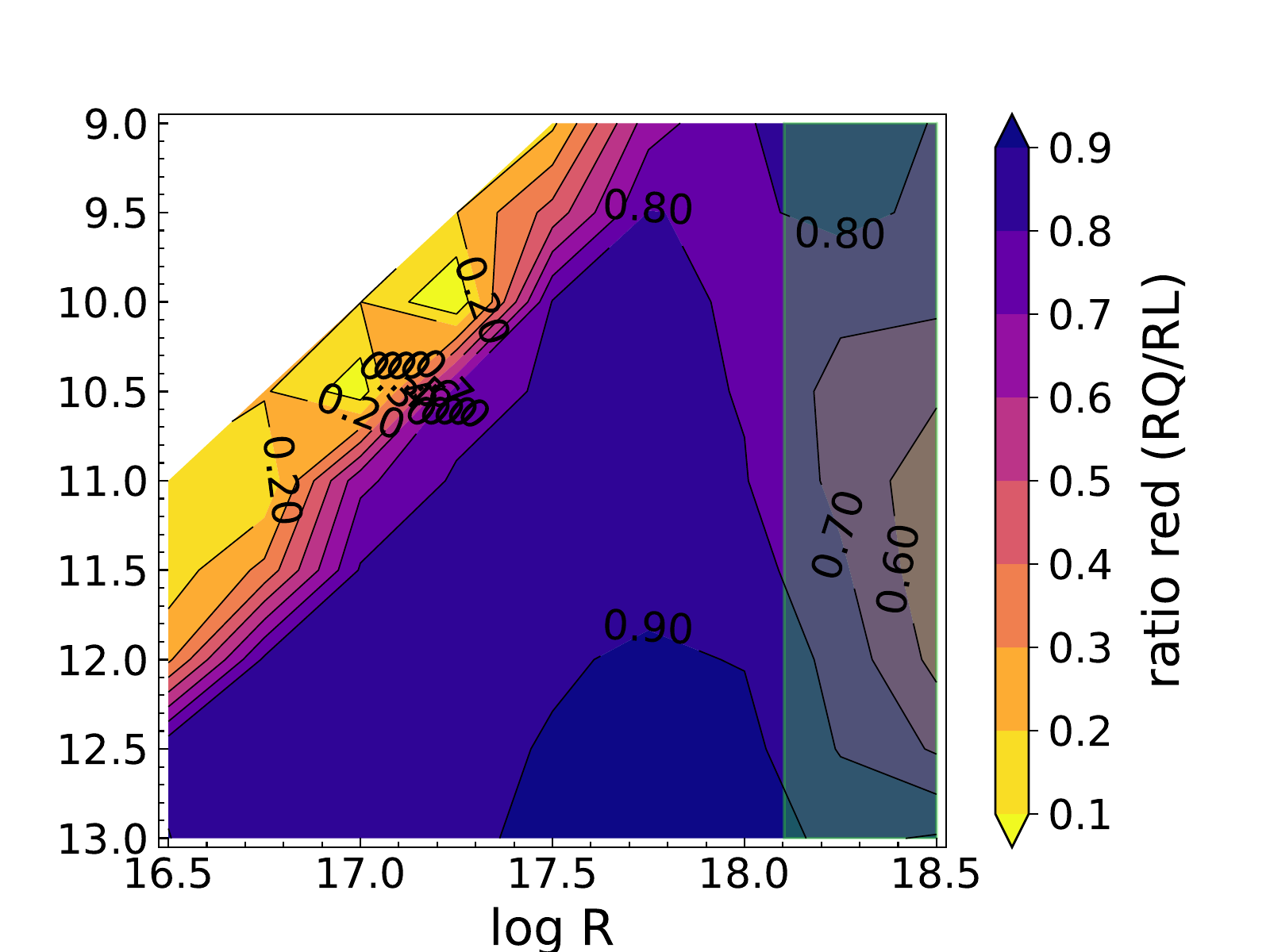}\\
\caption{Results of {\tt CLOUDY} simulations. First two rows: W(\feiiq) for RQ and RL (i.e., computed for the same physical parameters but with the two different SEDs of \citet{laoretal97b})  and their ratios;  third and fourth row from top: \rfe, for RQ and RL and their ratios. Last four rows: same, for the red (R) \feii\ blend peaked at $\lambda$ 5312 \AA.  \label{fig:ph_comp1}}
\end{figure}

Fig. \ref{fig:ph_comp1} shows the expected \rfe, the ratio between the red blend R and \hb, and the ratio B/R for the RQ and RL SEDs, as a function of \rb\ and \nh. {As mentioned above,} the range in \rb\ has been centered on the value expected from the correlation \rb\ - $L$\ by \citet{bentzetal13}. {The density range includes the most likely values for the BLR along the MS. Typical values are lower for Population B, and much higher for Population A.} \citep{marzianietal01,marzianietal18}. 

The most important result is that {\em the SED for RL sources is able to account for the modest \rfe\ and W(\feiiq) reported in Table \ref{tab:basic}}. The {observed} equivalent width $\sim 10 - 20$ \AA\ and \rfe $\sim $ 0.15\ are explained at the \rb\ expected from the \rb\ -- $L$\ relation, and by the moderate density \nh $\sim 10^{11}$ cm$^{-3}$. The result seems to be especially robust against changes in \nh\ and \rb\ (Fig. \ref{fig:ph_comp1}, middle column). Restricting the ranges to $\tilde{B}$\ and $\tilde{R}$\ confirms this conclusion (Fig. \ref{fig:ph_comp_res}; the result of the analysis for the restricted range are shown in Appendix \ref{app:restr}).  

\begin{figure}[t!]
\centering
\includegraphics[width=4.25  cm]{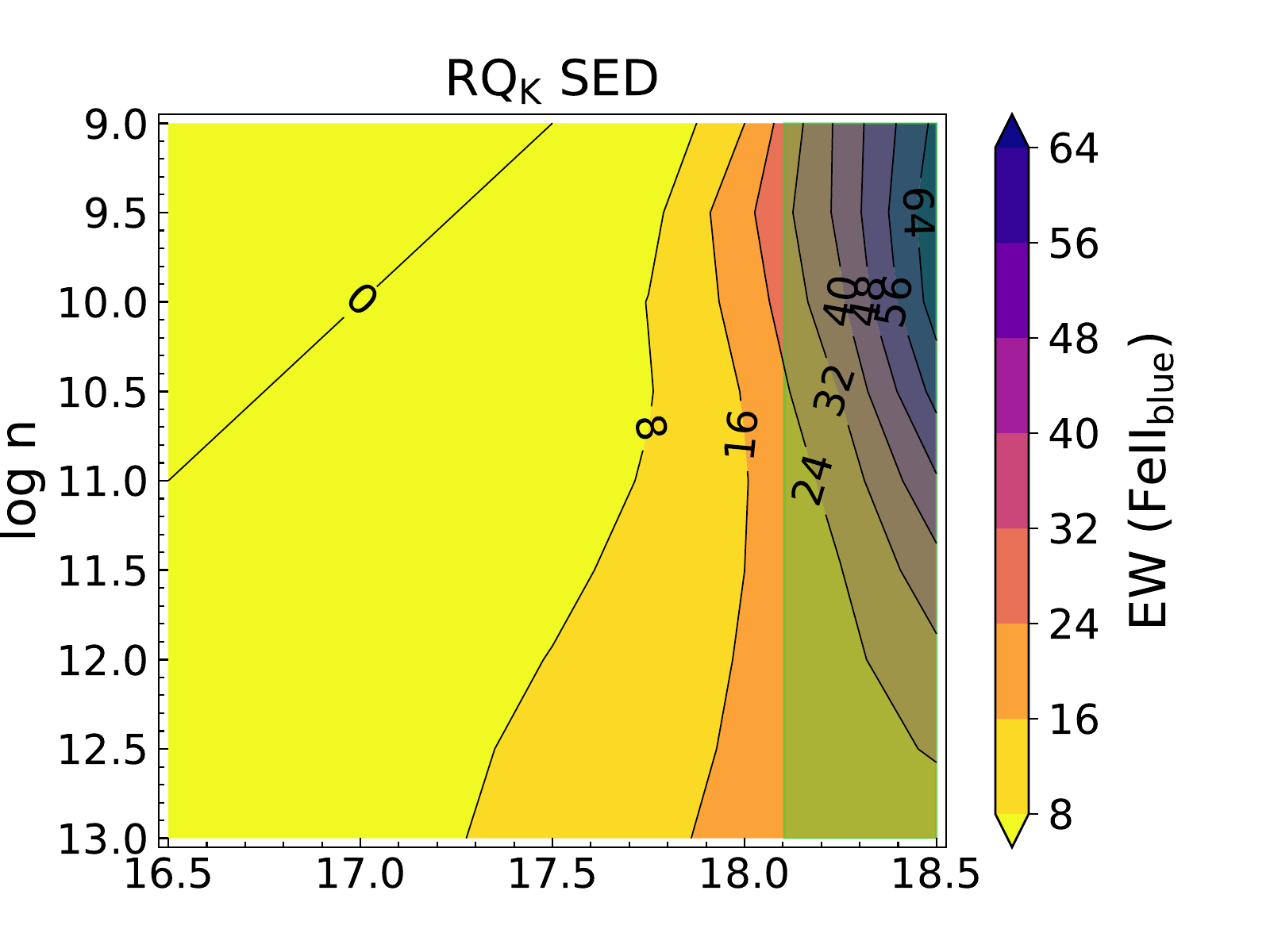}
\includegraphics[width=4.25  cm]{logR-logn_vs_EWFe_blue_RL.pdf}
\includegraphics[width=4.25  cm]{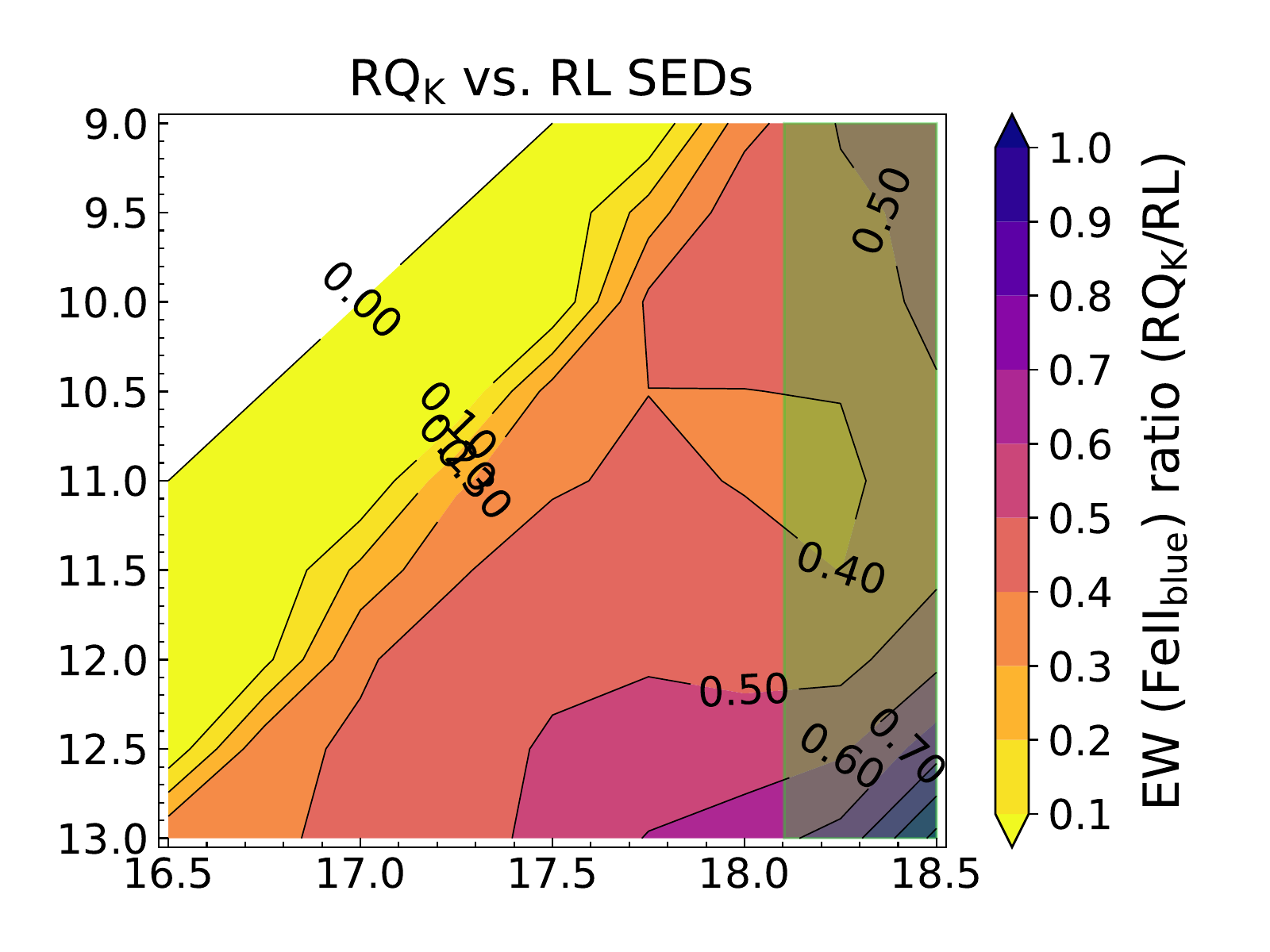}\\
\vspace{-0.25cm}
\includegraphics[width=4.25  cm]{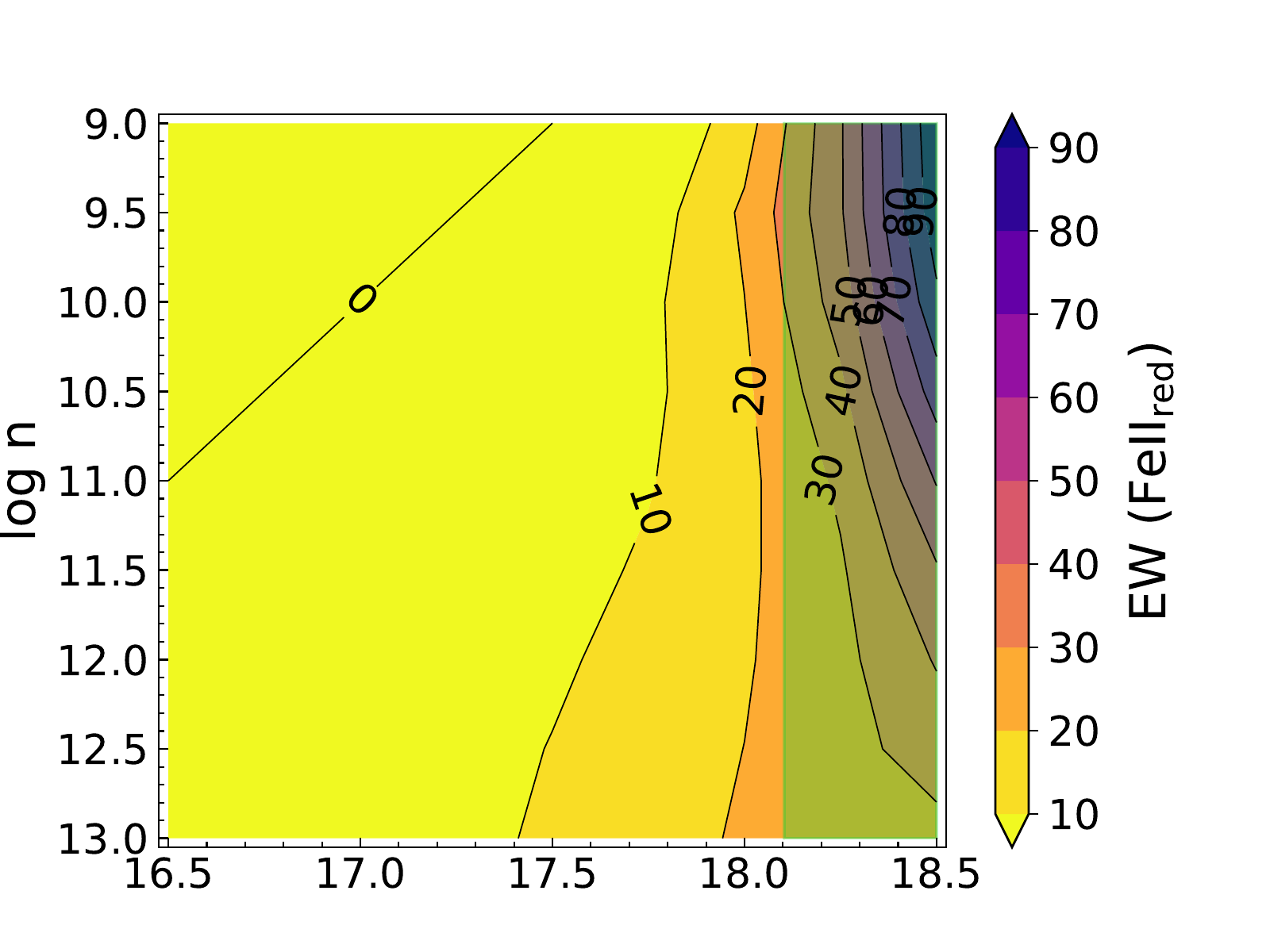}
\includegraphics[width=4.25  cm]{logR-logn_vs_EWFe_red_RL.pdf}
\includegraphics[width=4.25  cm]{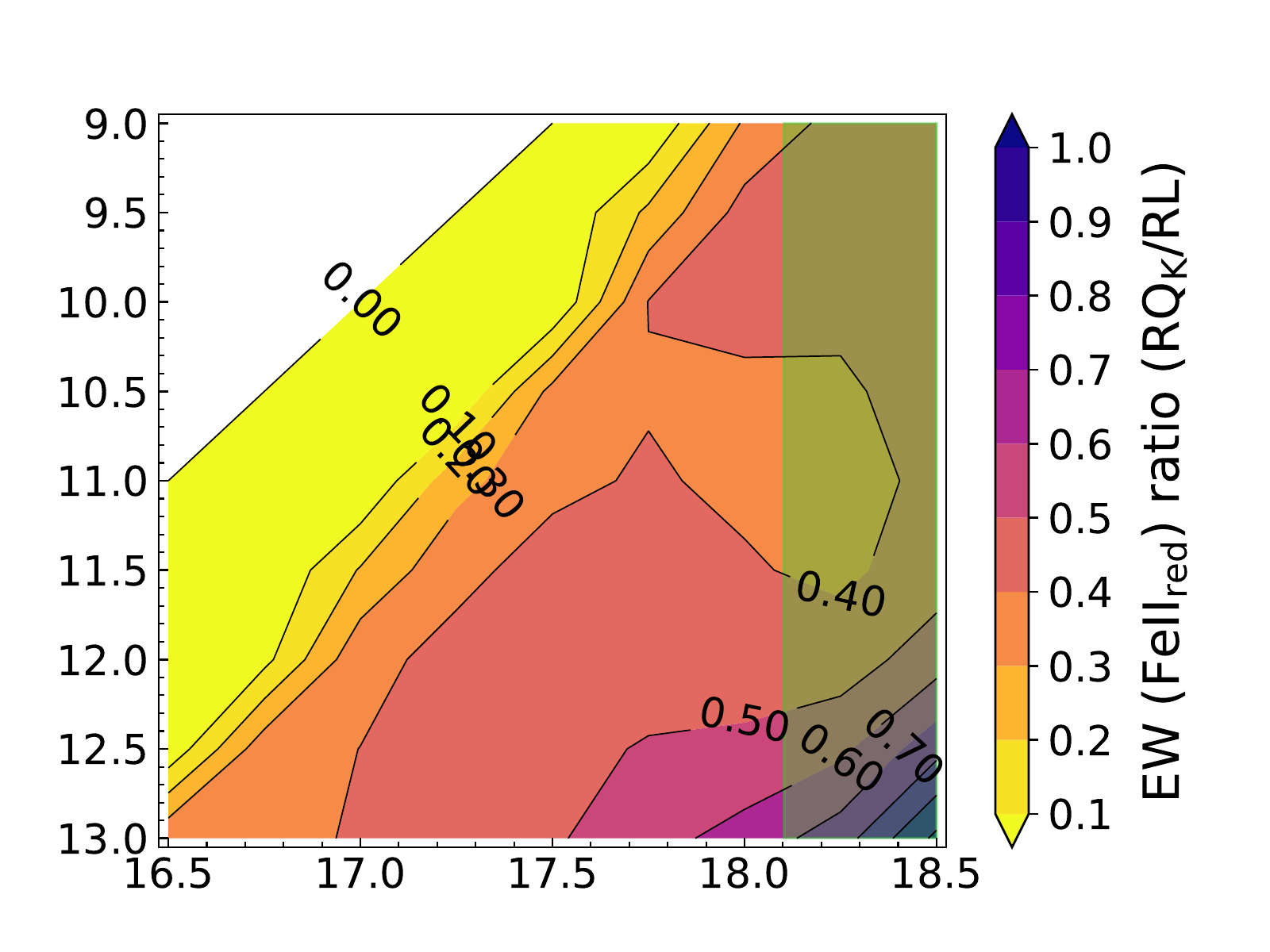}\\
\vspace{-0.25cm}
\includegraphics[width=4.25 cm]{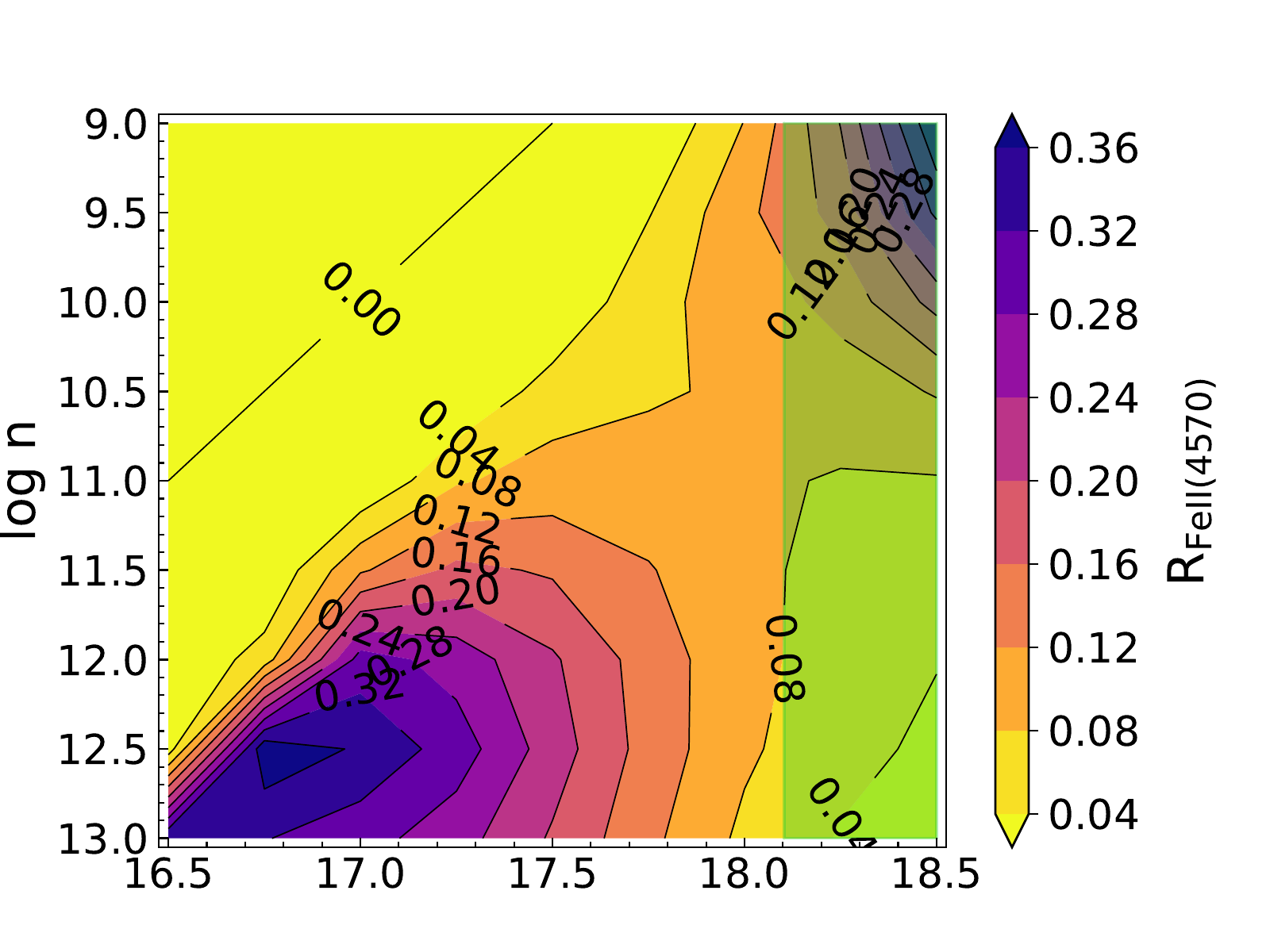}
\includegraphics[width=4.25  cm]{logR-logn_vs_RFe_blue_RL.pdf}
\includegraphics[width=4.25  cm]{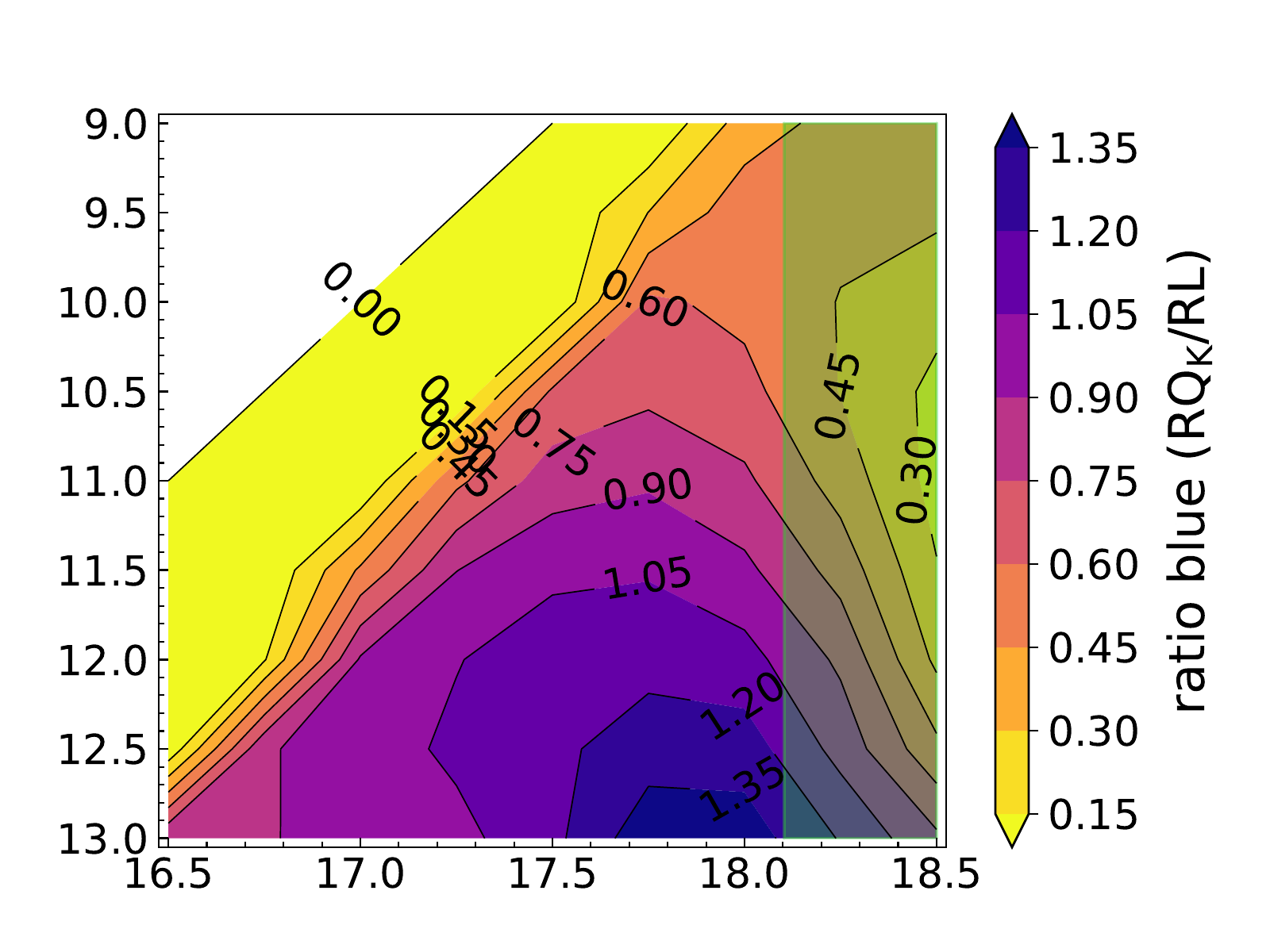}\\
\vspace{-0.25cm}
\includegraphics[width=4.25 cm]{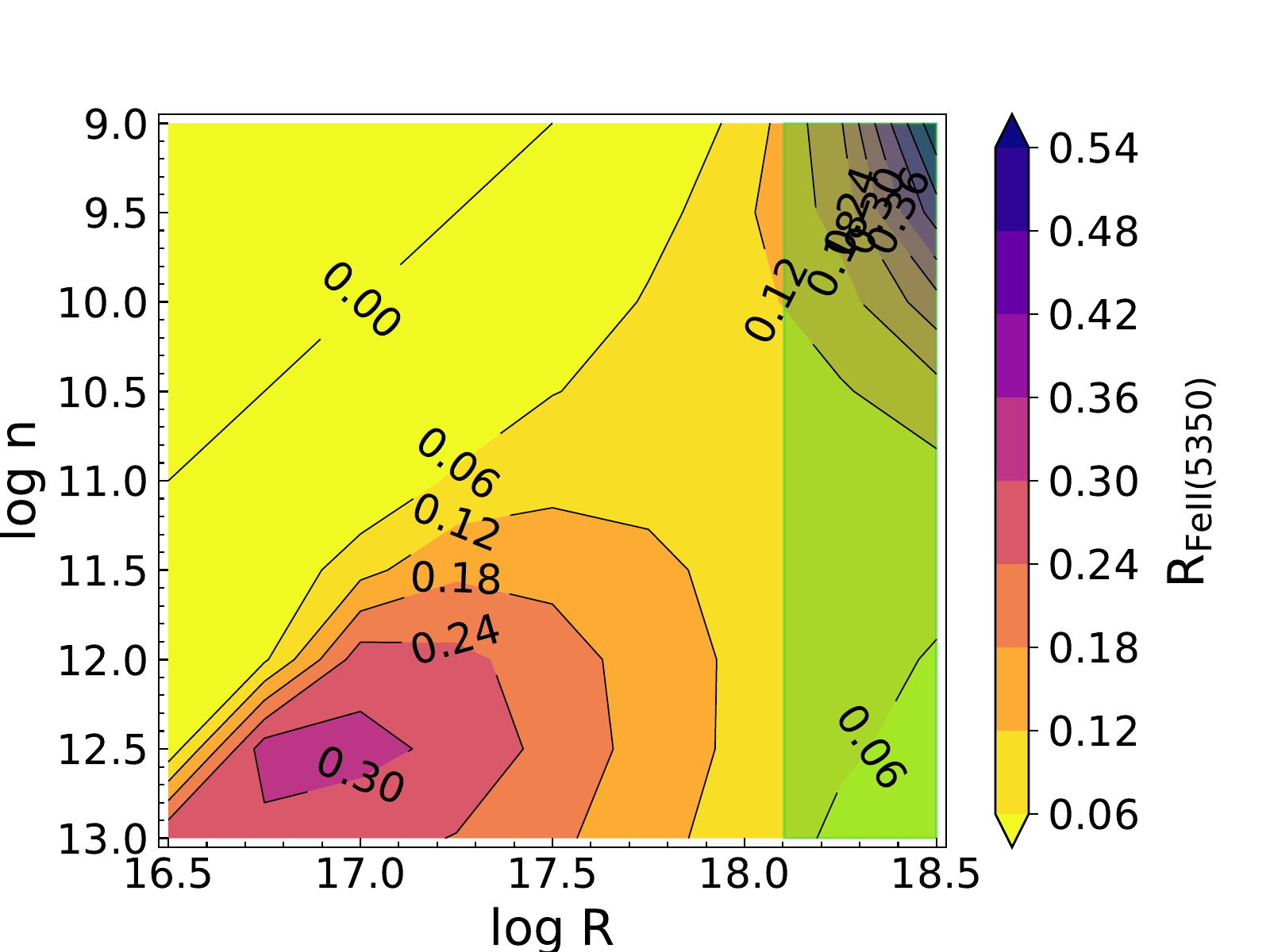}
\includegraphics[width=4.25  cm]{logR-logn_vs_RFe_red_RL.pdf}
\includegraphics[width=4.25  cm]{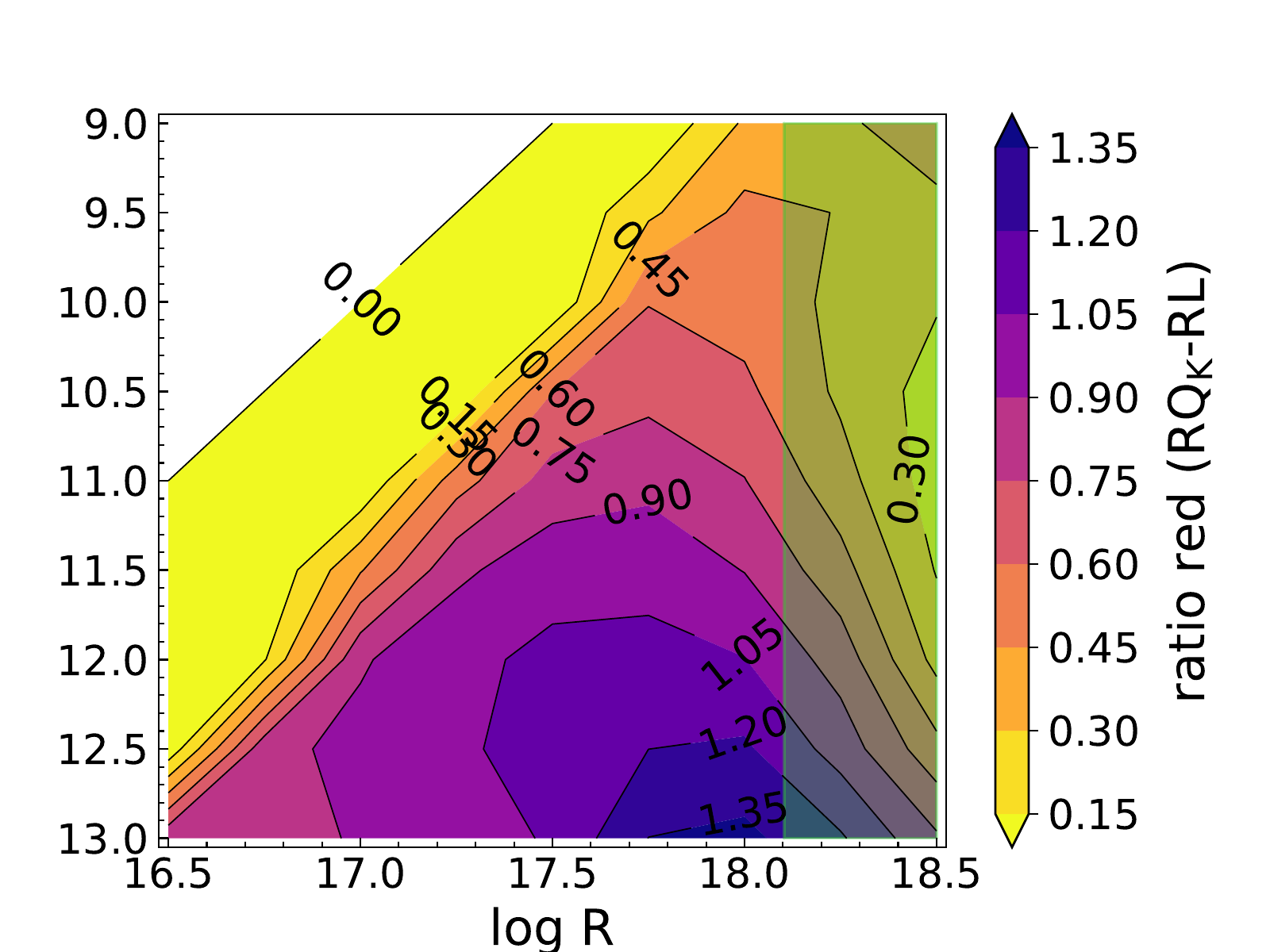}\\
\caption{Results of {\tt CLOUDY} simulations, same as in Fig. \ref{fig:ph_comp_rlrq} but for the RQ SED of \citet{koristaetal97}.  \label{fig:ph_comp_kor}}
\end{figure}

Regarding the interpretation of the composite spectra, a second important issue is whether the difference { in RQ and RL} SED can explain the differences in {their} \feii\ strength. We stress that the comparison is being carried out between sources that are in the same spectral types, meaning that they have similar \mbh, $L$, and Eddington ratio. The \rfe\ computed over a large area of the parameter plane \nh\ -- \rb\ shows differences that are at most 0.05, i.e., less than $15$\%\ (Fig. \ref{fig:ph_comp1}). This has the important implication that the parameter $k \approx 2$, {\em i.e., \rfe\ twice as strong in RQ than in RL AGNs, cannot be explained on the basis of the SED only.}
However, the RQ SED that we adopt is not necessarily equivalent to the one of the sources used to construct the {Population} A and B. Using the SED from \citet{koristaetal97} that has a stronger {\em big blue bump} (implying for fixed \rb\ and \nh\ an increase in ionization parameter by a factor 3) cannot lead to  a  higher \rfe\ (Fig. \ref{fig:ph_comp_kor} and Fig. \ref{fig:ph_comp_kor_res}).

\begin{figure}[t!]
\centering
\includegraphics[width=4.25  cm]{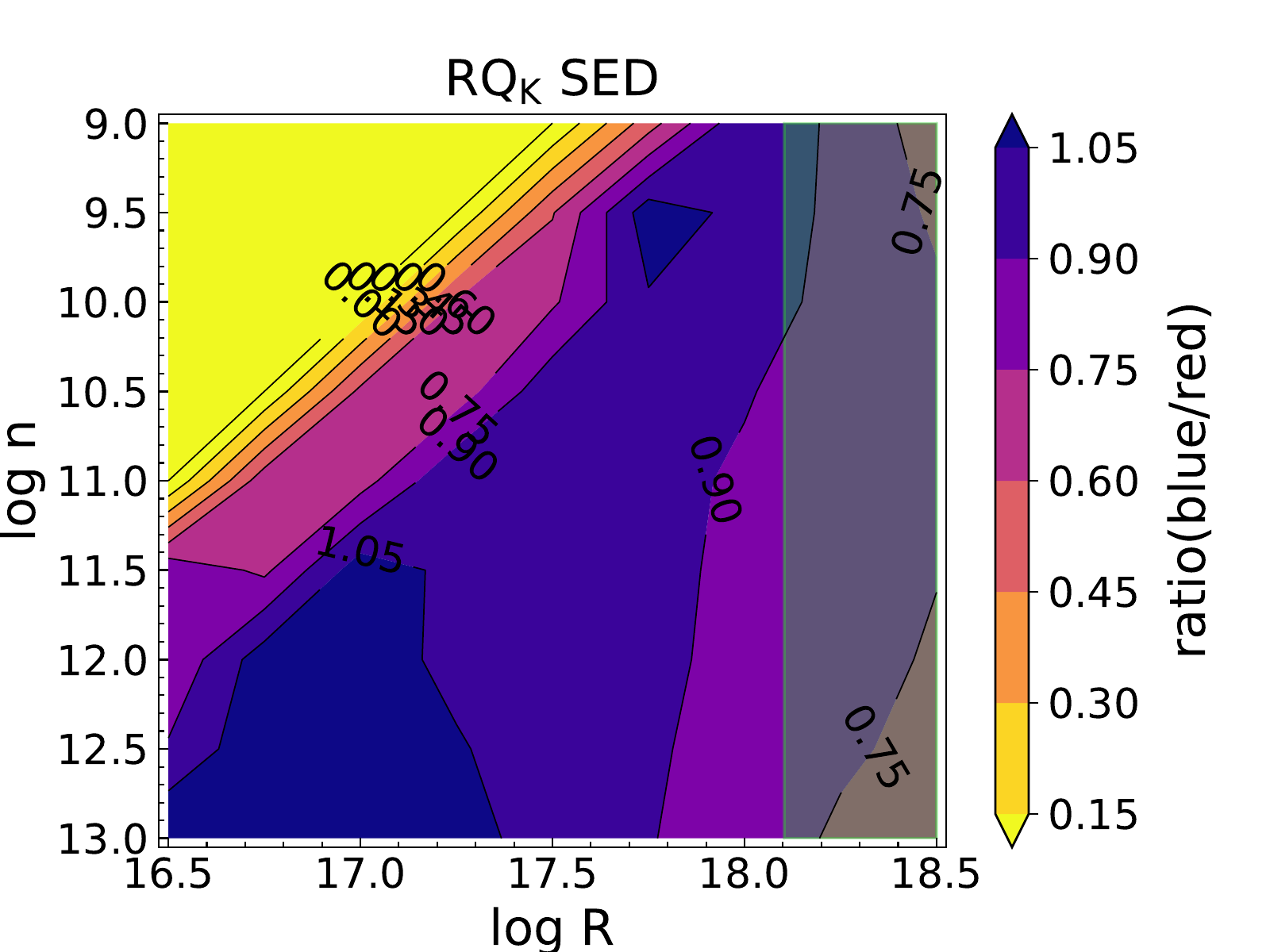}
\includegraphics[width=4.25  cm]{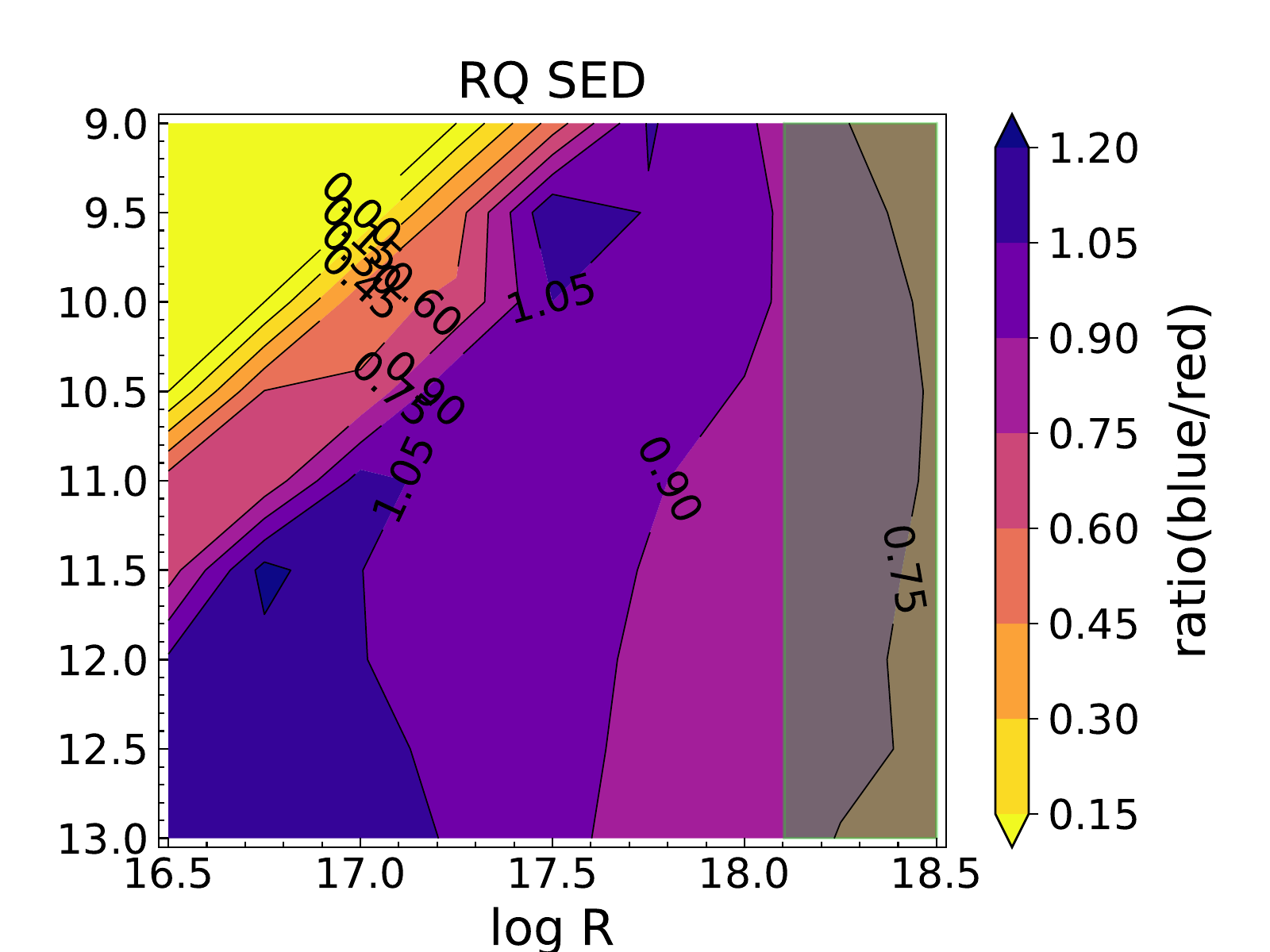}
\includegraphics[width=4.25 cm]{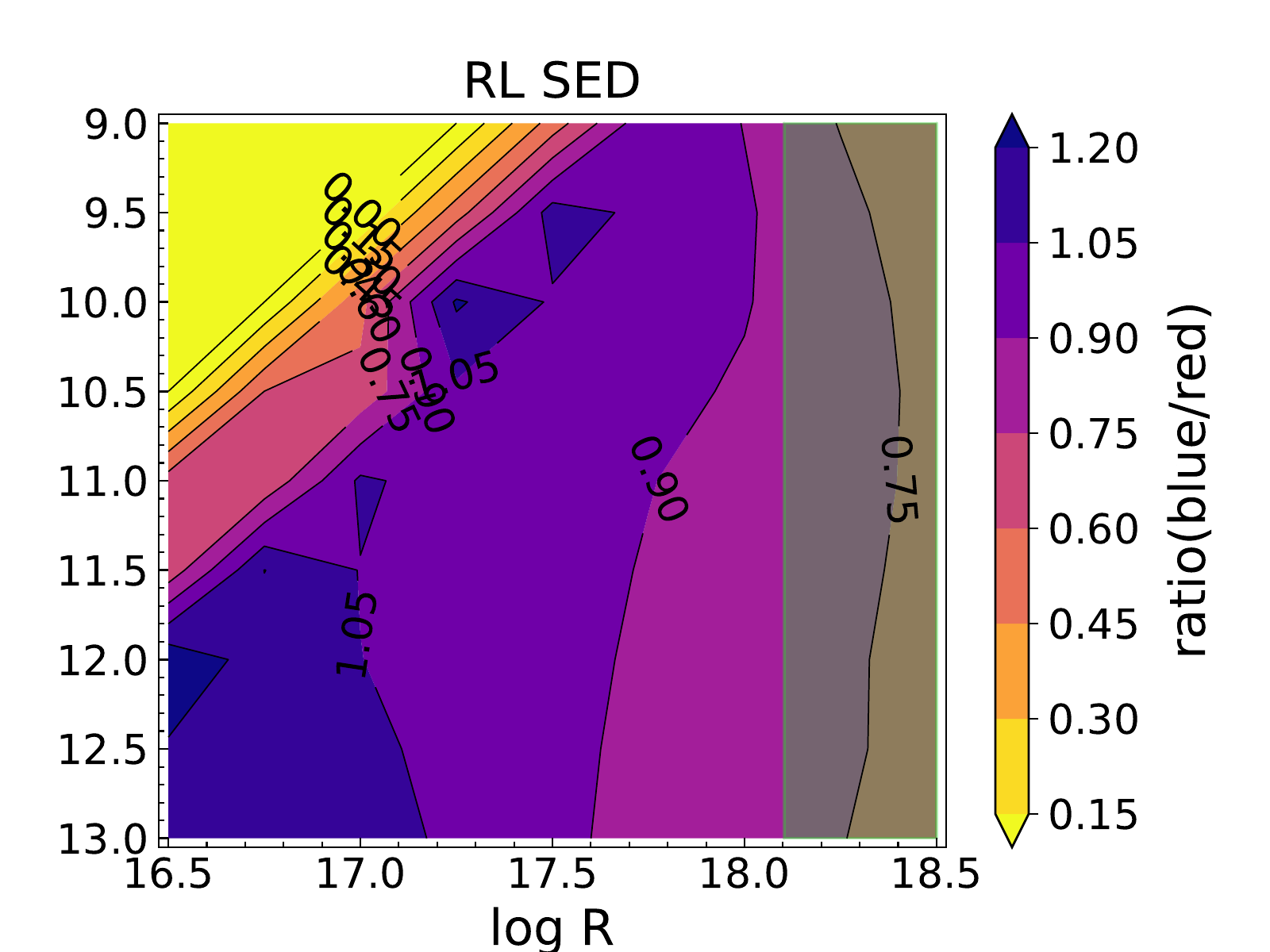}
\caption{Results of {\tt CLOUDY} simulations: the three   panels yield  the ratio  \rfe\ B to \rfe R (in practice B/R) as a function of density and BLR radius, for the RQ \citet{koristaetal97} (leftmost), the RQ \citet{laoretal97a} (middle) and RL SED (rightmost). \label{fig:ph_comp_rlrq}}
\end{figure}

The second issue investigated in the present paper is the possibility of a systematic difference in the B/R \feii\ ratio. This result should be viewed with care, and confirmatory data are needed, also considering that we are dealing mainly with weak \feii\ emitters. The agreement between the observed and photoionization-predicted B/R is good, and in most cases the B/R $\sim$ 1, with little dependence on the \rb\ and \nh\ parameters. Fig. \ref{fig:ph_comp_rlrq} (Fig. \ref{fig:ph_comp_rlrq_res} for the restricted ranges) shows that there is a possible lowering of the B/R ratio in the B1 RL (both CD and FR-II) with respect to the RQ.  Accepted at face value, it might imply a lower density, and a higher ionization parameter for the RL sources (Fig. \ref{fig:ph_comp_rlrq}), although Figs. \ref{fig:ph_comp1} and \ref{fig:ph_comp_kor} show that this implies fainter \feii\ emission than {what is expected for the typical values of} \rb. 

\citet{sulenticetal16} speculated that the cocoon associated with the shock wave due to the expansion of the jet could push outward the line emitting regions. It is known that the \civ\ high-ionization line reaches lower blueshift amplitudes in RL than in RQ sources \citep{sulenticetal07,richardsetal11}. The present paper confirms that an effect on the low-ionization lines is minor. Considering the {layout} of the atomic levels  in the Grotrian diagram of Fig. \ref{fig:grotrian}, the lower B in radio loud sources could be due a lower electron $T$\ in RL than in RQ, which could be, according to Fig. \ref{fig:ph_comp_rlrq}, in turn associated with a lower $U$ and a larger distance.  

\begin{figure}[t!]
\centering
\includegraphics[width=6.25 cm]{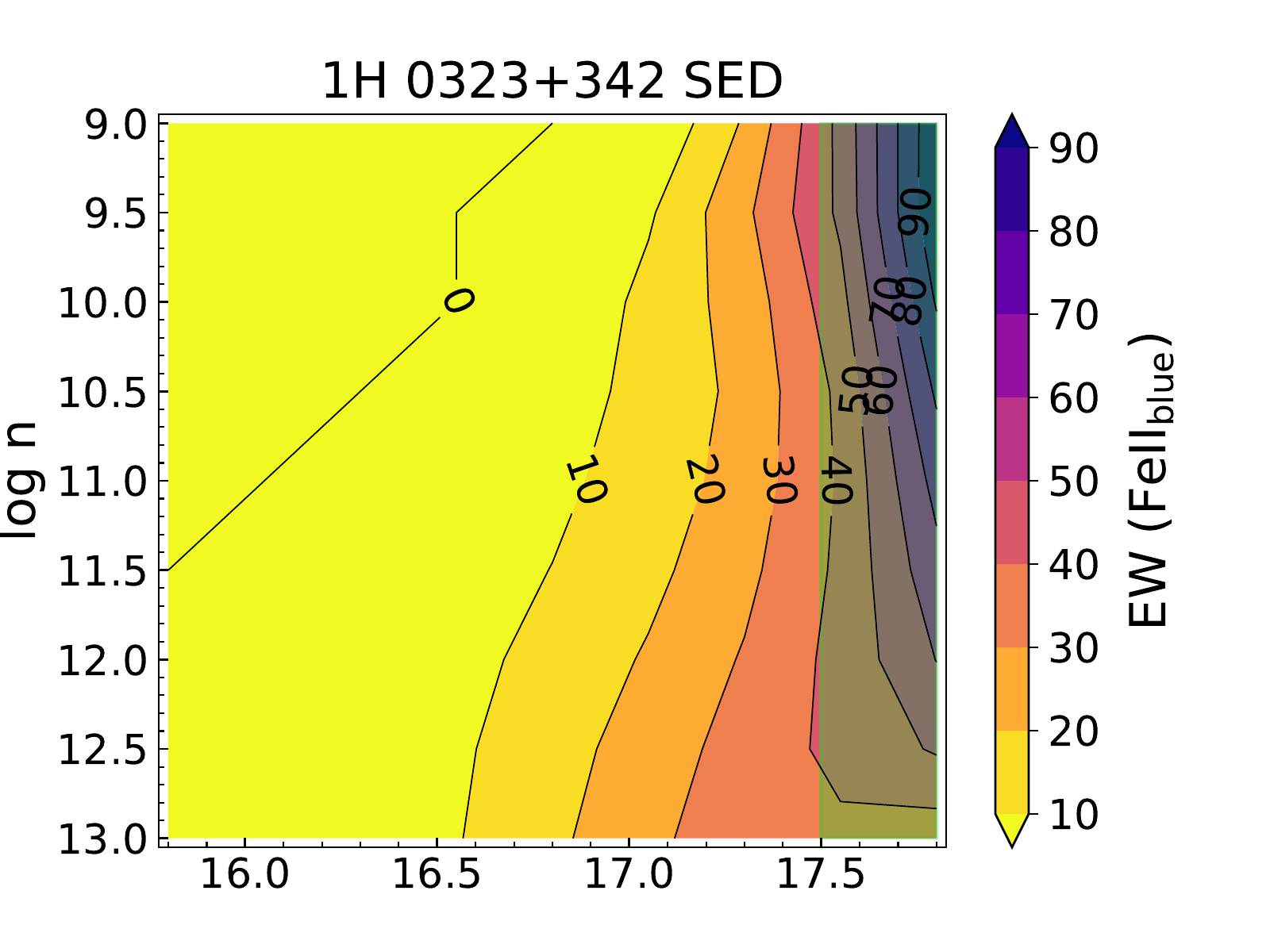}
\includegraphics[width=6.25 cm]{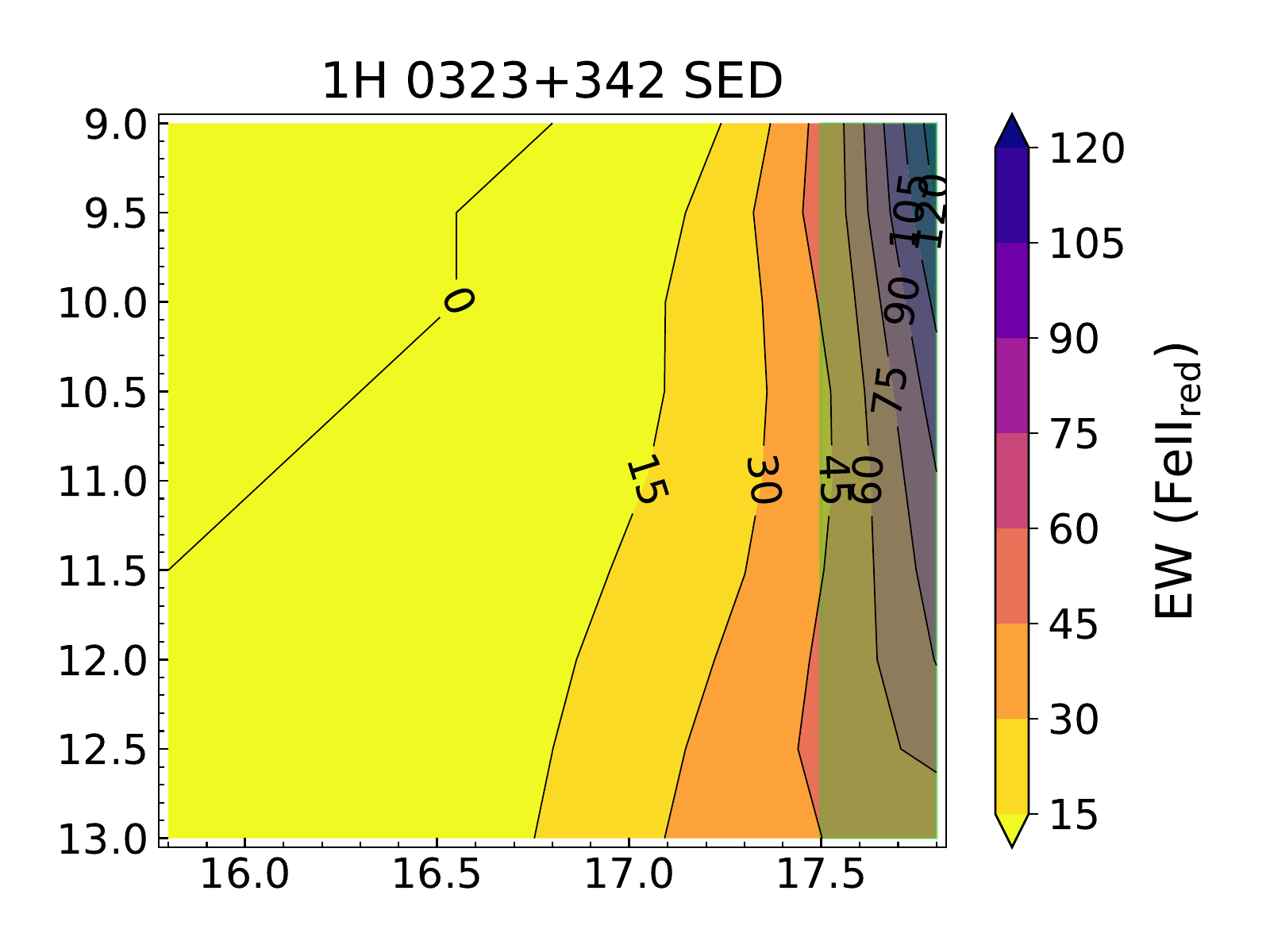}\\
\includegraphics[width=6.25 cm]{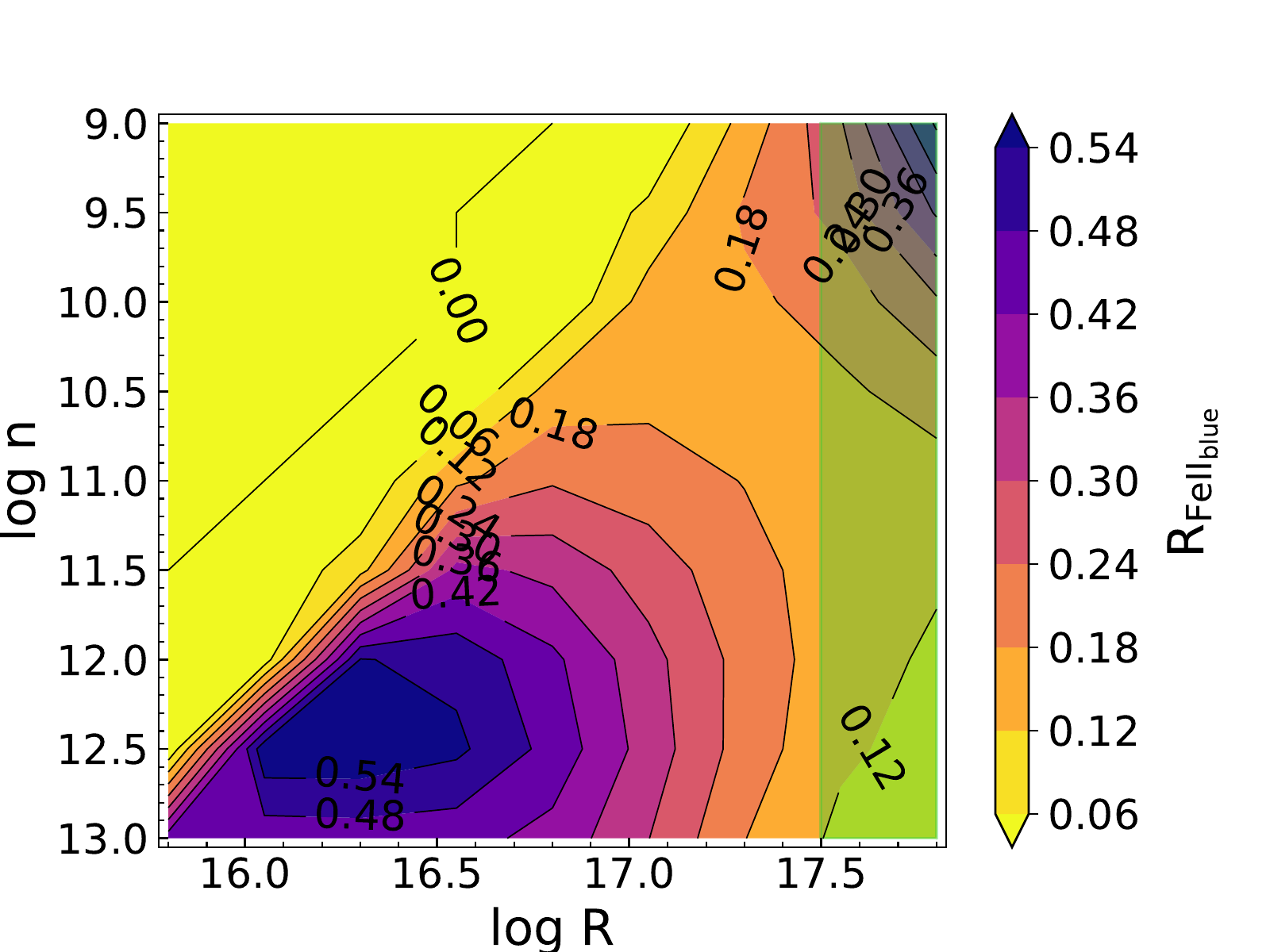}
\includegraphics[width=6.25 cm]{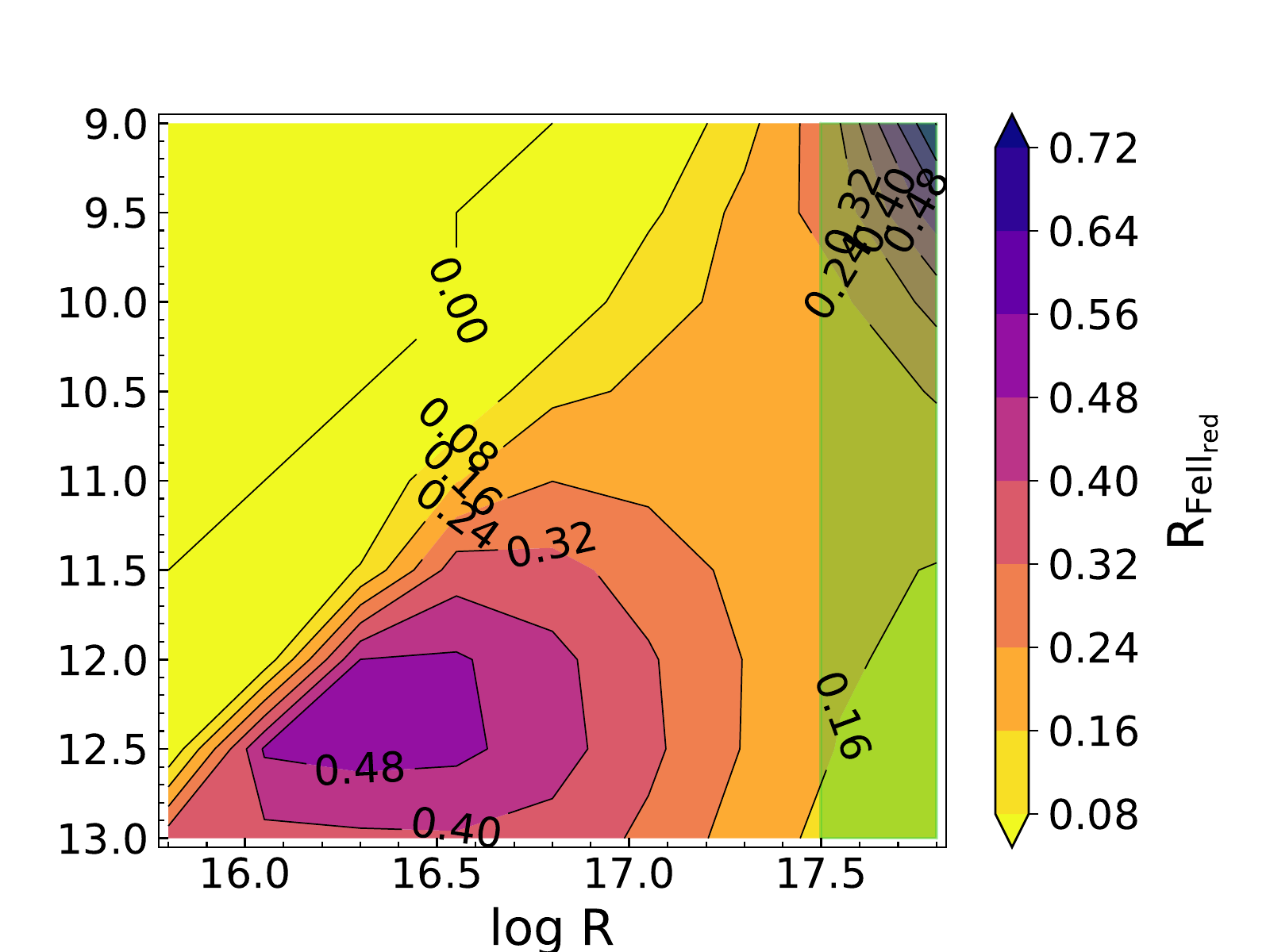}\\
\includegraphics[width=6.25 cm]{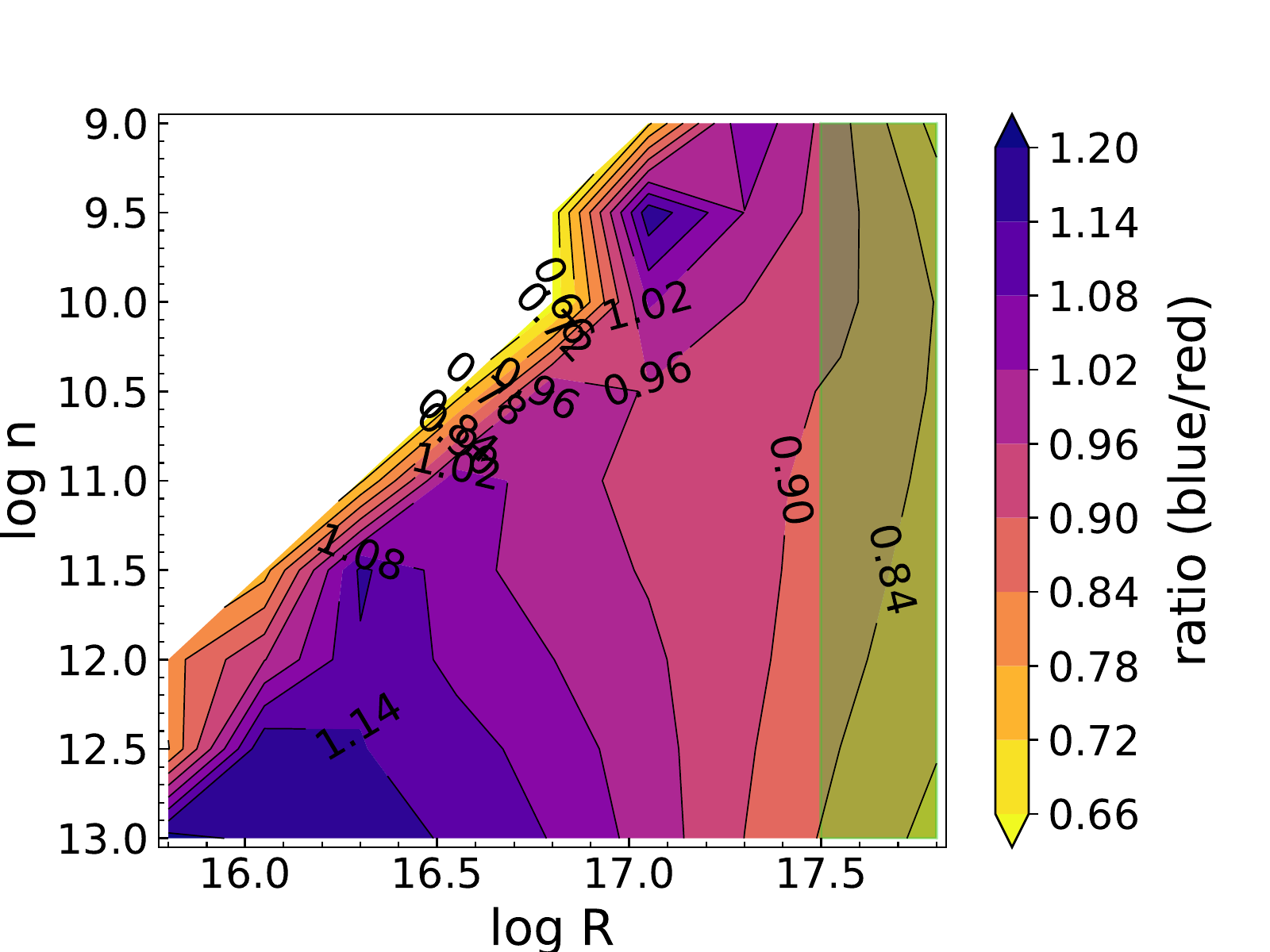}\\
\caption{Results of {\tt CLOUDY} simulations for the 1H 0323+342 SED: the two top panel yields the equivalent width of \feiiq\ and the two middle panels   \rfe\ for the blue and red blends, as a function of density and BLR radius. The bottom panel shows the ratio of the blue and red \feii\ emission.   \label{fig:ph_comp_1H0321}}
\end{figure} 

\subsection{$\gamma$-detected NLSy1s}

The photoionization calculations performed with the parameters appropriate to model 1H 0323+342 (Fig. \ref{fig:ph_comp_1H0321}, repeated in Appendix \ref{app:restr} for the restricted range as Fig. \ref{fig:ph_comp_1H0321_res}) reveal that is not possible to achieve \rfe $\sim 1$\ with the SED shown in Fig. \ref{fig:sed}. This raises an important issue about the \feii\ emission in these jetted sources. A change in SED within the limit of {Population} B, such as the one provided by \citet{koristaetal97} is likely not enough to explain the increase in \feii. This leaves several other possibilities to account for this intrinsic difference between jetted and non-jetted sources. They can be loosely grouped in "evolutionary" (chemical evolution of the host and of the circumnuclear regions, in turn related to the host morphology), and "intrinsic" (density, column density, ionization degree, covering fraction of the line emitting gas). The host morphologies of RL and RQ are systematically different, and recent work has validated the long-held paradigm that RL are hosted in earlier morphological types than RQ \citep{koziel-wierzbowskaetal17a,koziel-wierzbowskaetal17b}. This result might not be applicable to NLSy1s, and especially to the ones that are moderate-to-strong \feii\ emitters. Basically all of the recent studies on the host of jetted RL NLSy1s \citep{jarvelaetal18,bertonetal19,olguin-iglesiasetal20} showed that they are in late-type hosts. Merging is instead rather common, and it may constitute the biggest difference with respect to RQ NLSy1. In particular, \citet{olguin-iglesiasetal20} clearly pointed out on a rather large sample that disk galaxies dominate among jetted NLSy1s.  

The visual inspection of almost all known RL NLSy1s suggests that they never reach the extreme \rfe\ value obtained by RQ sources. This impression -- even if consistent with the difference found for weak \feii\ emitters --  should be validated by a systematic study that goes beyond the scope of the present paper. Enrichment  in chemical composition appears a likely possibility, as the strongest \feii\ emitters are also believed to have the highest metal content \citep{shinetal13,sulenticetal14,nagaoetal06b,sniegowskaetal21}. Photoionization analyses indicate that significant \feii\ emission is associated with super-solar metallicity, {  in addition to high density and high column density} \citep{pandaetal18,pandaetal19}, although a complete observational analysis with matching \rfe\ and UV line diagnostics for metallicity { and physical parameters} estimates along the main sequence is still lacking.

\section{Summary and conclusion}

This work presented an analysis of the \feii\ emission in the spectral bins along the main sequence where both RQ and RL coexist (A1, B1) with measurable \feii\ strength, and a comparison between the emission of RQ and RL composite template, along with the analysis of 3 RL sources found to have a stronger \feii\ emission.   
The main results can be summarized as follows:

\begin{itemize}
    \item the template based on the I Zw 1 spectrum works equally well for RQ and RL objects. Deviation in \rfe\ and equivalent width measurements due to differences between the template and the observed spectra have been found to be $\delta$\rfe$\lesssim$0.03, and $\delta W \lesssim$ a few \AA. In particular, no correction is needed for the placement of RL in the MS built from RQ-dominated samples. 
    \item The RL SED of {\citet{laoretal97b}} can account for the modest emission in RL sources, for conditions that are very likely in {Population} B objects \citep{pandaetal19}. 
    \item Somewhat surprisingly, SED differences between RQ and RL cannot account for the stronger \feii\ emission in RQ\footnote{Note that the equivalent width is increased, as shown in Fig. \ref{fig:ph_comp_kor} and Fig. \ref{fig:ph_comp_kor_res} but, since there is a corresponding increase in \hb, the \rfe\ is left unaffected.}. We suggest that other factors related to the evolutionary pattern of the circumnuclear regions of the active nucleus should be investigated. 
    \item Last, the SED shape does not seem so important in determining the optical \feii\ prominence { with respect to \hb} {for such sources with intermediate Eddington ratios ($\sim$0.1)}, as the change from the {\citet{laoretal97b}} to the \citet{koristaetal97} SED produces no significant change in \rfe, even if the number of ionizing photons is increased by a factor $\approx 3$. {We remark that, for sources at higher Eddington ratios, this might be different, as has been tested and shown in \citet{pandaetal19}. Those authors found  that the SED shape matters for pushing the \rfe\ higher, especially when the contribution from a soft-X-ray excess is accounted for. }
    \item The ratio between the B and the R blend is found to be $\approx$ 0.9 --  1.0 and to be consistent with the predictions of the photoionization computation. A slightly lower value ($\approx 0.7$) might be possible for the B1 RL composites. However, the measurement is excruciatingly  difficult, considering the uncertain influence of the \heiiopt\ emission that is known to be very strong in cases where \feii\ is negligible \citep[e.g., ][]{marzianisulentic93}, and of several other factors that should play a lesser role. More importantly, the B/R value around $\approx$ 1 is found also in  higher Eddington ratio sources which are stronger \feii\ emitters such as the 3 $\gamma$-detected NLSy1s considered in this study, all with Eddington ratio $\approx$ 0.5, typical of Population A quasars.  Since \rfe\ is   correlated with Eddington ratio \citep{sunshen15,duetal16a}, { and the B/R is apparently independent from \rfe, }  a lower B/R might be a genuine radio loudness effect { or, perhaps more likely,  an effect dependent on other RQ/RL sample differences.} 
\end{itemize}
In summary, we find that the \feii\ emission is stable, { and we confirm} that a solid template based on I Zw 1 provides an accurate representation of the optical \feii\ emission in the spectral region of \hb\ within the limit of precision allowed by moderate dispersion spectroscopy.  
 


\authorcontributions{{All authors contributed equally to this paper.}}


\acknowledgments{SP would like to acknowledge the financial support by the Polish Funding Agency National Science Centre, project 2017/26/\-A/ST9/\-00756 (MAESTRO  9), and from CNPq Fellowship (164753/2020-6). EB acknowledge the support of Serbian Ministry of Education, Science and Technological Development, through the contract number 451-03-68/2020-14/200002. {Based on observations made with ESO Telescopes at the La Silla Paranal Observatory under program ID 096.B-0256. This work is partially based on observations made with the Galileo 1.22 m telescope of the Asiago Astrophysical Observatory operated by the Department of Physics and Astronomy “G. Galilei” of the University of Padova. Funding for the Sloan Digital Sky Survey has been provided by the Alfred P. Sloan Foundation, and the U.S. Department of Energy Office of Science. The SDSS web site is \texttt{http://www.sdss.org}. SDSS-III is managed by the Astrophysical Research Consortium for the Participating Institutions of the SDSS-III Collaboration including the University of Arizona, the Brazilian Participation Group, Brookhaven National Laboratory, Carnegie Mellon University, University of Florida, the French Participation Group, the German Participation Group, Harvard University, the Instituto de Astrofisica de Canarias, the Michigan State/Notre Dame/JINA Participation Group, Johns Hopkins University, Lawrence Berkeley National Laboratory, Max Planck Institute for Astrophysics, Max Planck Institute for Extraterrestrial Physics, New Mexico State University, University of Portsmouth, Princeton University, the Spanish Participation Group, University of Tokyo, University of Utah, Vanderbilt University, University of Virginia, University of Washington, and Yale University.}}
\conflictsofinterest{The authors declare no conflict of interest.} 

 
\abbreviations{The following abbreviations are used in this manuscript:\\
\noindent 
\begin{tabular}{@{}ll}
AGN& Active Galactic Nucleus\\
BLR& Broad Line Region\\
{CD}& {Core dominated} \\
{FR-I}& {Fanaroff-Riley I} \\
{FR-II}& {Fanaroff-Riley II} \\
FWHM & Full Width Half-Maximum\\
MS& Main Sequence\\
NLSy1 & Narrow-Line Seyfert 1\\
{RL} & {Radio loud} \\
{RQ} & {Radio quiet} \\
{SDSS} & {Sloan Digital Sky Survey}\\
{SED} & {Spectral energy distribution}\\
{S/N} & {Signal-to-noise ratio}\\
\end{tabular} }
\vfill
\eject

\vfill\eject\pagebreak\newpage


%

\pagebreak\vfill\eject
\setcounter{section}{0}
\appendix

\setcounter{figure}{0}
The Appendix shows the results of the {\tt CLOUDY} simulations in the case of the restricted ranges 4500 -- 4590 \AA\ and 5200 -- 5300 \AA\ ($\tilde{B}$\ and $\tilde{R}$\ through the paper). 

\section{Restricted range}
\label{app:restr}
\begin{figure}[htp]
\centering
\includegraphics[width=4.25  cm]{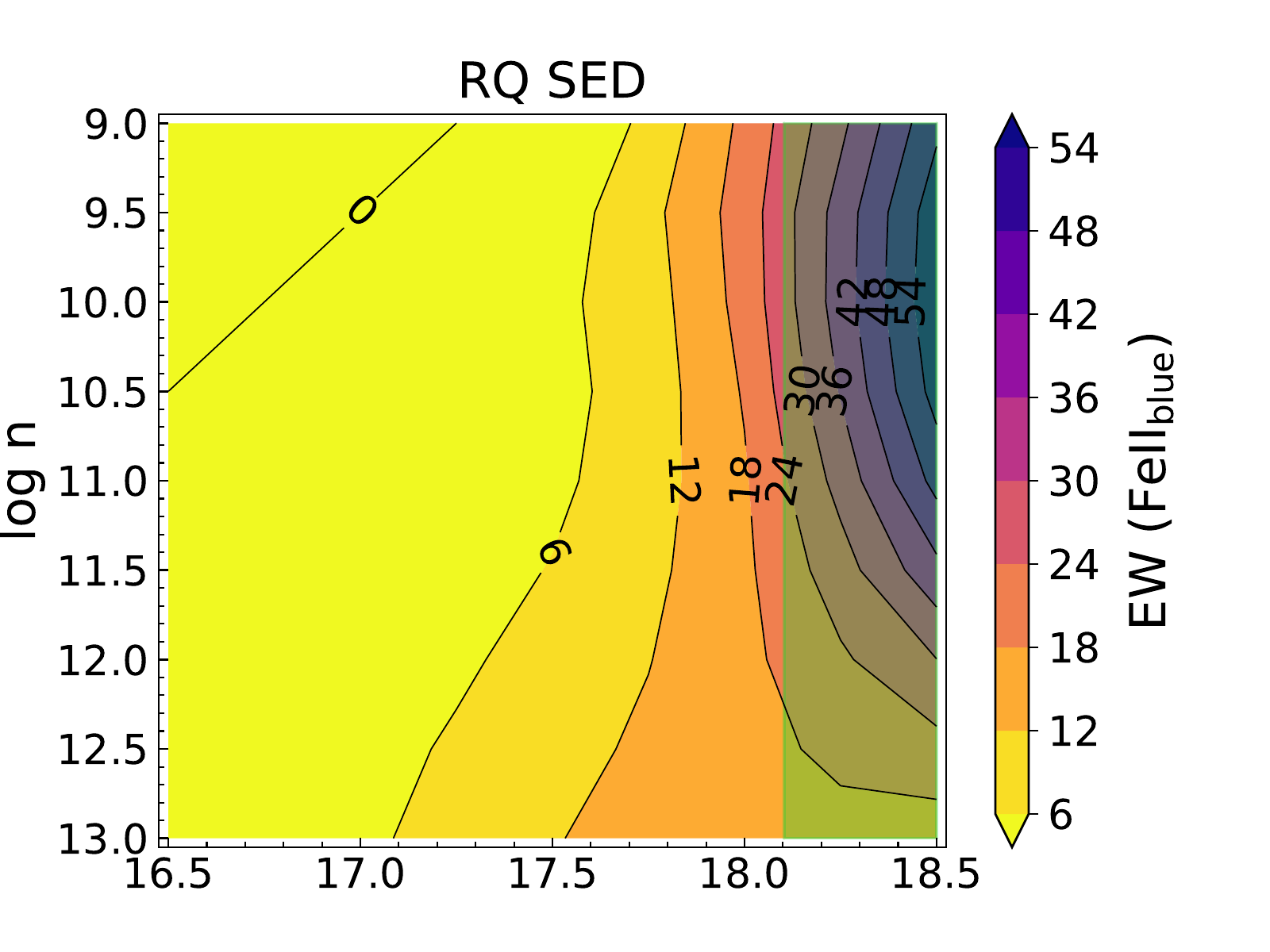}
\includegraphics[width=4.25  cm]{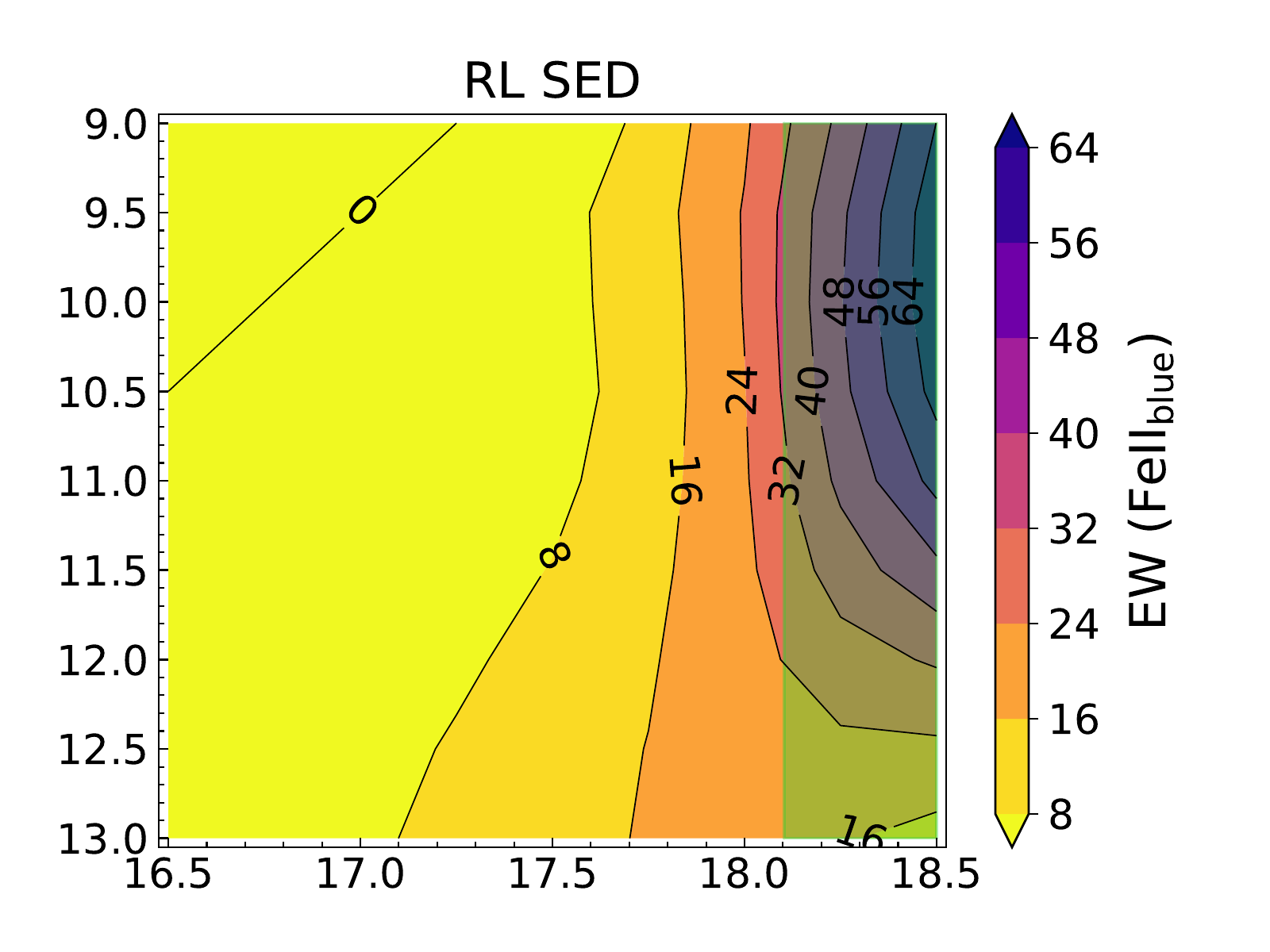}
\includegraphics[width=4.25  cm]{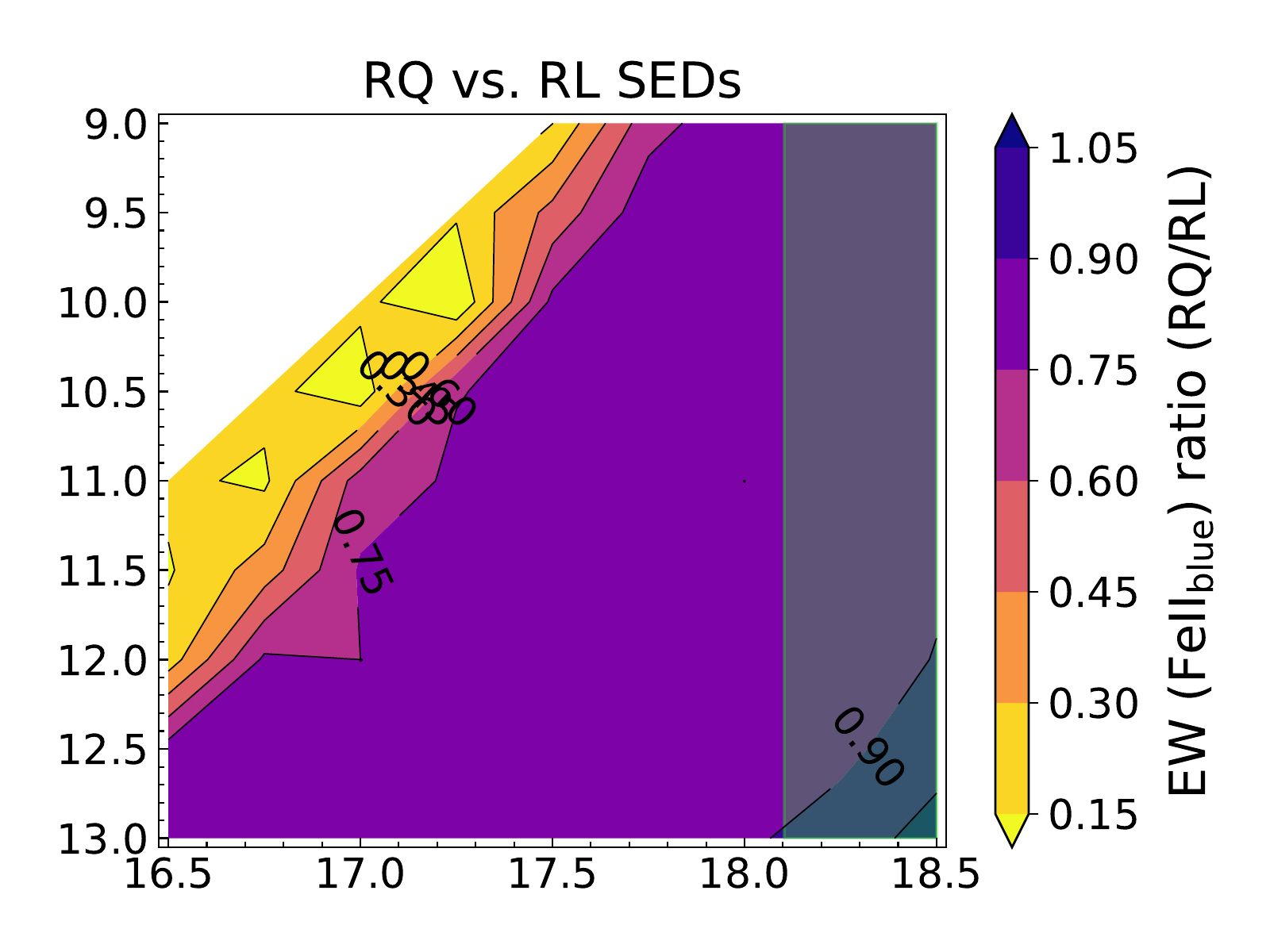}\\
\vspace{-0.25cm}
\includegraphics[width=4.25  cm]{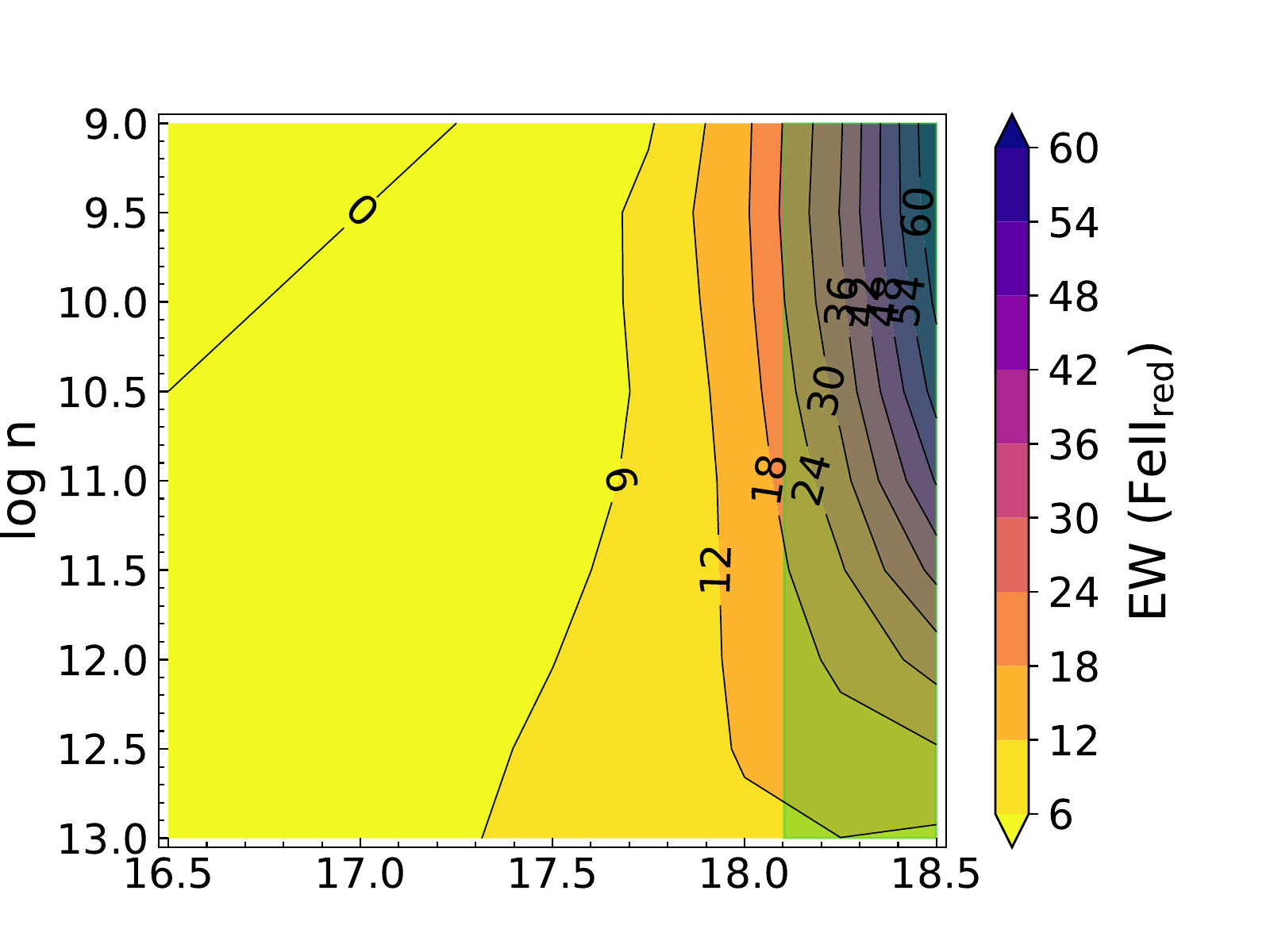}
\includegraphics[width=4.25  cm]{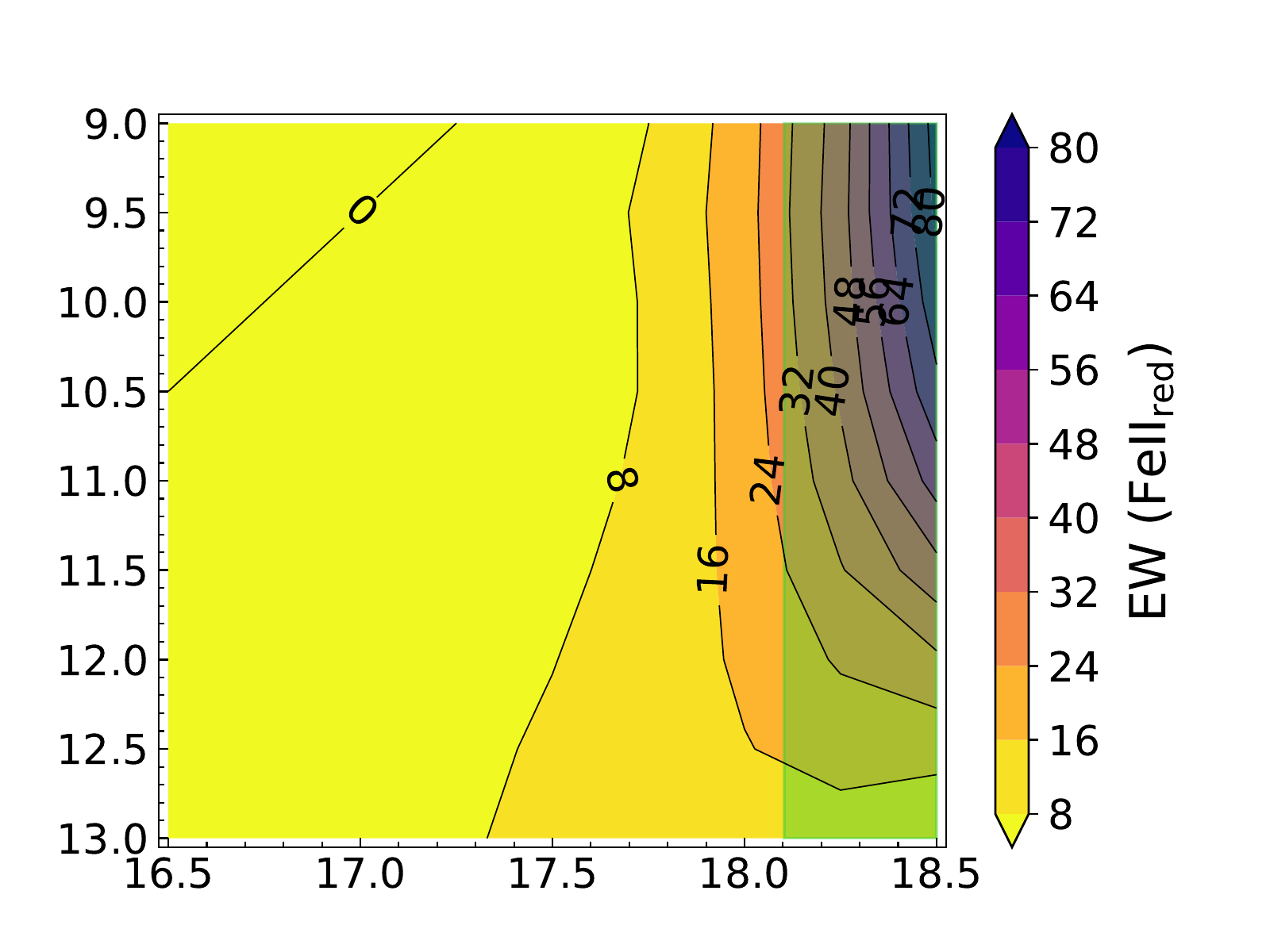}
\includegraphics[width=4.25  cm]{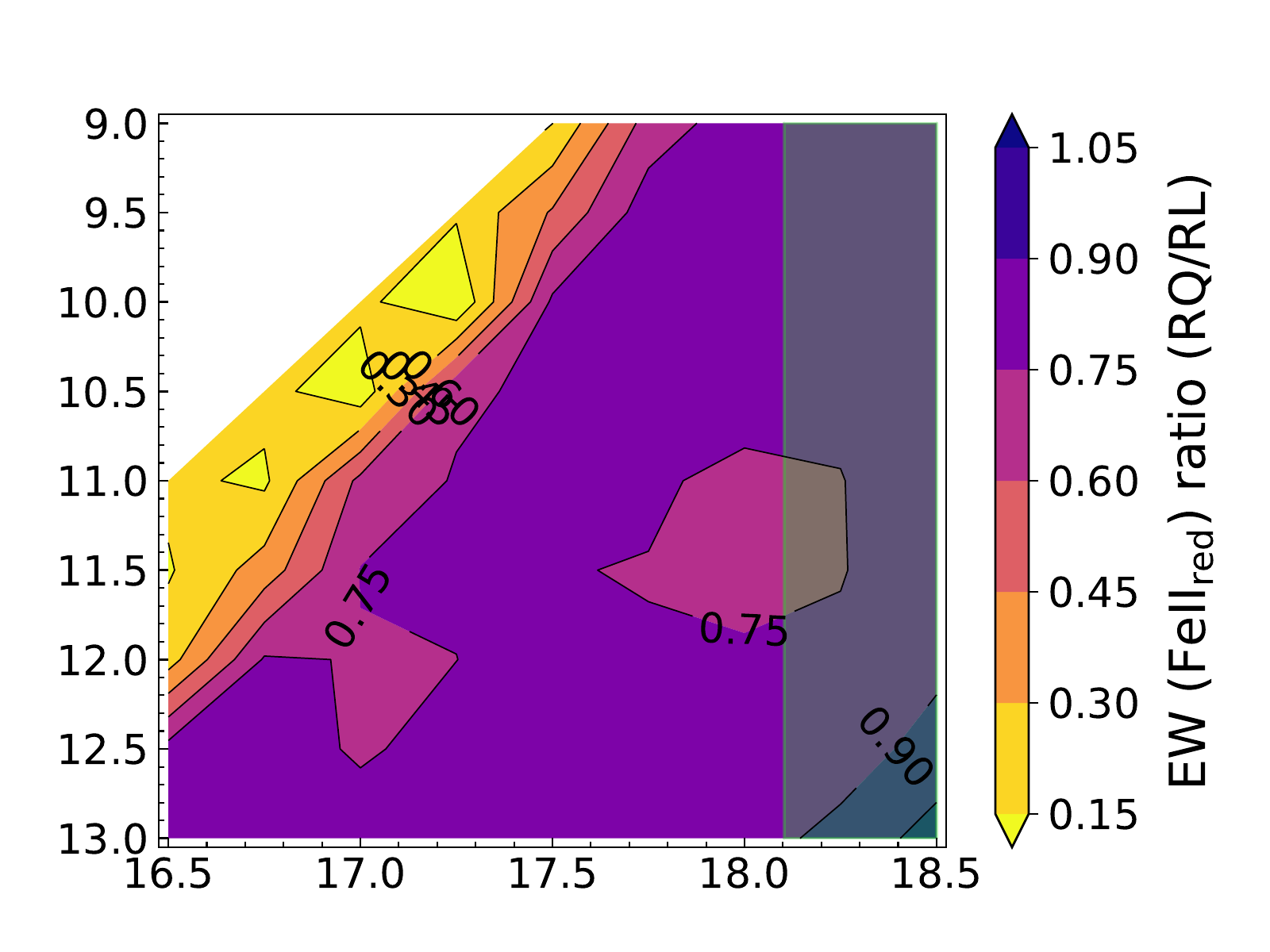}\\
\vspace{-0.25cm}
\includegraphics[width=4.25 cm]{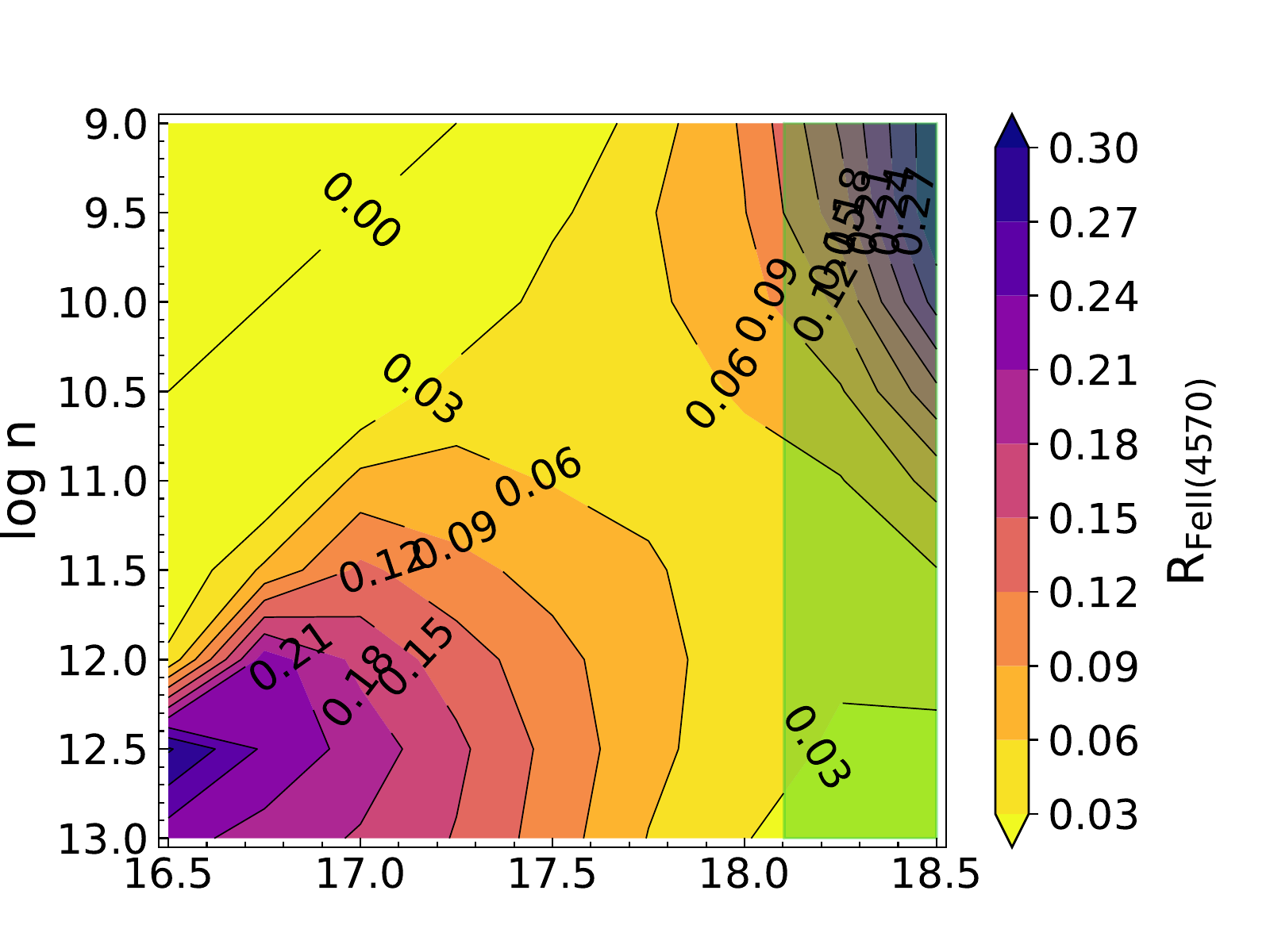}
\includegraphics[width=4.25  cm]{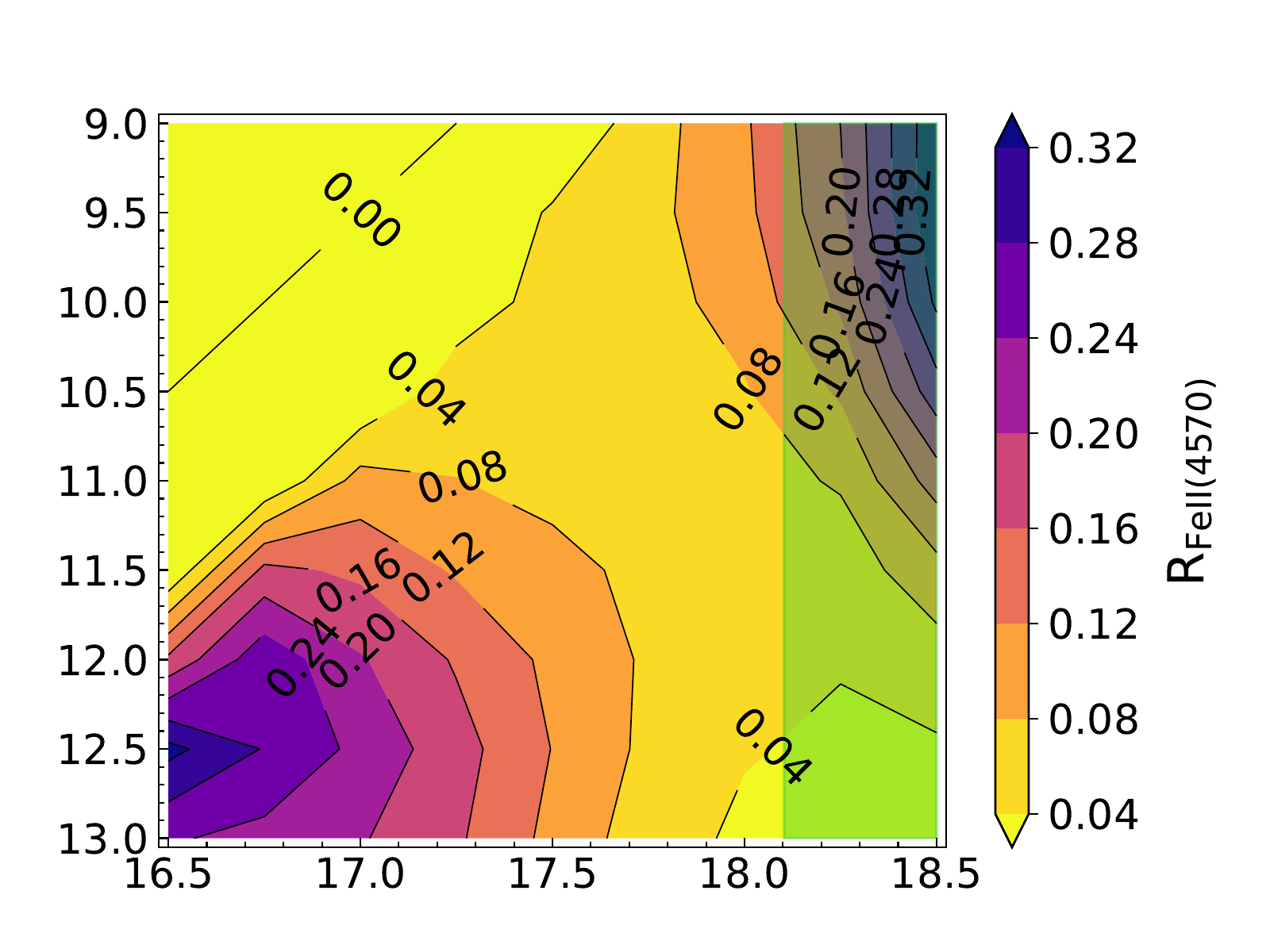}
\includegraphics[width=4.25  cm]{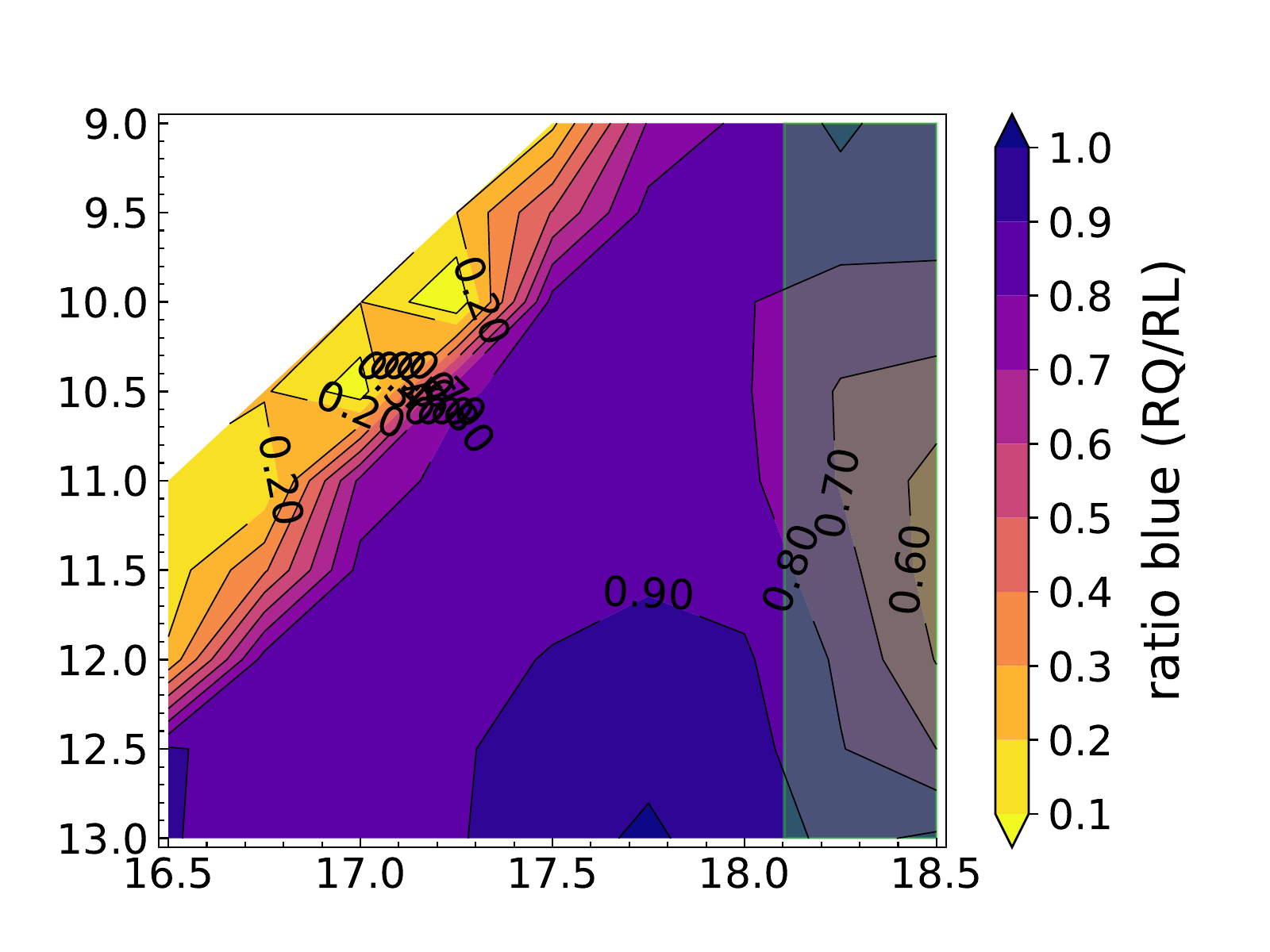}\\
\vspace{-0.25cm}
\includegraphics[width=4.25 cm]{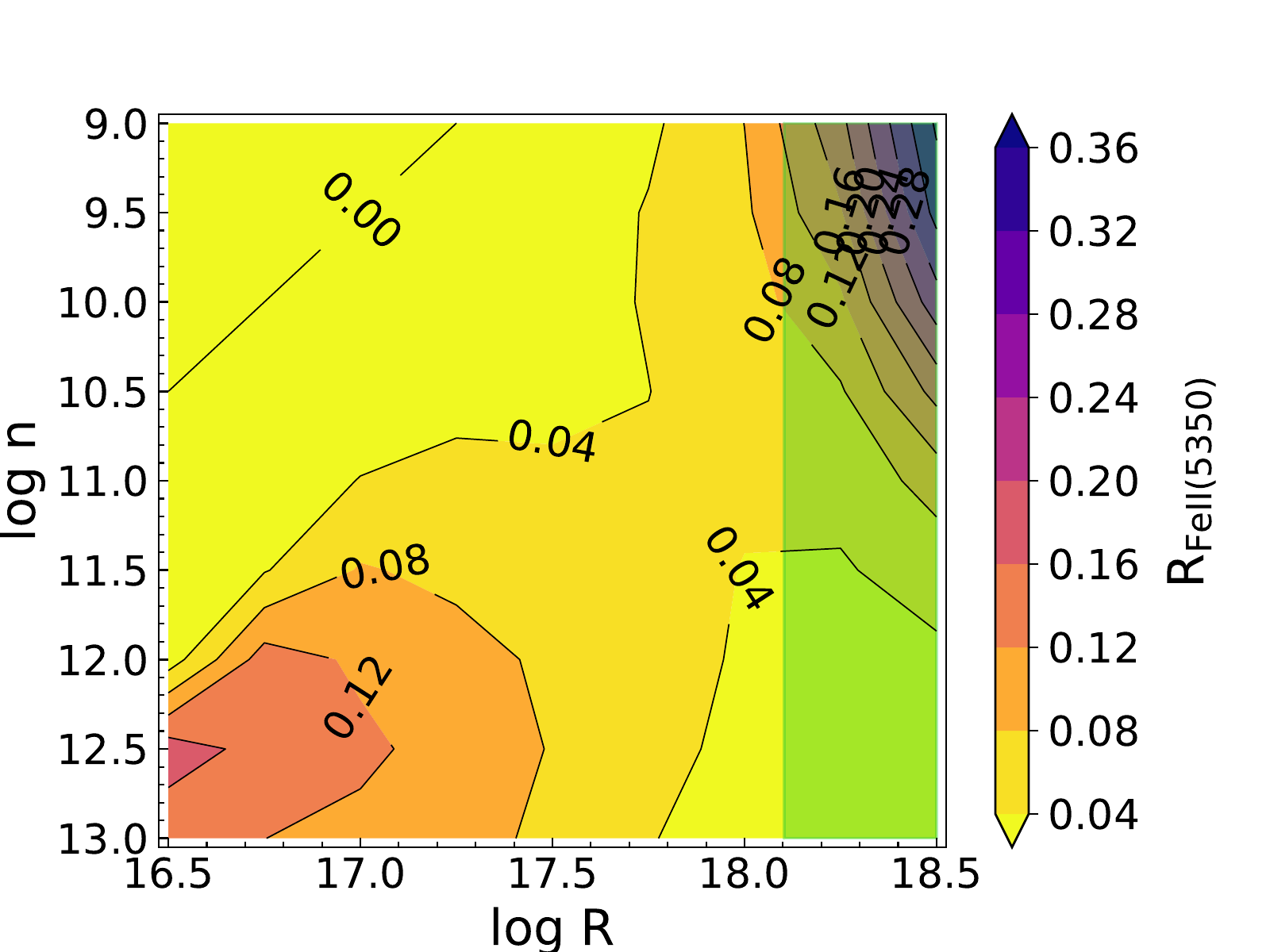}
\includegraphics[width=4.25  cm]{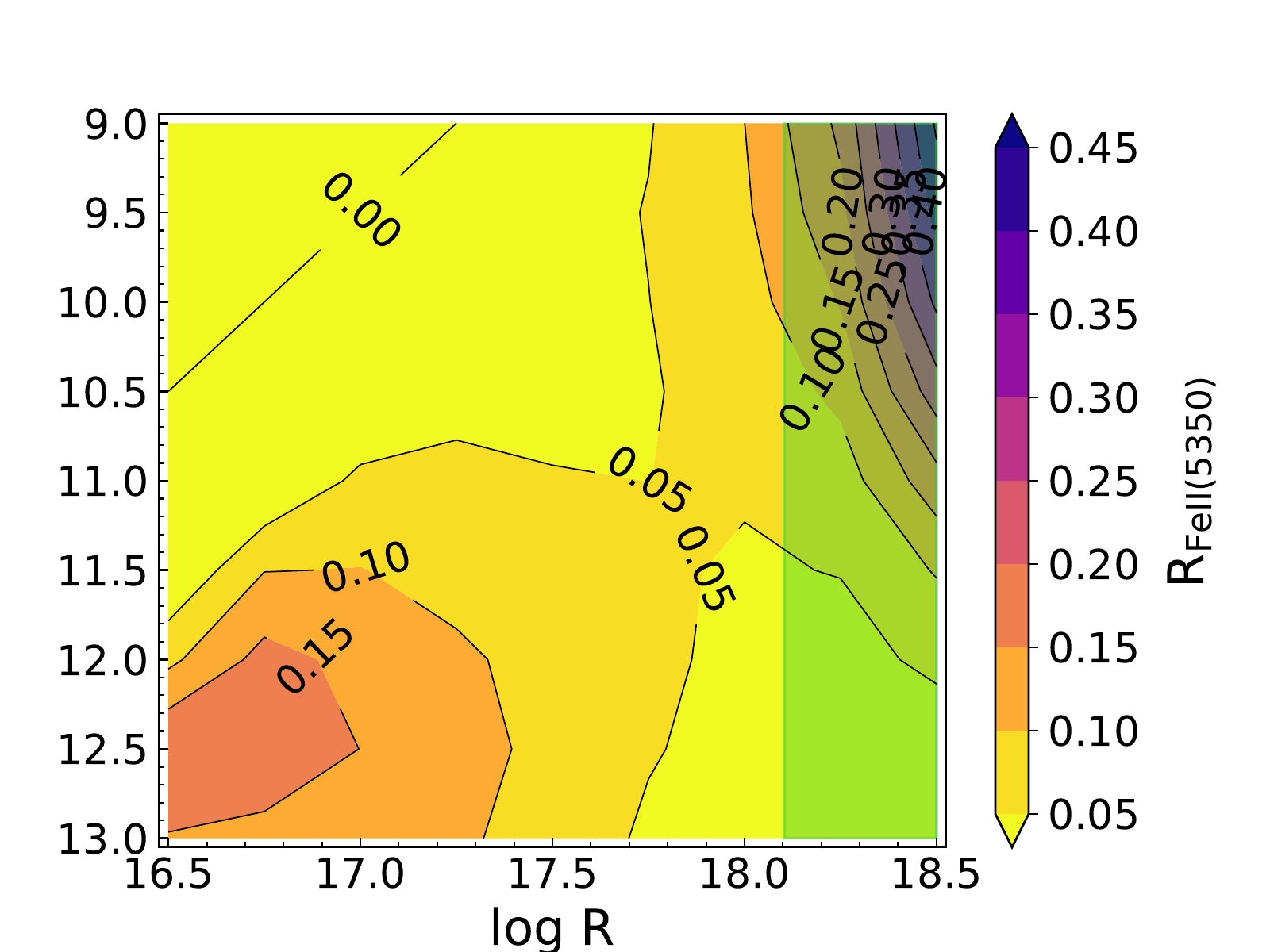}
\includegraphics[width=4.25  cm]{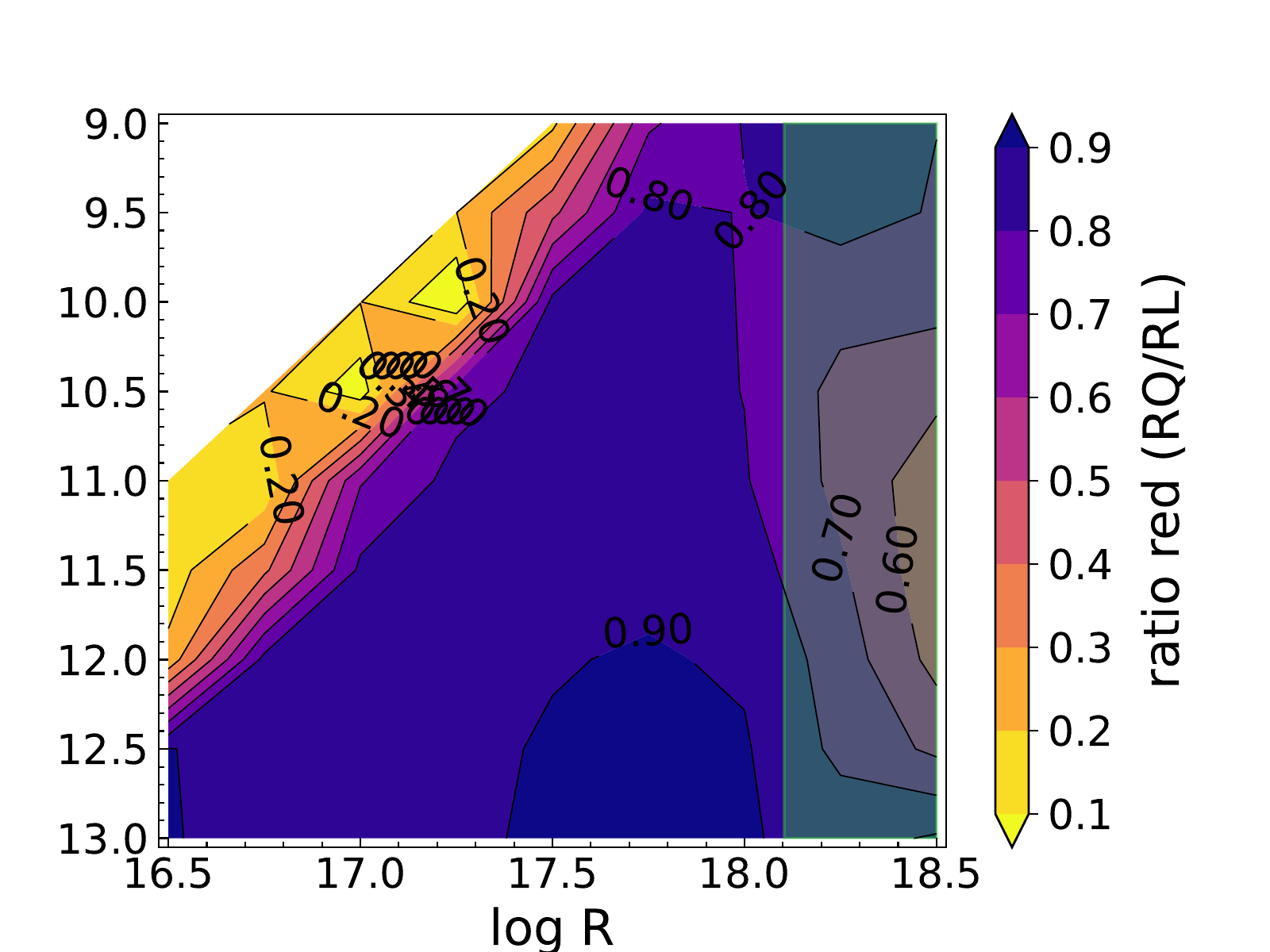}\\
\caption{Results of {\tt CLOUDY} simulations. First two rows: W(\feiiq) for RQ and RL (i.e., computed for the same physical parameters but with the two different SEDs of \citet{laoretal97b})  and their ratios;  third and fourth row from top: \rfe, for RQ and RL and their ratios. Last four rows: same, for the red (R) \feii\ blend peaked at $\lambda$ 5312 \AA.  \label{fig:ph_comp_res}}
\end{figure}

\begin{figure}[t!]
\centering
\includegraphics[width=4.25  cm]{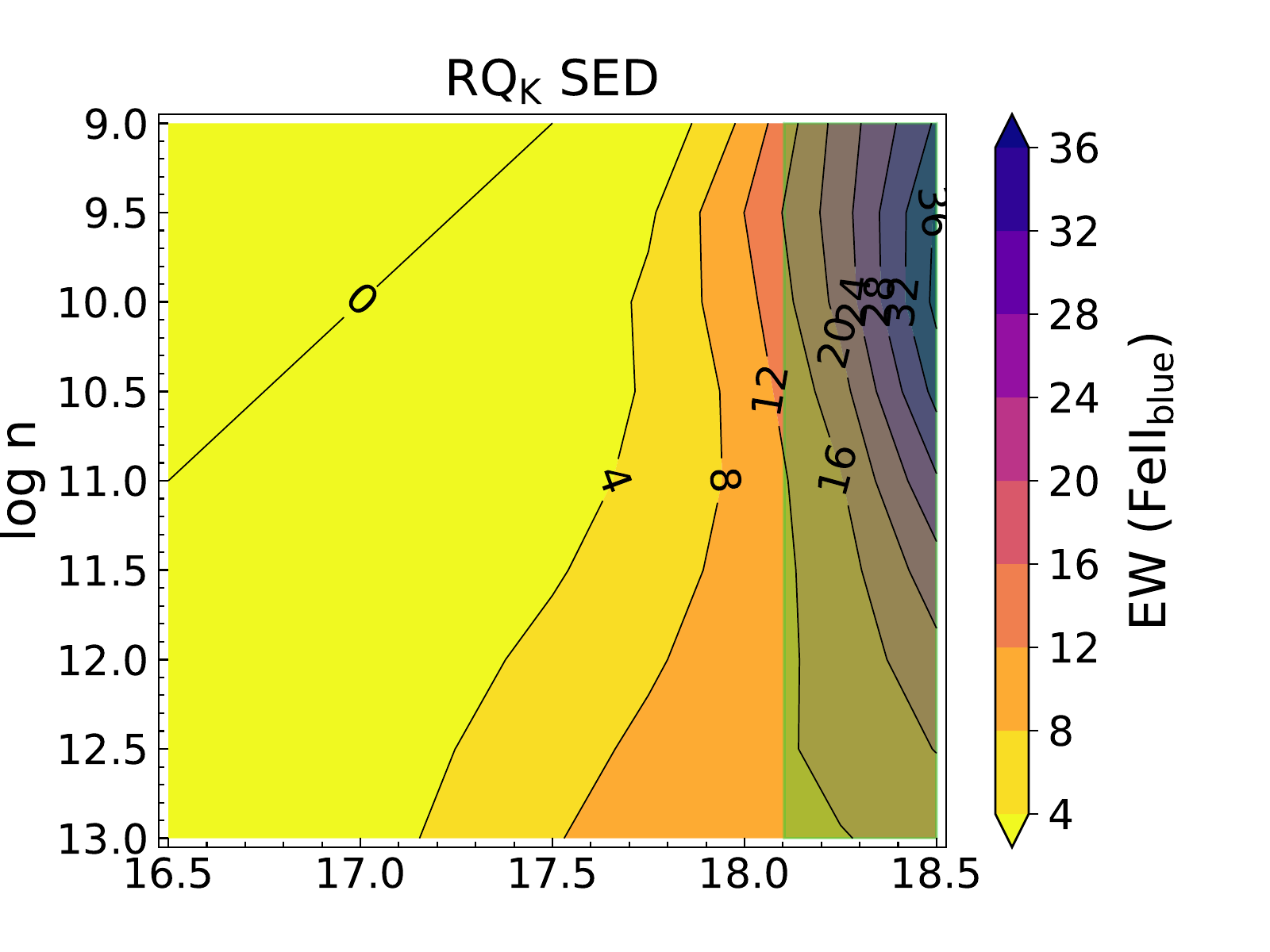}
\includegraphics[width=4.25  cm]{logR-logn_vs_EWFe_blue_RL_restricted.pdf}
\includegraphics[width=4.25  cm]{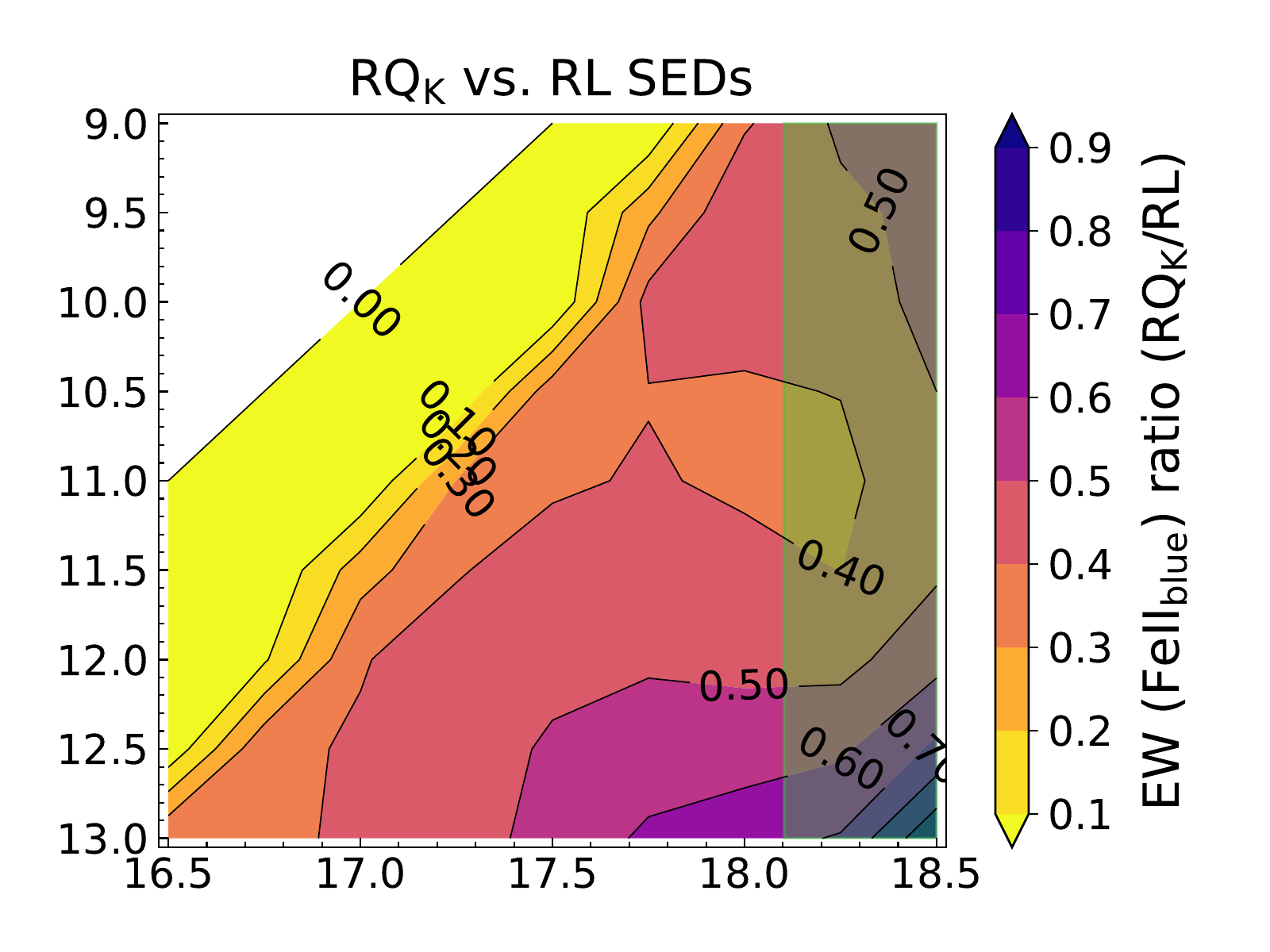}\\
\vspace{-0.25cm}
\includegraphics[width=4.25  cm]{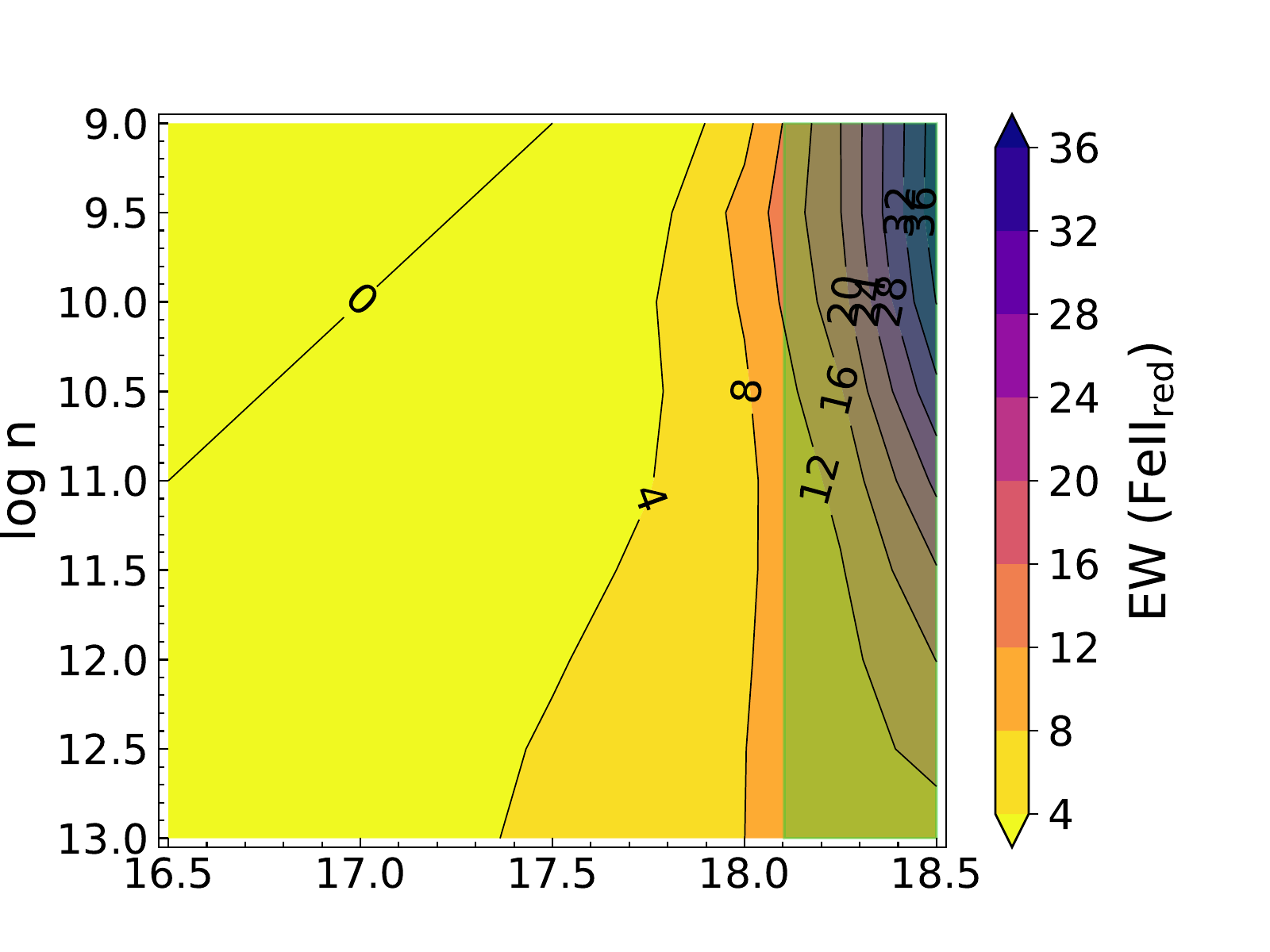}
\includegraphics[width=4.25  cm]{logR-logn_vs_EWFe_red_RL_restricted.pdf}
\includegraphics[width=4.25  cm]{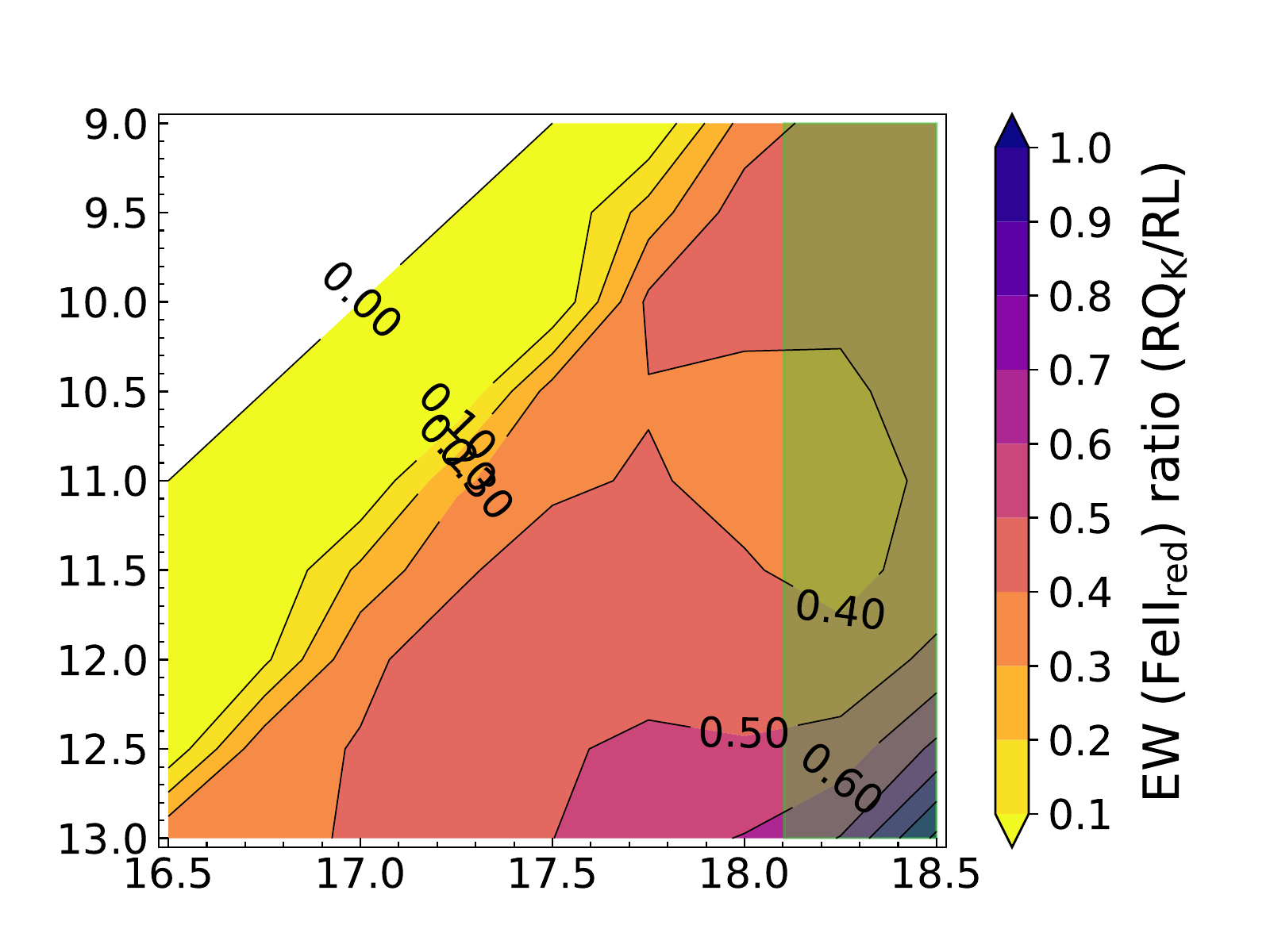}\\
\vspace{-0.25cm}
\includegraphics[width=4.25 cm]{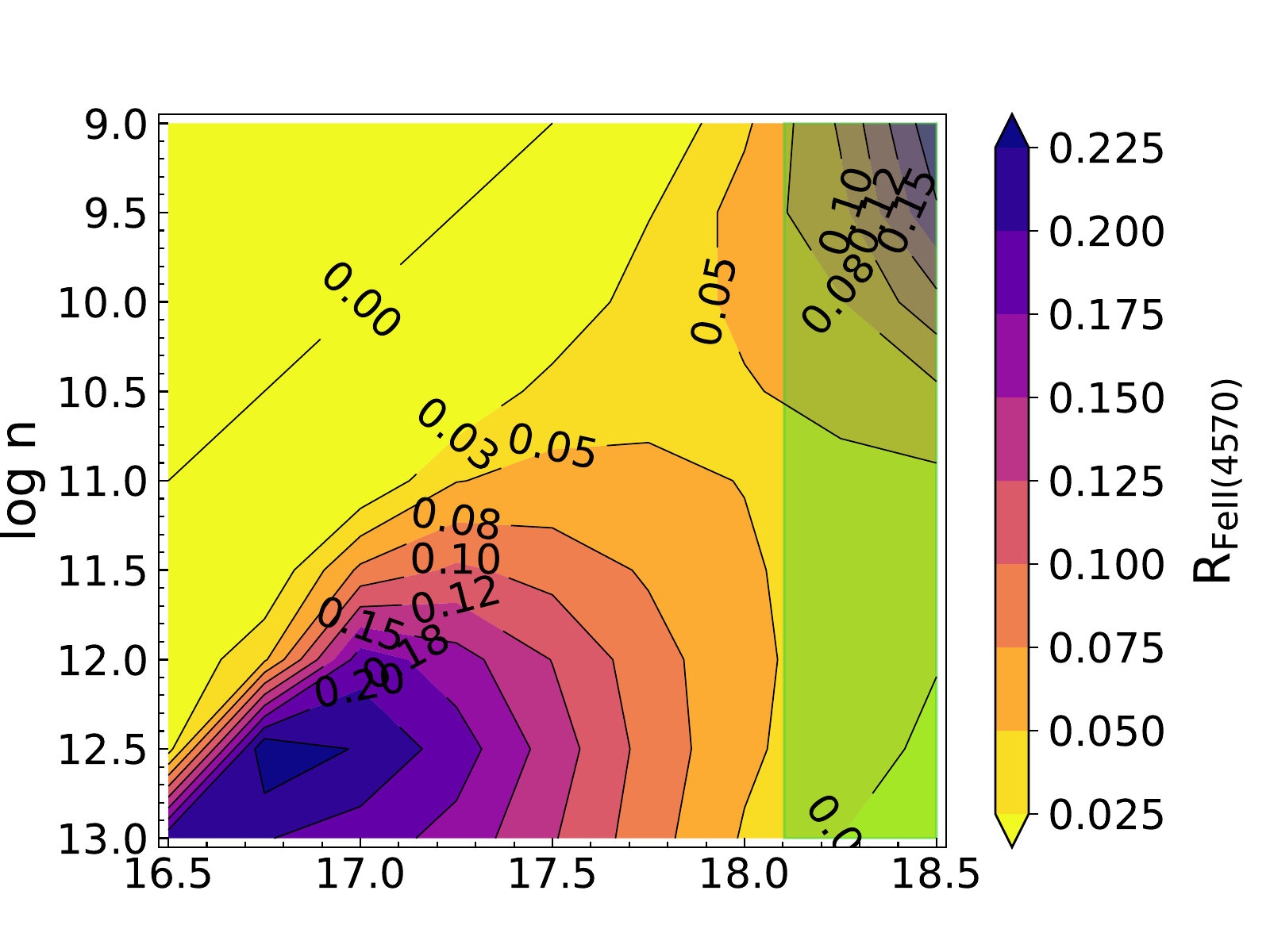}
\includegraphics[width=4.25  cm]{logR-logn_vs_RFe_blue_RL_restricted.pdf}
\includegraphics[width=4.25  cm]{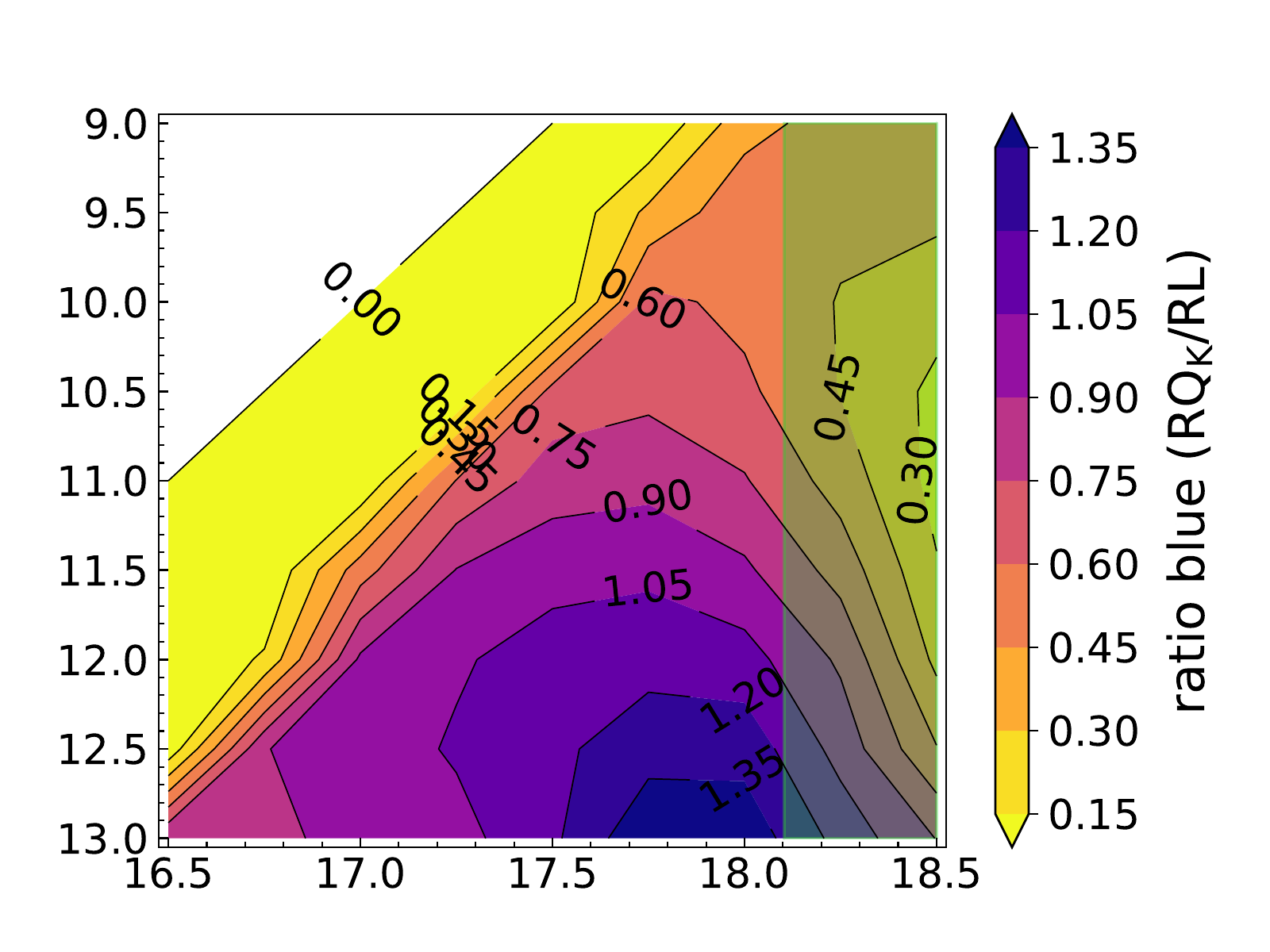}\\
\vspace{-0.25cm}
\includegraphics[width=4.25 cm]{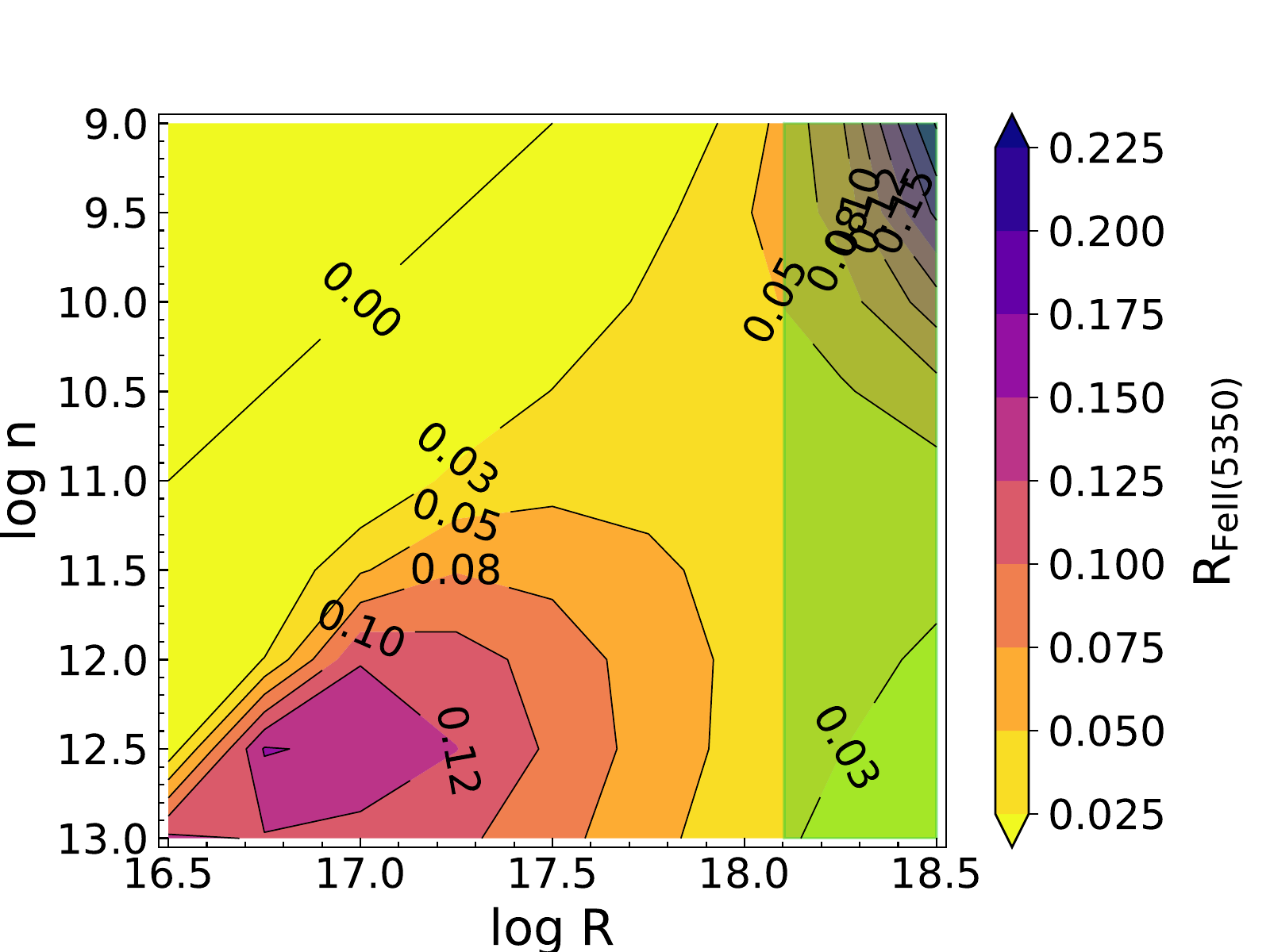}
\includegraphics[width=4.25  cm]{logR-logn_vs_RFe_red_RL_restricted.pdf}
\includegraphics[width=4.25  cm]{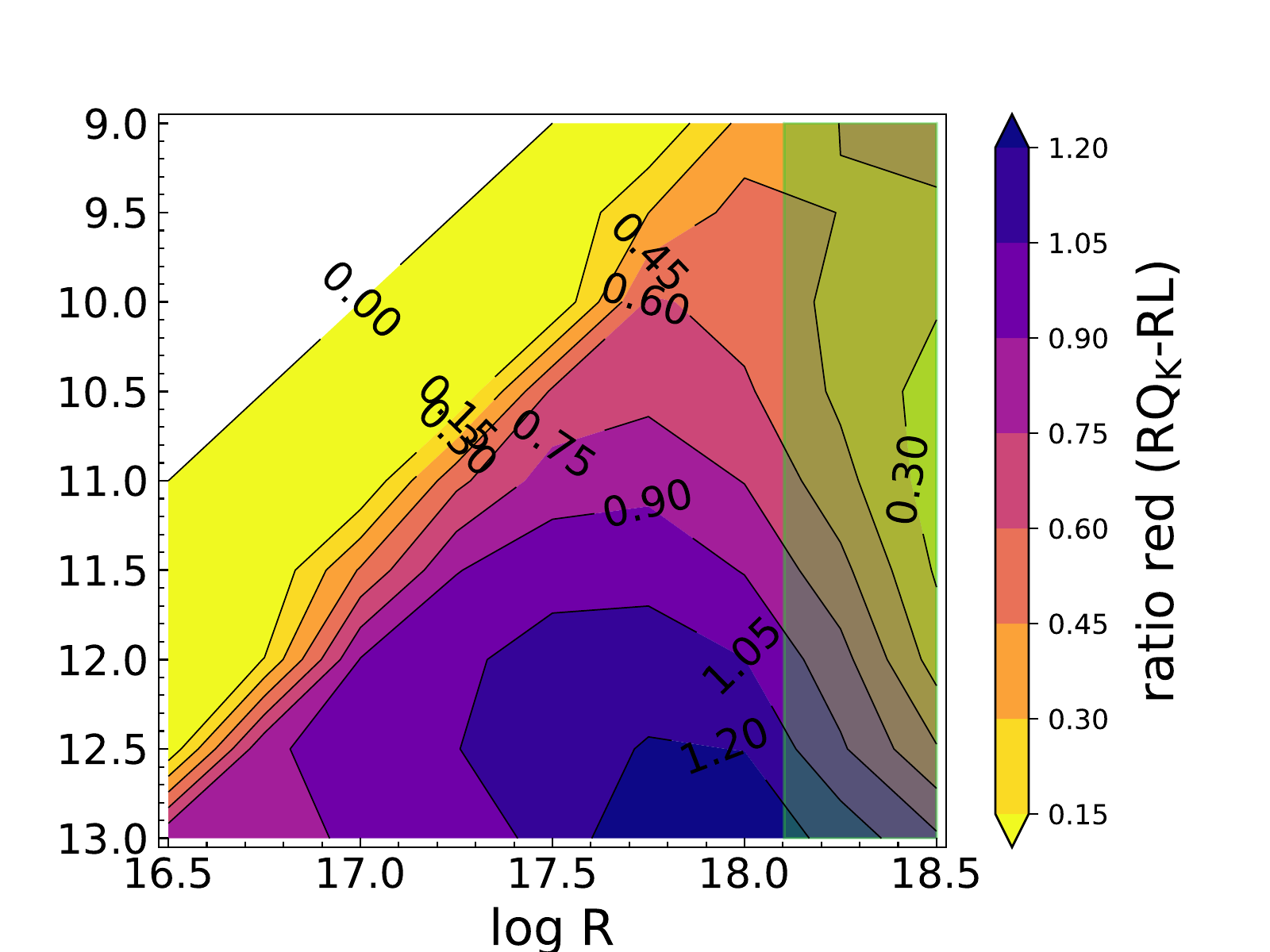}\\
\caption{Results of {\tt CLOUDY} simulations, same as in Fig. \ref{fig:ph_comp_rlrq} but for the RQ SED of \citet{koristaetal97}.  \label{fig:ph_comp_kor_res}}
\end{figure}

\begin{figure}[t!]
\centering
\includegraphics[width=4.25  cm]{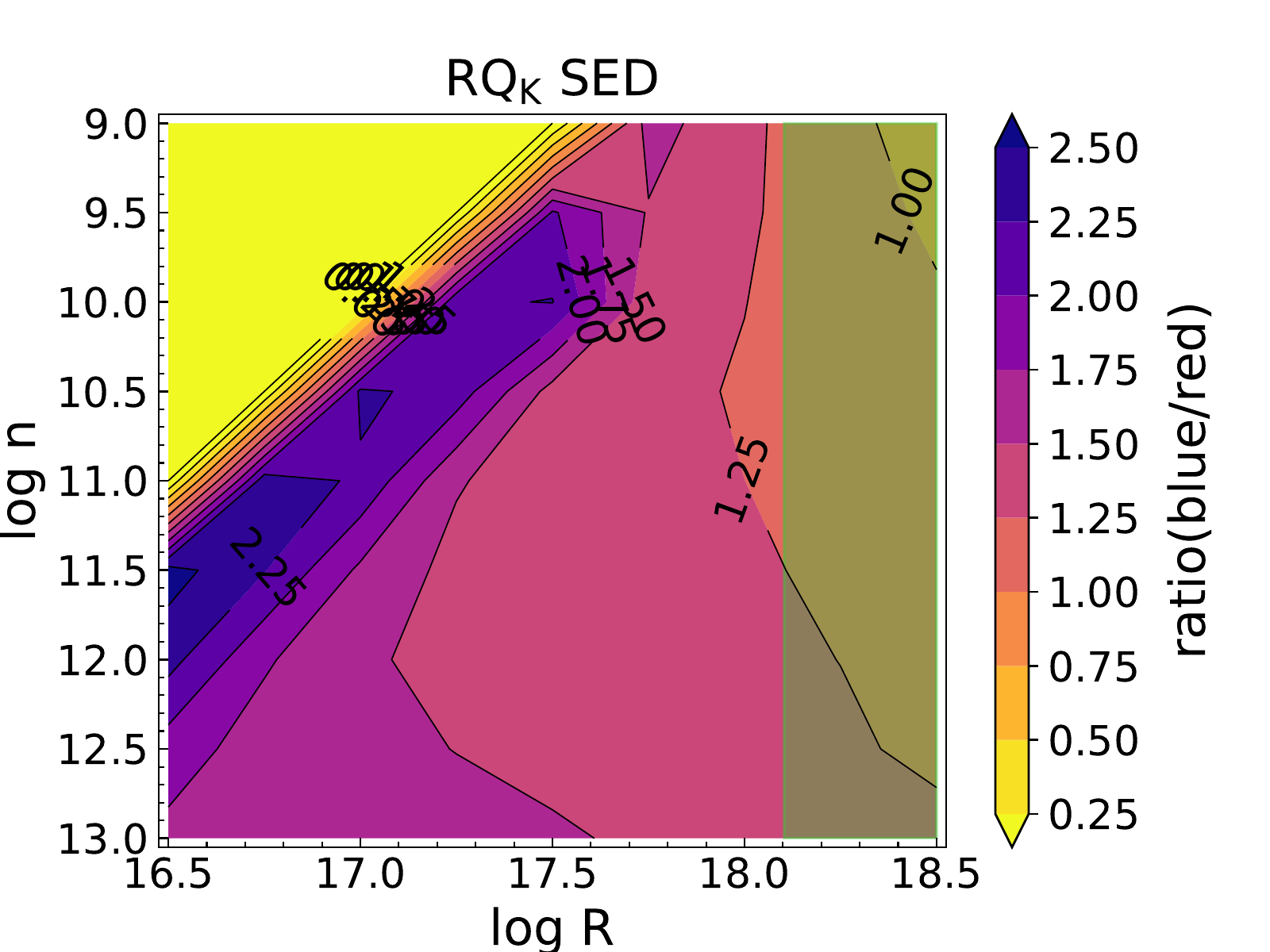}
\includegraphics[width=4.25  cm]{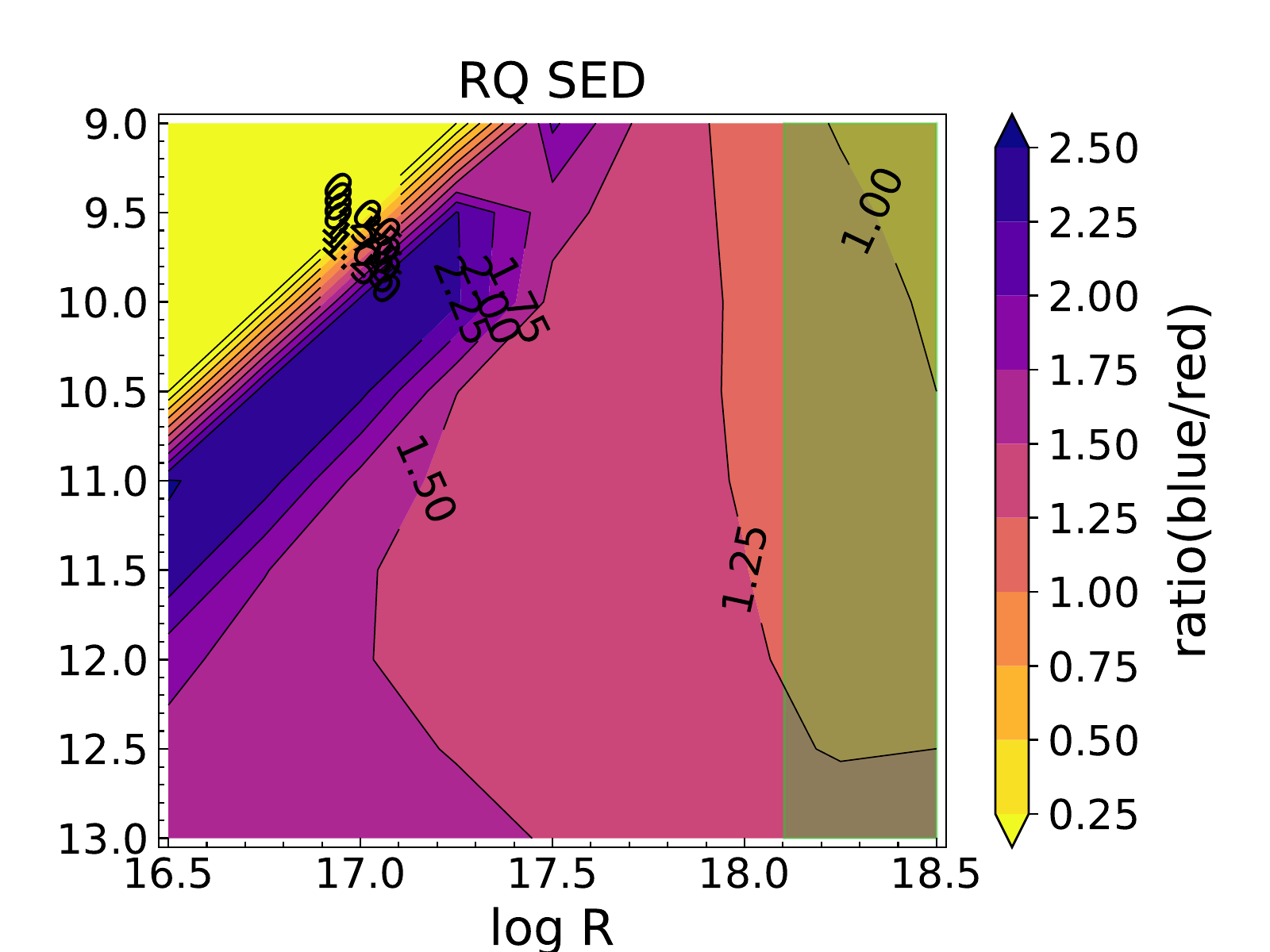}
\includegraphics[width=4.25 cm]{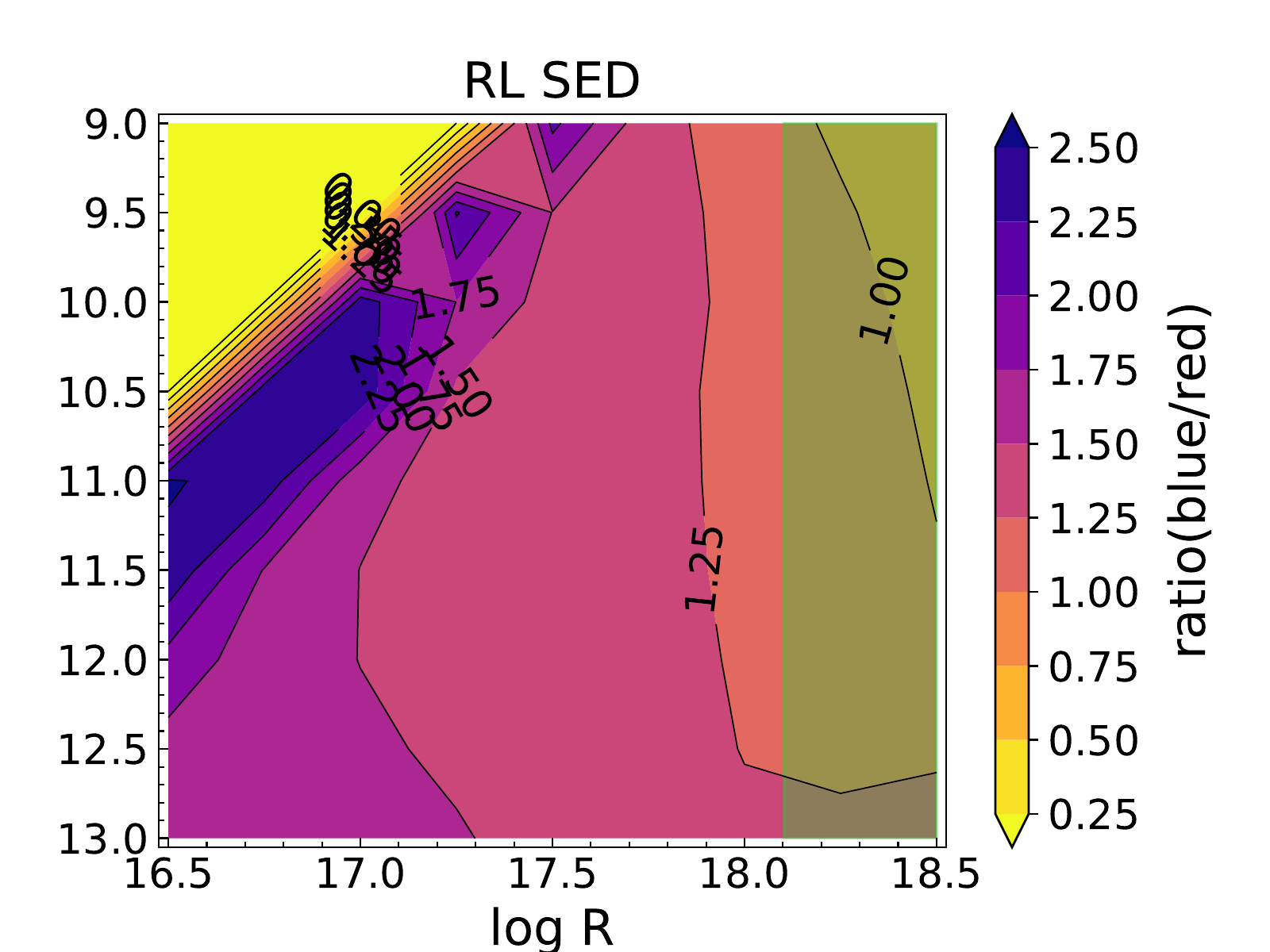}
\caption{Results of {\tt CLOUDY} simulations: the three   panels yield  the ratio  \rfe\ B to \rfe R (in practice B/R) as a function of density and BLR radius, for the RQ \citet{koristaetal97} (leftmost), the RQ \citet{laoretal97a} (middle) and RL SED (rightmost). \label{fig:ph_comp_rlrq_res}}
\end{figure} 

\begin{figure}[t!]
\centering
\includegraphics[width=6.25 cm]{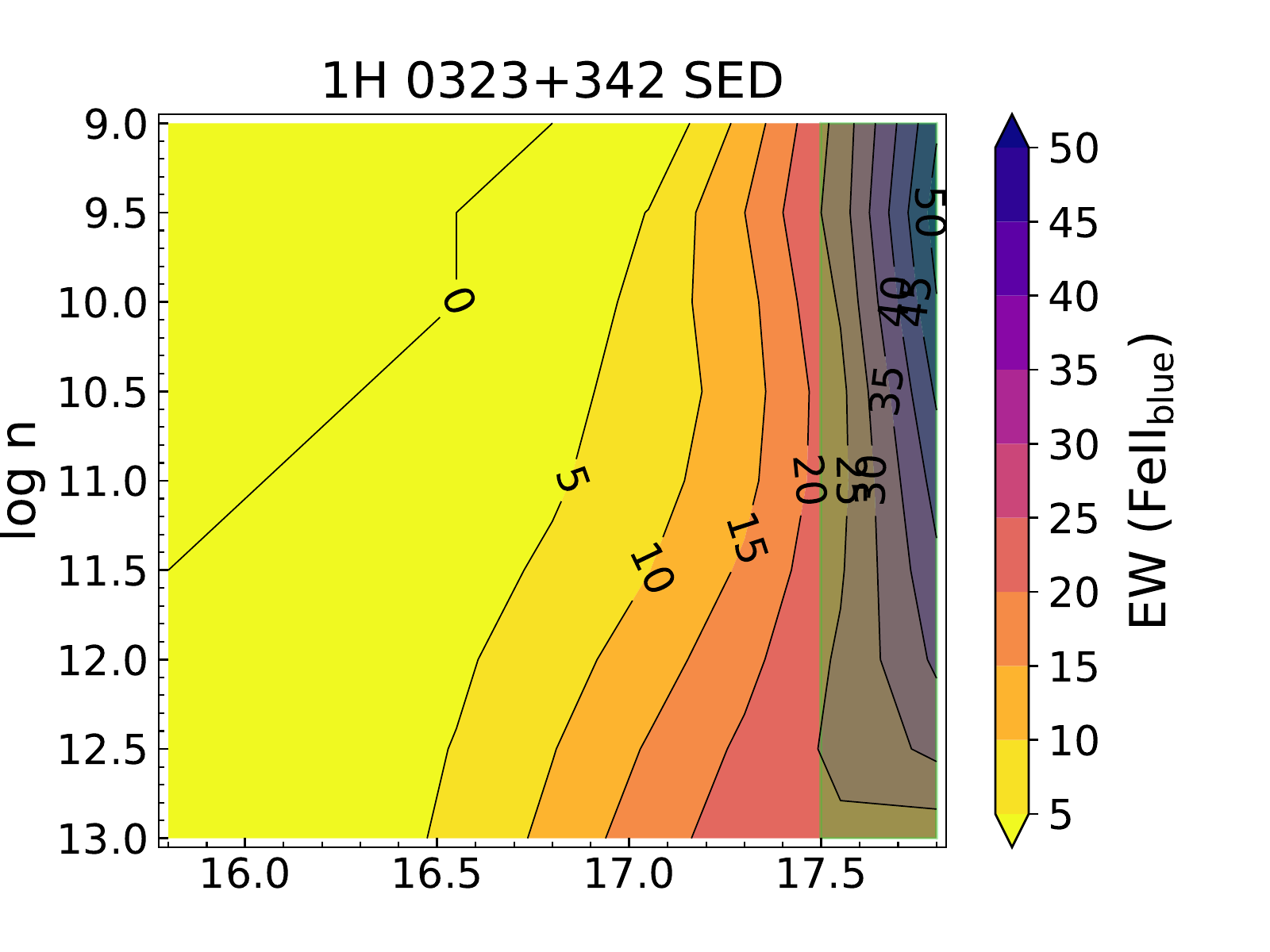}
\includegraphics[width=6.25 cm]{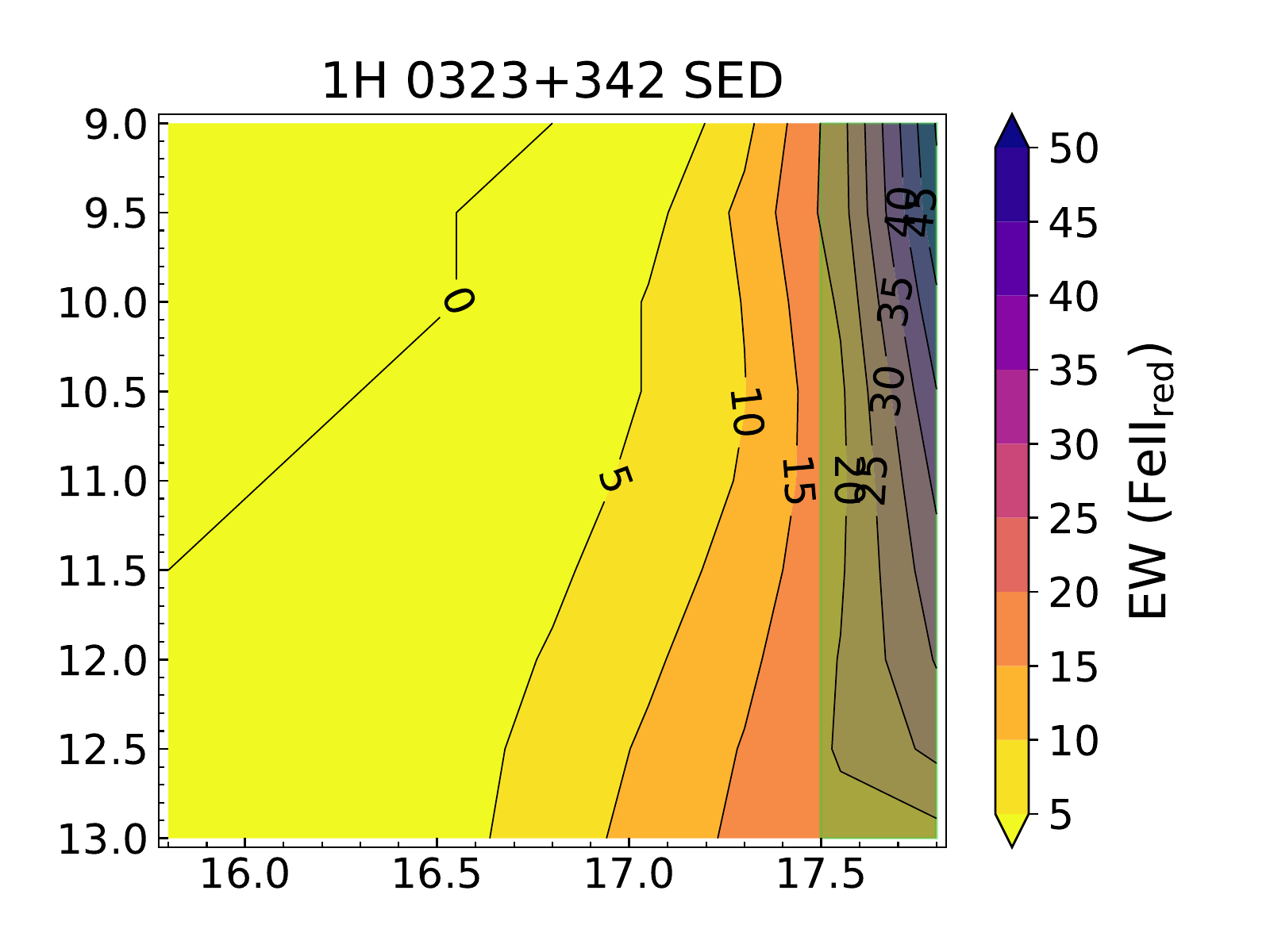}\\
\includegraphics[width=6.25 cm]{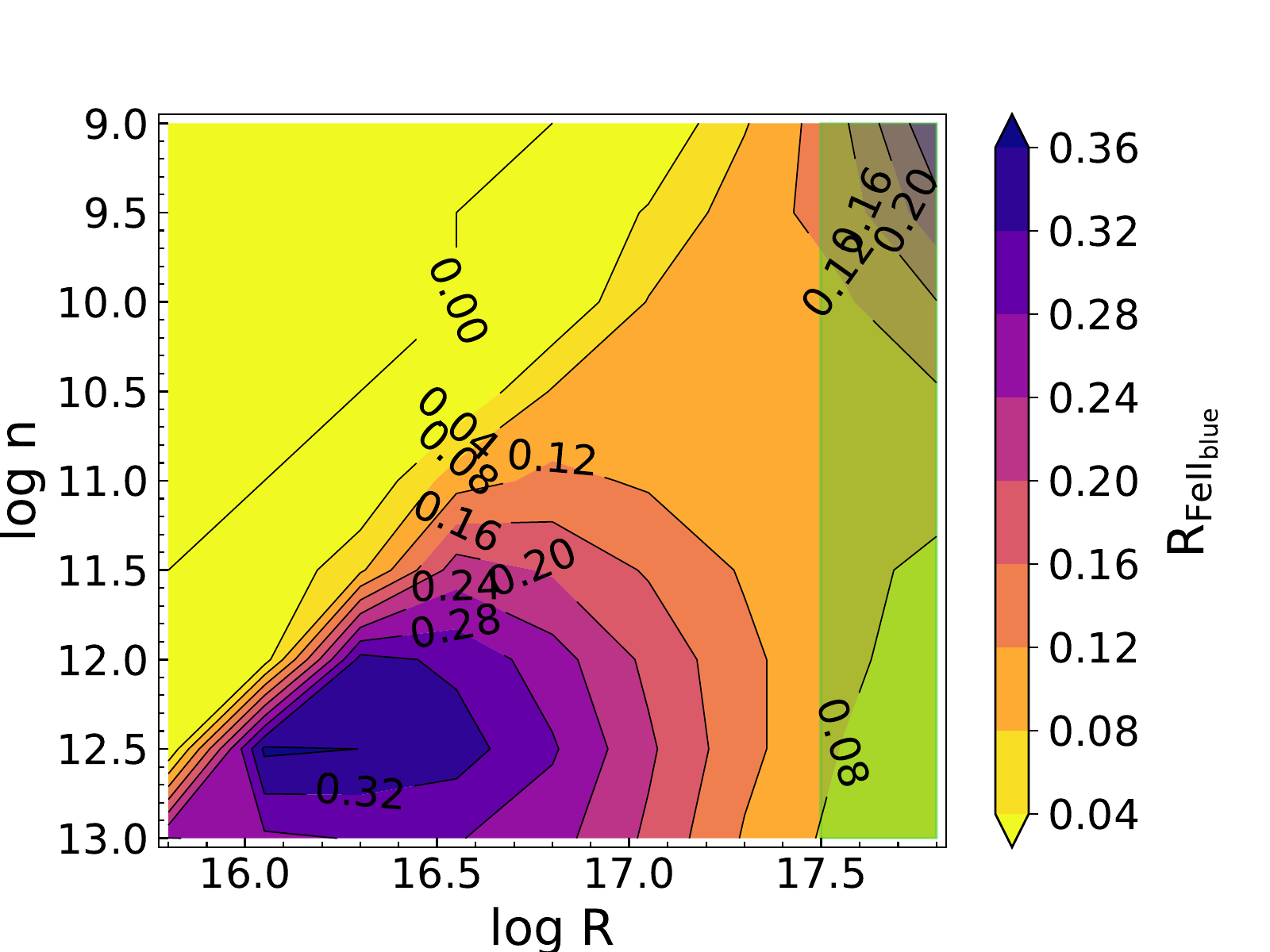}
\includegraphics[width=6.25 cm]{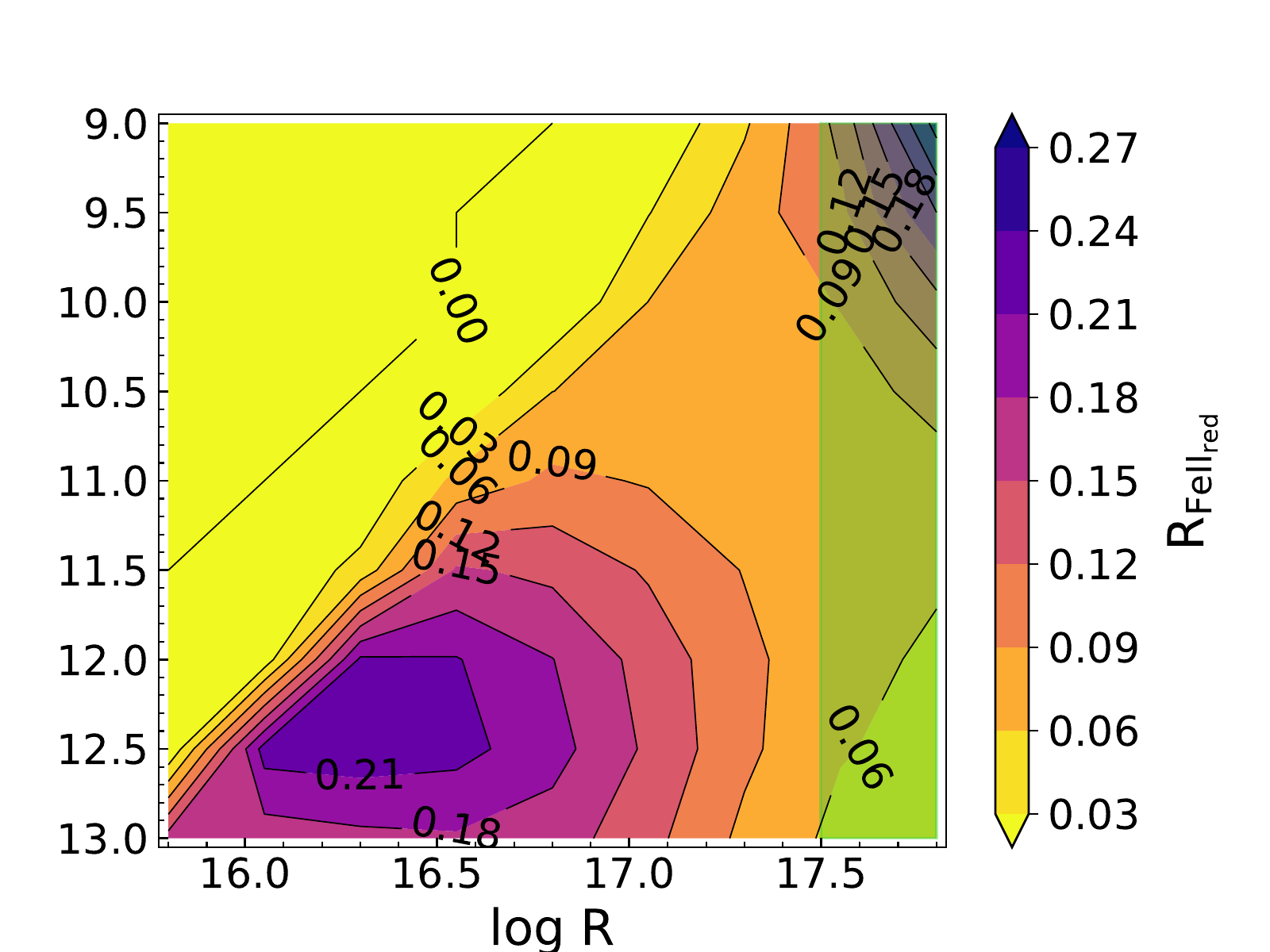}\\
\includegraphics[width=6.25 cm]{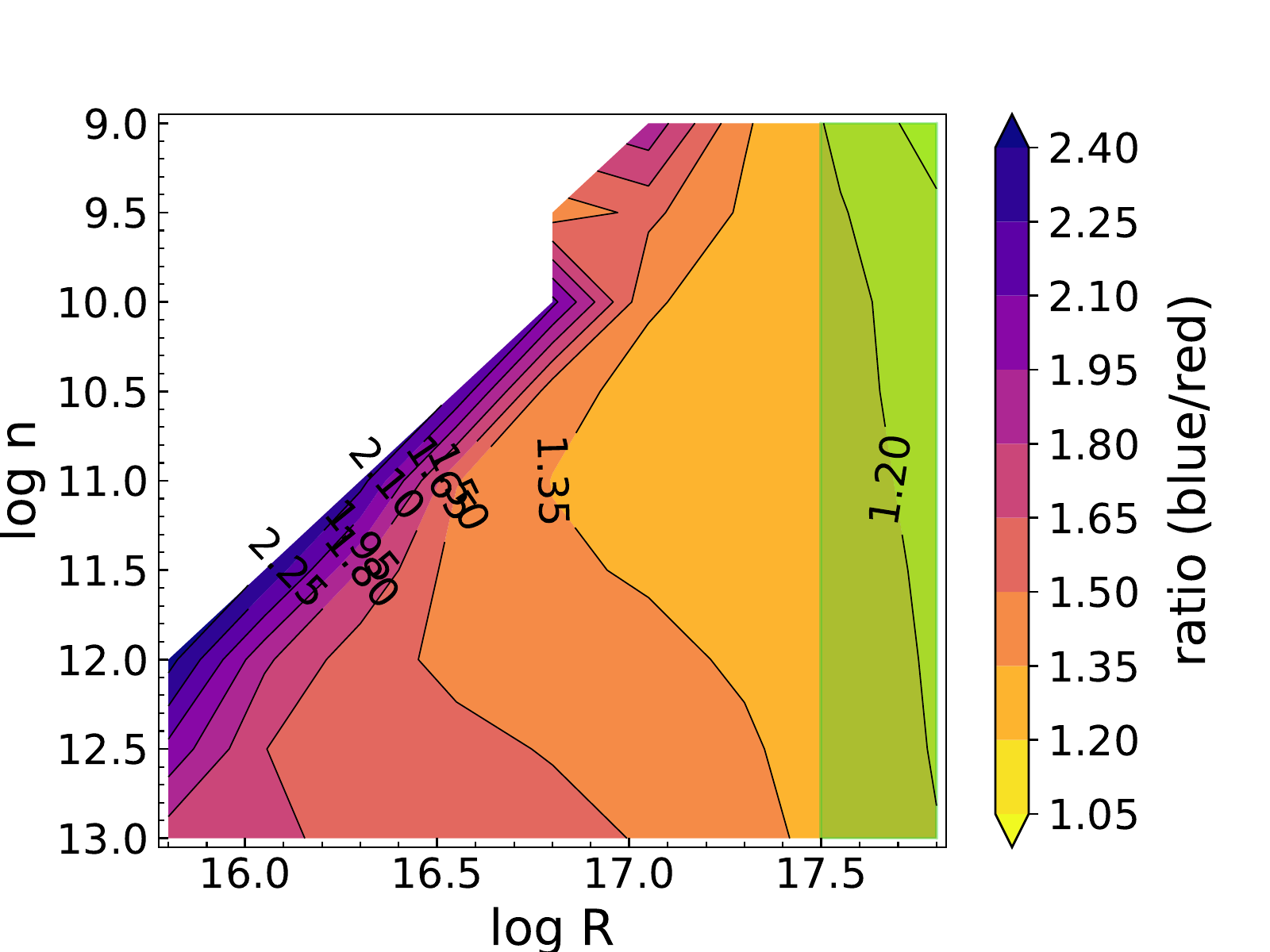}\\
\caption{Results of {\tt CLOUDY} simulations for the 1H0323+342 SED: the two top panel yields the equivalent width of \feiiq\ and the two middle panels   \rfe\ for the blue and red blends, as a function of density and BLR radius. The bottom panel shows the ratio of the blue and red \feii\ emission.   \label{fig:ph_comp_1H0321_res}}
\end{figure} 
\vfill

\end{document}